\documentclass[11pt]{article}
\usepackage{color,amsfonts,amsmath,amssymb,mathrsfs,verbatim,epsf}

\def\inn{\in}
\def\innf{\in}
\def\nsubst{\mbox{$\:\subset \!\!\!\!\! / \,\:$}}

\def\scirc{\mbox{\raisebox{0.2ex}{\scriptsize {$\circ$}}}}

\def\TM{\mbox{\it TM}}
\def\FM{\mbox{\it FM}}

\def\rrr{{\mathbb{R}}}
\def\ccc{{\mathbb{C}}}

\def\ooo{{\mathbb{O}}}

\def\b1{\mbox{\boldmath $1$}}
\def\v{\mbox{\boldmath $v$}}

\def\bh{\mbox{\boldmath $h$}}

\def\br{\mbox{\boldmath $r$}}
\def\bv{\mbox{\boldmath $v$}}
\def\bvh{\hat{\bv}}
\def\hG{\hat{G}}

\def\bx{\mbox{\boldmath $x$}}

\def\bu{\mbox{\boldmath $u$}}

\def\chc{\check c}
\def\che{\check e}
\def\chg{\check g}

\def\bR{\mbox{\boldmath $R$}}
\def\bT{\mbox{\boldmath $T$}}

\def\bsig{\boldsymbol {\sigma}}

\def\lvh{L(\hat{\v})=1}

\def\lvth{L(\v_3)=1}
\def\lvf{L(\v_4)=1}
\def\lvfh{L(\v_4) =h^2}

\def\lvni{L(\v_9)=1}
\def\lvte{L(\v_{10})=1}
\def\lvt{L(\v_{27})=1}
\def\lvfs{L(\v_{56})=1}
\def\lvtfe{L(\v_{248})=1}
\def\lvn{L(\v_n)=1}

\def\htwc{\mbox{h}_2\ccc}
\def\hthc{\mbox{h}_3\ccc}

\def\htwo{\mbox{h}_2\ooo}
\def\htho{\mbox{h}_3\ooo}

\def\glfrp{\mbox{GL}^+(4,\rrr)}

\def\sltc{\mbox{SL}(2,\ccc)}
\def\sltca{\mbox{sl}(2,\ccc)}

\def\slthc{\mbox{SL}(3,\ccc)}

\def\sltho{\mbox{SL}(3,\ooo)}

\def\sltwoo{\mbox{SL}(2,\ooo)}

\def\soth{\mbox{SO}(3)}

\def\uo{\mbox{U}(1)}

\def\sutw{\mbox{SU}(2)}

\def\suth{\mbox{SU}(3)}

\def\soot{\mbox{SO}^+(1,3)}
\def\soota{\mbox{so}^+(1,3)}

\def\sootn{\mbox{SO}^+(1,9)}

\def\ee{\mbox{E}_8}

\def\ese{\mbox{E}_7}

\def\esi{\mbox{E}_6}

\def\stab{\mbox{Stab}(\TM_4)}

\def\teta{e^{a}_{\phantom{i}\mu}(x)}

\def\ph{\phantom}

\def\lag{{\mathcal L}}

\def\mcX{{\mathcal X}}

\def\pal{\partial}
\def\fh{\frac{1}{2}}
\def\fhs{\mbox{\small{$\fh$}}}

\def\cstr{c^{\alpha}_{\ph{\alpha}\beta\gamma}}
\def\Gamabc{\Gamma^a_{\ph{a}bc}}
\def\Gamalp{\Gamma^{\alpha}_{\ph{\alpha}\beta\gamma}}

\def\gpath{ }

\def\setb{\setlength{\baselineskip}{0.625\baselineskip}}

\begin{document} 

{\setlength{\baselineskip}{0.625\baselineskip}

\begin{center}

 {\LARGE{\bf  Construction of a Kaluza-Klein type Theory}} \\
 \vspace{3pt}
 {\LARGE{\bf  from One Dimension}} \\
  
\bigskip

 \mbox {{\Large David J. Jackson}}\footnote{email: david.jackson.th@gmail.com}  \\
  
  \vspace{10pt}
 
 { \large September 5, 2016 }

 \vspace{12pt}

{\bf  Abstract}

\vspace{-2pt} 
 
\end{center}

       We describe how a physical theory incorporating the properties of fields deriving from 
extra-dimensional structures over a four-dimensional spacetime manifold can in principle be obtained 
through the analysis of a simple initial structure consisting of the one dimension of time alone, as 
represented by the real line. The simplicity of this starting point leads to symmetries of 
multi-dimensional forms of time, from which a geometrical structure can be derived which is similar to the 
framework employed in non-Abelian Kaluza-Klein theories. This leads to a relationship between the external 
and internal curvature on the spacetime manifold unified through the underlying constraint of the one 
dimension of time for the theory presented here. We also describe how the symmetry breaking structure is 
compatible with the Coleman-Mandula theorem for the subsequent quantisation of the theory.
 
{ 

\vspace{-0.1cm}

\tableofcontents

\vspace{0.9cm}

}

\section{Introduction and Motivation}
\label{one1}

    For many theories which aim to describe elementary empirical phenomena, such as 
summarised in the Standard Model of particle physics, a typical approach is to begin 
by postulating additional entities or structures on top of, or as an extension of, a 
4-dimensional spacetime background. For example a wide ranging class of models invokes  
the introduction of extra spatial dimensions, as have developed from the original 
proposal of Kaluza and Klein~\cite{Kaluza,Klein} and some of which we shall review in 
this paper.

   While one goal of any unification scheme is to incorporate a broad range of 
empirical phenomena collectively within a single framework, it is also generally 
desirable that the initial framework itself should be as simple as possible. That is, 
the theory should ideally be largely devoid of apparently arbitrary assumptions or 
postulated features. This is the point of view adopted for the present work in which 
we argue that a unified framework, sharing many of the properties of Kaluza-Klein 
theories, can be founded simply upon the structure of 
 the one dimension of time alone, as represented by
the real line.
   While the full development of this theory has been presented in~\cite{Unifi}
 with the uncovering of properties of the Standard Model described in \cite{Novel},
  which summarises (\cite{Unifi} chapters~6--9), here we elaborate upon the 
foundations of the theory and the elementary geometric structures involved as arising 
from the simple starting point of one dimension.
  Being largely self-contained this paper both summarises and expands upon the 
contents of (\cite{Unifi} chapters~2--5), with further discussion of the underlying 
conception of the theory in section~\ref{one2} and a more direct and explicit 
construction of the relation between the external and internal curvature presented 
here in section~\ref{one4}. The structure of the paper is further outlined below.

  In the following section we describe how analysis of simple arithmetic 
decompositions implicit in intervals of the real line itself leads directly to 
elementary  structures which exhibit a geometrical and spatial interpretation in 
several dimensions. Generalising from these observations the resulting 
multi-dimensional forms for the flow of time exhibit symmetry structures that allow 
both the identification of a 4-dimensional spacetime manifold together with apparent 
`extra dimensions'. This geometrical framework closely resembles that of non-Abelian 
Kaluza-Klein theories as constructed on the space of a principle fibre bundle. These 
latter structures are hence reviewed in section~\ref{one3}, in which much of our 
notation and general conventions will also be established. Combining the motivations 
of section~\ref{one2}
 with geometric arguments adapted from the  Kaluza-Klein theories of 
section~\ref{one3}, a means of constructing a relationship between the external 
spacetime geometry and the  internal curvature for the present theory is proposed in 
section~\ref{one4}. We further compare and contrast the new approach with Kaluza-Klein 
theory  in  section~\ref{one5}, where we also relate this work to 
references~\cite{Unifi} and \cite{Novel} and allude to the further development of the 
theory.

  While not describing details of the quantisation of the theory in this paper, in 
subsection~\ref{one53}, with reference to section~\ref{one2}, we also address the 
compatibility of the elementary symmetry breaking structures in the theory with the 
Coleman-Mandula theorem, which concerns the possible symmetries for a relativistic 
theory of interacting particles. It is necessary to address this question since both 
the external Lorentz symmetry and the internal gauge symmetry originate from a common  
unifying simple group in the mathematical construction of the theory, with physical 
structures derived through the breaking of the full symmetry.  

\section{Elementary Structure of the Theory}
\label{one2}

\subsection{Time and Spatial Dimensions}
\label{one21}

   We begin by considering a finite interval of time represented by the real number $s 
\inn \rrr$. Amongst the myriad of ways of expressing $s$ in terms of other real 
numbers in accordance with the basic rules of arithmetic one possibility is the 
quadratic composition:
\begin{equation}
 \label{sfin}
   s^2 = (x^1)^2  + (x^2)^2 + (x^3)^2 = \eta_{ab} x^a x^b
\end{equation}
  Here $s^2$ is the square of $s \inn \rrr$ while $x^1$ $x^2$ and $x^3$ are three 
further real numbers and $\eta_{ab}$, with $a,b = 1\ldots 3$, are the components of 
the $3 \times 3$ diagonal unit matrix, with the conventional summation over repeated 
indices implied. In the above expression the left-hand side $s^2$ is invariant under 
the action of the orthogonal group $\soth$  on the ordered set of  numerical 
components 
 $(x^1,x^2,x^3)$, which can be \textit{interpreted} as a 3-vector
 sweeping out a spherical shell
  in a 3-dimensional geometrical space.

  This geometrical structure provides a simple example demonstrating how  possible 
arithmetic compositions of a real interval $s \inn \rrr$ can exhibit a form and 
symmetry with a multi-dimensional \textit{spatial} interpretation. A natural 
generalisation of equation~\ref{sfin} can also be considered with $s^2 = \eta_{ab} x^a 
x^b$ for $a,b = 1\ldots n$ for arbitrary $n$ and unit $n \times n$ matrix $\eta$. In 
this case the corresponding symmetry transformations SO($n$) on the components $(x^1, 
\ldots, x^n)$ describe an implicit $n$-dimensional spatial structure. The case  
 with a metric $\eta$ of arbitrary signature and an SO($p,q$) symmetry, with $p+q=n$, 
marks a further generalisation.
  We can also consider the case for the limit of an infinitesimal interval denoted by 
$s \to \delta s \inn \rrr$ for which equation~\ref{sfin} becomes simply:
\begin{equation}
 \label{sfind}
  (\delta s)^2 = (\delta x^1)^2  + (\delta x^2)^2
         + (\delta x^3)^2 = \eta_{ab} \delta x^a \delta x^b
\end{equation}
 In this case the uniformity of the quadratic order of the expression is now 
\textit{required}, to balance the order of the infinitesimal quantities in each term, 
while the property of the invariance of  $(\delta s)^2$ under  $\soth$ transformations 
as described for equation~\ref{sfin}, now applied to the components $\delta x^a$ in 
equation~\ref{sfind},  is retained. 

  However, a further symmetry is also apparent for the new expression, namely an 
invariance under the translation $x^a \to x^a +r^a$ for any constant vector 
$\br_{\;\!\!3} = (r^1,r^2,r^3) \inn \rrr^3$; with $\delta x^a \to \delta (x^a + r^a) = 
\delta x^a$ under this simple transformation.
   That is, while  for the original case with only one dimension  the quantity $\delta 
s$ can be conceived of as an infinitesimal interval \textit{anywhere} on the real line 
by the symmetry of $\rrr$, as depicted in figure~\ref{spillout}(a),
 similarly each $\delta x^a$ expresses an infinitesimal interval \textit{anywhere} on 
the real line with each $x^a \inn \rrr$. Collectively the translation symmetry of the 
right-hand side of equation~\ref{sfind}
 over $\rrr^3$ describes the parameter space depicted in figure~\ref{spillout}(b), 
with the interval $\delta s$ and the relation 
 $\delta s^2 = \eta_{ab} \delta x^a \delta x^b$ composed at any point
  $(x^1,x^2,x^3) \inn \rrr^3 \equiv M_3$. Through this $\rrr^3$ translation symmetry 
both the metric structure with 
  $\eta = \mbox{diag}(+1,+1,+1)$ and the  $\soth$ rotation symmetry of 
equation~\ref{sfind} are exhibited locally throughout the extended manifold $M_3$. 

\begin{figure}[htbp]  
\centering
\hspace*{-16pt}
\epsfxsize=15.4cm
\leavevmode
\epsffile[0 0 2503 820]{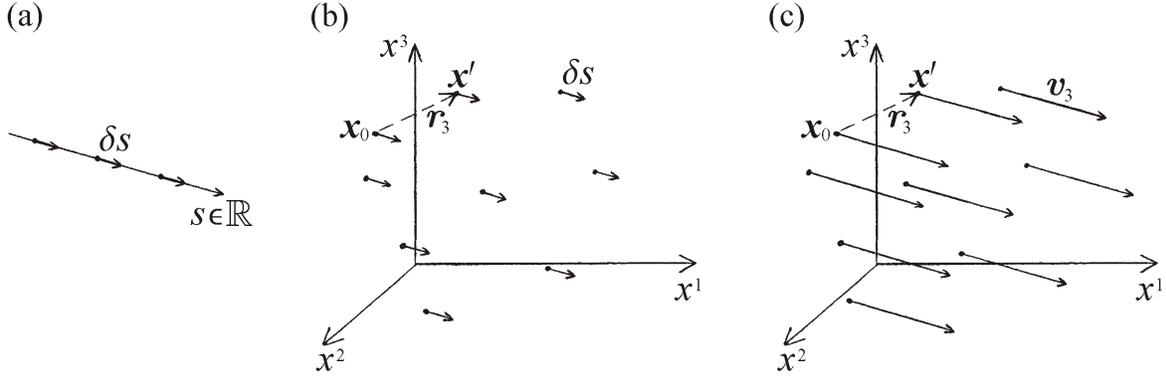}
\caption{\setb  (a) The infinitesimal interval $\delta s$, represented by the short 
arrows, resides at any location on the real line $\rrr$. (b) Similarly, since the  
real variables $\{ x^a \} \innf  \rrr^3$ of equation~\ref{sfind} are arbitrary
 this equation applies equally for the particular value $\bx_0 \innf \rrr^3$ as  for 
$\bx^{\prime} = \bx_0 + \br_{\;\!\!3}$ and over the 
range $-\infty < r^a < \infty$ for each of $a = 1,2,3$. (c) An equivalent observation 
is made for equation~\ref{lvthree}, with the finite 3-vector $\bv_3$ in $\lvth$ 
represented here by the longer arrows.  It is through this translation symmetry that a 
`base manifold' $M_3 \equiv \rrr^3$ may be identified.}
\label{spillout}
\end{figure}

    Generalising from equation~\ref{sfin} 
	a finite duration of time,
	represented by an interval of the real line $s \inn \rrr$,  can be equated with an 
arithmetic composition of $n$ further real numbers 
$x^a$ $(a = 1\dots n)$ in a large variety of ways, including inhomogeneous polynomial 
expressions. However on taking the infinitesimal limit such expressions are 
constrained to homogeneous $p^{\mathrm{th}}$ order polynomials of the form:
\begin{equation}
  \label{seqxxx}
   (\delta s)^{p} = \alpha_{abc\ldots} \delta x^a \delta x^b \delta x^c \ldots 
\end{equation}
  where $p$ is a power, $a,b,c\ldots$ are $p$ indices with values $1\ldots n$, and 
  with each coefficient $\alpha_{abc\ldots} \inn \{-1,0,1\}$, such that each non-zero 
term is of the same order in infinitesimal elements, generalising from 
equation~\ref{sfind}. Further, it is possible to avoid dealing directly with such 
infinitesimal quantities and express  
 equation~\ref{seqxxx}  itself  in terms of generally finite quantities by dividing 
both sides by $(\delta s)^p$ and defining the $n$-dimensional vector $\bv_n$ with 
components $v^a = \delta x^a / \delta s$ for the limit $\delta s \to 0$.
   This leads directly to the general homogeneous polynomial form (as described for 
\cite{Unifi}~equation~2.9 and \cite{Novel}~equation~11):
\begin{equation}
 \label{lvn}
  L(\bv_n) := \alpha_{abc\ldots} v^a v^b v^c \ldots  = 1
\end{equation}

 For the trivial 1-dimensional case with $s=x^1$ and $v^1= \frac{dx^1}{ds} \equiv 
\frac{\delta x^1}{\delta s}{\vert}_{\delta s \to 0}  = 1$  the symmetry $x^1 \to 
x^1+r^1$  can be readily visualised as a flow $v^1$ present everywhere on the real 
line parametrised by $x^1\inn\rrr$, by close analogy with figure~\ref{spillout}(a).  
In the general case for $n$ dimensions the vector $\bv_n$ is invariant under 
translations of the form:
\begin{equation}
  \label{rspill}
      \bv_n \; = \;  \bigg\{\frac{d(x^1+r^1)}{ds},\frac{d(x^2+r^2)}{ds},\ldots  
\frac{d(x^n+r^n)}{ds}\bigg\} 
\end{equation} 
  Since this equation is equally valid for all possible constant
   $\br_{\! n} = (r^1, \ldots, r^n) \inn \rrr^n$  the condition $L(\bv_n)=1$ of 
equation~\ref{lvn} implicitly holds over the entire $\rrr^n \equiv M_n$ manifold. This 
structure is represented in figure~\ref{spillout}(c) for
 the 3-dimensional case with
\begin{equation}
\label{lvthree} 
  L(\bv_3) = (v^1)^2 + (v^2)^2 + (v^3)^2 = \eta_{ab}v^av^b = 1
\end{equation}  
   which is equivalent to equation~\ref{sfind} and with which it shares the same 
translation symmetry  
 as described for figure~\ref{spillout}(b).
  These figures exemplify  the elementary structure of the present theory, here for 
the particular case with a base space $M_3$ arising from the symmetries of the 
quadratic form $\lvth$ 
  as a particular multi-dimensional expression for the original one-dimensional flow 
of time. 

  Further, consistent with  the generalisation described immediately before 
equation~\ref{sfind} (and with the general form of equation~\ref{seqxxx}) we can write 
a quadratic expression with Minkowski metric
 $\eta = \mbox{diag}(+1,-1,-1,-1)$ in four dimensions (with $a,b = 0\ldots 3$):   
\begin{eqnarray}
 \label{minko}
  (\delta s)^2 & = & (\delta x^0)^2  - (\delta x^1)^2  - (\delta x^2)^2
         - (\delta x^3)^2 \; = \; \eta_{ab} \delta x^a \delta x^b 
		  \label{lors} \\
 \mbox{that is:} \quad
  L(\bv_4) & = & (v^0)^2 \: - \: (v^1)^2 \: - \: (v^2)^2
        \: - \: (v^3)^2 \; = \; \eta_{ab} v^a v^b  \; = \; 1 	 \quad
		 \label{lorl}	 
\end{eqnarray}
  For the corresponding 4-dimensional Lorentzian extension of figure~\ref{spillout}(c) 
the Minkowski metric $\eta$ is imported throughout the space $M_4 \equiv \rrr^4$ 
through the translation symmetry $x^a \to x^a + r^a$ for
  $\br_{\;\!\! 4} =  (r^0,r^1,r^2,r^3) \inn \rrr^4$ of  $\lvf$. The $\soot$ symmetry 
of equations~\ref{lors} and \ref{lorl} is also everywhere imported locally onto the 
manifold $M_4$.

  However, rather than taking quadratic space or spacetime forms as fundamental we 
take a different perspective and treat the time interval $\delta s$ on the 
\textit{left-hand} side of equation~\ref{lors} as the underlying basic entity of the 
theory. From this point of view equation~\ref{lors} is interpreted as  one of many 
possible arithmetic decompositions of a one-dimensional temporal interval
 within the more general form of equation~\ref{seqxxx}.    
For any homogeneous form $\lvn$, including the cubic and higher polynomial expressions 
implied in this generalisation to equation~\ref{lvn}, the translation symmetry in 
$\br_{\! n} \inn \rrr^n$ of equation~\ref{rspill} can also be identified and  
represented  by an $n$-dimensional parameter space, similarly as depicted in 
figure~\ref{spillout}(c), however in general \textit{without} a quadratic `metric' 
structure as for the case of equations~\ref{lvthree} and \ref{lorl}.

 On the other hand even for cubic or higher degree polynomial  forms $\lvn$ a subspace 
structure for a subset of the components of $\bv_n$ may exhibit a metrical form, and 
this observation will be closely associated with the breaking of the symmetry of the 
full form $\lvn$. Indeed if an underlying metric with components $\eta_{ab}$ can be 
identified within the coefficients $\alpha_{abc\ldots}$ of equation~\ref{lvn} the 
manifold constructed from the translation symmetry associated with corresponding 
subcomponents of $\bv_n$ can be directly identified as an extended geometrical 
manifold with local metric $\eta$, or even in principle as a physical `spacetime' 
manifold -- similarly as for the full translation symmetry of the form $\lvf$ with the 
metric of equation~\ref{lorl}.
 We next further analyse the geometric structure of $M_4$ for this simplest case 
before considering a higher-dimensional extension in subsection~\ref{one23}.

\subsection{Geometry of the Spacetime Manifold}
\label{one22}

 The `base manifold' $M_4$ originates from the continuous $4$-dimensional parameter 
space of the $4$-dimensional translational freedom of the form $\lvf$ in 
equation~\ref{lorl},
 which is trivially invariant under $x^a \to x^a + r^a$ for the four components $v^a = 
dx^a/ds$ with $a = 0\ldots 3$, as described more generally for equation~\ref{rspill}.
The set of four variables $\{r^a\} \inn \rrr^4$ can be identified with an initial set 
of four coordinates $\{x^{\mu}\} \inn \rrr^4$, with \mbox{$x^{\mu} = 
\delta^{\mu}_{\ph{\mu}a}r^a$}
 (Greek indices $\{\mu,\nu,\ldots\}$  denote general coordinates while Latin indices 
$\{a,b,\ldots\}$  denote any frame in general or an orthonormal frame in particular
 on $M_4$ depending on the context).

 The Lorentzian structure of the vector space of $\bv_4 \inn \rrr^{1,3}$  is 
transferred onto the tangent space of the manifold $M_4$, via the translation 
symmetry,  and hence this space acquires the properties of a 4-dimensional 
pseudo-Riemannian manifold. That is, with
 the vector field $\bv_4(x)$ naturally residing in the tangent space $\TM_4$ to the 
manifold, with components $v^a = dx^a/ds$, a metric on $M_4$  derives locally  from 
the pseudo-Euclidean form $L(\bv_4) = \eta_{ab}v^a v^b = 1$ of equation~\ref{lorl}, as 
can be described by the metric components $g_{\mu\nu}$ in a general coordinate system 
on $M_4$ via a tetrad field $\teta$:
 \begin{equation}
    g_{\mu\nu} = e^{a}_{\ph{a}\mu} e^{b}_{\ph{a}\nu} \eta_{ab}  \label{guneen}
  \end{equation}
  Hence the manifold $M_4$ inherits its pseudo-Riemannian structure \textit{from} the 
Lorentz symmetry of $L(\bv_4) = 1$.
 Incorporating a spatial $\soth$ symmetry, similarly as described for 
figure~\ref{spillout}(c),  
  the  structure of the form $\lvf$ and its symmetries  contains the skeletal form of 
a mathematical framework for the description of an apparently external and extended 
geometrical structure with the Minkowski metric $\eta_{ab}$ imported from 
equation~\ref{lorl}.

  For such a manifold in which there exist global coordinates such that $g_{\mu\nu}(x) 
= \mbox{diag}(+1,-1,-1,-1)$ for all $x \inn M_4$, that is the constant Minkowski 
metric, we have the 4-dimensional spacetime of special relativity. In this case the 
local metric $\eta_{ab}$  has been drawn out globally through the existence of large 
scale coordinates with respect to which the tetrad field can be  expressed
 simply as $\teta = \delta^a_{\ph{a}\mu}$ in equation~\ref{guneen}.
  In the following we formalise this notion of a flat manifold by considering the 
geometry of  the Lorentz symmetry group $\soot$ of the form $\lvf$ in relation to that 
of the manifold $M_4$, which has been identified through the translation symmetry of 
the same 4-dimensional form of time. 
 
  We first note that
  in contrast to the `translational' symmetries of a form $\lvn$, acting directly on 
the components of $\bx_n$ as described in equation~\ref{rspill}, groups such as 
$\soot$ acting directly on the components of $\bv_n$ will be referred to generically 
as `isochronal' symmetries,  which encompass both rotations and boosts as well as 
other transformations for higher symmetry groups leaving a higher-dimensional form of 
time $\lvn$ invariant. 
  In general the highest-dimensional form under consideration will be denoted by 
$\lvh$ with the full isochronal symmetry $\hat{G}$. The full set of isochronal 
$\sigma_h$, with $h \inn \hat{G}$,  and translational
  $\{r^a\}$, with $\br_{\! n} \inn \rrr^n$,   symmetries act on the components of 
$\lvh$
   as:
  \begin{equation}
     \label{lv2syms}
	   L\left(\sigma_h \left\{ \frac{d(x^a + r^a)}{ds} \right\} \right) = 1
  \end{equation}

  In this subsection we are considering the relation between the full isochronal 
$\hat{G}= \soot$ and full translational $\{r^a\} \inn \rrr^4$ symmetries of the form 
$\lvf$.
 The metric structure on the base manifold $(M_4, g)$ arising from the translation 
symmetry of equation~\ref{lorl}, as described above for equation~\ref{guneen}, is 
represented by the rectangular box employed for the base space in 
figures~\ref{mtogmap}(a) and (b).  It is the local Minkowski metric on $M_4$ that 
allows this structure to be interpreted as an extended \textit{spacetime} manifold 
rather than simply as a real parameter space $\rrr^4$ alone.

 The  manifold of a Lie group, such as $\hat{G}= \soot$, itself also exhibits a 
characteristic geometrical structure.  
The Maurer-Cartan 1-form $\theta$ is a canonical object on any Lie group manifold  
that, as a Lie algebra-valued 1-form, satisfies the Maurer-Cartan structure equation
 (see for example~\cite{kob,amp,fecko}): 
\begin{equation}
  \label{maca2}
  \mbox{d}\theta + \frac{1}{2} [\theta,\theta]    =  0
\end{equation}
   where `d' denotes the exterior derivative and the square brackets denote  the 
exterior product for Lie algebra-valued 1-forms.

The two manifolds $M_4$ and $\hat{G} = \soot$, respectively representing the 
translational and isochronal symmetries of the form $\lvf$, are linked through the 
mapping \mbox{$h(x): M_4 \to \hat{G}$} as depicted in figure~\ref{mtogmap}(a). An 
initial orthonormal frame field $\{e_a(x)\}$, with respect to the Minkowski metric 
$\eta$ on $M_4$, can be transformed to any other orthonormal frame field $\{e'_b(x)\}$ 
by the matrix action $e'_b(x) = e_a(x) h^a_{\ph{a}b}(x)$  via the group elements 
\mbox{$h(x)\inn\soot$}, which can be considered as a `gauge' freedom, at every $x\inn 
M$. Hence the map $h(x): M_4 \to \hat{G}$ expresses the local choice of an orthonormal 
frame field $\{e_a(x)\}$ which, since it can be chosen arbitrarily,  in general  will 
not represent parallelism on the base manifold.
\vspace{-3pt}
\begin{figure}[htbp]  
\centering
\epsfxsize=12.5cm
\leavevmode
\epsffile[0 0 1776 825]{\gpath aKfig2e}
\vspace{-6pt}
\caption{\setb (a) With $M_4 \equiv \rrr^4$ deriving from the full translational 
symmetry and $\hat{G} = \soot$ as the full isochronal symmetry of the form $\lvf$, the 
local choice of an orthonormal frame at each $x\in M_4$ can be described by a map
 \mbox{$h(x): M_4 \to \hat{G}$} between elements of the two manifolds. (b) Here the 
full group $\hat{G} = \soot$ acts upon $\bv_4 \in \rrr^{1,3}$ and objects defined on 
the tangent space $\TM_4$ generally.}
\label{mtogmap}
\end{figure}

  Since the operations of the exterior algebra of $p$-forms are preserved under the 
pull-back of forms through smooth maps between manifolds the Lie algebra-valued 
1-form:
\begin{equation}
  \label{aegasth}
 A(x) = h^{\ast}\theta(h)
\end{equation}
  on $M_4$ captures the structural properties of the Maurer-Cartan 1-form $\theta$ on 
$\hat{G}$ relative to the map $h(x): M_4 \to \hat{G}$. While on $\hat{G}$ we have the 
linear map $\langle \theta, V(h)\rangle \inn L(\hG)$ from $V(h)\inn T_h\hat{G}$ into 
the Lie algebra of $\hG$, on $M_4$ we have the linear map $\langle A(x), \bu(x) 
\rangle \inn L(\hG)$ from $\bu(x) \inn T_xM_4$ into the same Lie algebra. The Lie 
algebra-valued 1-form $A(x)$  may be written as $A(x) = A^{\alpha}_{\ph{\alpha}\mu}(x) 
\, X_{\alpha} \,\mbox{d}x^{\mu} $ where $\{ \mbox{d}x^{\mu} \}$ is a coordinate basis 
of 1-forms on $M_4$ and $\{X_{\alpha}\}$ is a basis for $L(\hG)$.

   Unlike the canonical 1-form on $\hG$ (which can be written $\theta =  
X_{\alpha}\theta^{\alpha}$ where $\{\theta^{\alpha}\}$ is the basis of 1-forms on 
$\hG$ dual to the basis $\{X_{\alpha}\}$) the 1-form $A$ of equation~\ref{aegasth} on 
$M_4$ has variable real coefficients $A^{\alpha}_{\ph{\alpha}\mu}(x)$ which, however, 
are not arbitrary but depend upon the choice of gauge function $h(x)$ as well as upon 
the choice of coordinates $\{x^{\mu}\}$ on $M_4$. Explicitly, for the matrix group 
$\hG = \soot$, the 1-form $A = h^{\ast}\theta$ on $M_4$ can be written in terms of the 
matrices $h(x) \in \hG$ as:
\begin{equation}
  \label{puga}
     A(x) = h^{-1} \mbox{d}h = h^{-1} \frac {\pal h}{\pal x^{\mu}} \mbox{d}x^{\mu}
\end{equation}
   This canonical mathematical object can be interpreted as a \textit{connection 
1-form} on the base manifold $M_4$, to be described more generally in 
subsection~\ref{one32}, formalising the notion of parallelism in a manner which will 
naturally generalise for the case of finite curvature. Here it is possible to choose a 
gauge with $A(x) = 0$ everywhere on $M_4$, simply by taking $h(x)$ to be constant in 
equation~\ref{puga}, and hence we have a flat connection. Indeed, this connection can 
always be written in terms of `pure gauge', as it is in equation~\ref{puga}, which is 
one way of defining a flat connection.

 By the homomorphism of exterior algebra relations across the pull-back map the 
Lorentz Lie algebra-valued 1-form $A = h^{\ast}\theta$ is also subject to a structure 
equation corresponding to equation~\ref{maca2}, that is:
\begin{equation}
 \label{maca3}
   \mbox{d}A + \frac{1}{2} [A,A]    =  0  
\end{equation}
In general the curvature 2-form $F$ on the base manifold can be expressed as:
\begin{equation}  
   F =  \mbox{d}A + \frac{1}{2} [A,A]    \label{fdefaaa}
\end{equation}
    which transforms under a gauge change $h(x)$ as $F\to F' = h^{-1} F h = 
\mbox{Ad}(h^{-1})F$, that is under the adjoint representation. Equations~\ref{maca3} 
and \ref{fdefaaa} then immediately show that the curvature is equal to zero, with 
$F=0$ in any gauge, and further expresses the global parallelism implied by the 
canonical flat connection of equation~\ref{aegasth}.
  Since here $F$ is the external Riemann curvature expressed in an orthonormal frame 
field, the full Riemann tensor vanishes in any general coordinate frame on the 
manifold $M_4$.

 The group $\hG=\soot$ was introduced as the isochronal symmetry action on the form 
$\lvf$ and hence the Lie algebra values of $A(x)$ and $F(x)$ are composed from a basis 
of $4 \times 4$ matrices
 $\{E_{\alpha}\}$ 
 in a representation of $L(\hG)$ acting naturally upon the vectors $\bu \inn \TM_4$, 
that is on the tangent space of the base manifold, and in particular on the vector 
$\bv_4 \inn \TM_4$ originating in the form $\lvf$ of equation~\ref{lorl},
 as depicted in figure~\ref{mtogmap}(b).
  The parameter space $x\inn M_4$ itself arose from the translational symmetry of 
$\lvf$ as described in equation~\ref{rspill}.
  The mathematical objects involved are hence intimately associated with each other, 
deriving from the symmetries summarised in equation~\ref{lv2syms}.

If the 4-dimensional form of time is embedded in a higher-dimensional form $\lvh$ the 
full symmetry $\hat{G}$ of the larger form will be broken on employing only the 
4-dimensional component of the full translational symmetry  to generate the manifold 
$M_4$, exhibiting a structure that can be interpreted as an external spacetime arena 
with a local Minkowski metric.
 In either case
 the flow of time is \textit{not} considered to be projected onto a pre-existing 
4-dimensional spacetime manifold, rather the extended structure $M_4$ itself is 
implicit \textit{within} the symmetries identified for a multi-dimensional form of 
temporal flow, originally deriving from the interval $\delta s \inn \rrr$ itself as 
described for equation~\ref{lvn}. 
 That is, the underlying geometrical structure derives from the isochronal and 
translational symmetries of the possible forms $\lvh$ in the infinitesimal limit 
 $s \to \delta s$ for an interval of time.

  For the case of $M_4$ deriving from the symmetries of the form $\lvf$  while the 
connection form $A(x)$ is gauge-dependent the external curvature is zero in any frame, 
as described for equations~\ref{puga}--\ref{fdefaaa}. However for the case in which 
the Lorentzian manifold is identified through a 4-dimensional subset of the 
translational symmetry of a higher-dimensional form $\lvh$ it will be possible, given 
the extra degrees of freedom, to obtain a finite external curvature. Further,
the components of the higher dimensions and symmetries of $\lvh$, over and above those 
required to identify the Lorentz vector $\bv_4$ and the 4-dimensional spacetime 
manifold, will give rise to fields on the extended space $M_4$ which can be 
interpreted as a `matter' content. In this paper we focus upon the gauge field 
component arising from the \textit{internal} symmetry $G$ of the additional dimensions 
arising from the breaking of the full symmetry $\hat{G}$ of $\lvh$ over the 
\textit{external} spacetime $M_4$.  
   For this extended case the geometry of the  internal gauge curvature can also be 
investigated, and in fact exhibits a structure correlated with the finite external 
curvature as we explore in  this paper.

\subsection{Extra Dimensions and Symmetry Breaking}
\label{one23}
 
  On considering a higher-dimensional temporal form $\lvh$
the mathematical basis for obtaining  an extended base manifold is found in the 
application of the symmetry described in equation~\ref{rspill} and 
figure~\ref{spillout}(c) to a 4-dimensional  subset of the translational degrees of 
freedom exhibiting a Minkowski metrical form, as alluded to at the end of the previous 
subsection.
  This translation symmetry of such  a multi-dimensional form 
  is the underlying means through which the one-dimensional flow of time can in 
principle  be exhibited  simultaneously as a  flow of physical fields in an extended 
spacetime, as we consider in this subsection.

 We approach the generalisation for $\lvh$
   via the group $\sltc$ as the double cover of $\soot$, which may be introduced by 
first mapping a Lorentz vector $\bv_4 \inn \rrr^{1,3}$ into the space of $2\times 2$ 
complex Hermitian matrices $\htwc$ as:
\begin{equation}
\label{vtoh}
  \v_4 = (v^0,v^1,v^2,v^3)\; \to\; \bh_2 = \bv_4 \! \cdot \! \bsig =
   \left( \begin{array}{cc} v^0+v^3 & v^1-v^2i \\ v^1+v^2i & v^0-v^3 \end{array}  
\right) \in \htwc
\end{equation}
 where $\bsig$ denotes the $2\times 2$ identity matrix $\sigma^0$ together with the 
three Pauli matrices, that is $\sigma^1 = \binom{0 \;\; 1}{1 \;\; 0}$,  $\sigma^2 = 
\binom{0 \,\:\! -\;\!\! i}{i \;\,\; 0}$, $\sigma^3 = \binom{1 \;\,\; 0}{0 \,\, -\! 
1}$,
 as described for (\cite{Novel} equation 17).
  Under the mapping of equation~\ref{vtoh} we have $L(\bv_4) \to \det(\bh_2)$ for the 
quadratic form of equation~\ref{lorl}.
 While the fundamental representation of $\sltc$ acts on the spinor space $\ccc^2$ the 
group action for elements $S\inn \sltc$ on the space $\htwc$ describes the vector 
representation given by: 
\begin{equation}
 \label{hshs}
  \bh_2 \to  \bh_2^{\prime} =  S \, \bh_2 \, S^{\dagger}
\end{equation} 
  This maps $\bh_2 \to \bh_2^{\prime}$ onto a new $2\times 2$ complex Hermitian matrix 
while preserving the value of the determinant; hence mapping the corresponding 
components $v^a \to v^{\prime a}$ according to a Lorentz transformation of the real 
4-vector $\v_4\inn\rrr^{1,3}$.
 This $\sltc$ action expresses the symmetry of $\lvf$ of equation~\ref{lorl} in a 
manner that naturally extends to an $\slthc$ symmetry of the cubic polynomial form:
\begin{equation} 
 \label{lvnine}
  L(\bv_9) = \det(\bv_9) = 1 \quad \mbox{with} \quad \bv_9  \inn \hthc
\end{equation}

  The extension to this cubic expression, together with the set of symmetry 
transformations under which it is invariant, places  the emphasis on generalising
 expressions for $\delta s$ on
 the left-hand side of  equation~\ref{minko}, \textit{without} being restricted to  
quadratic compositions  as for the right-hand side of that equation.
 That is, equation~\ref{lvnine}
is considered necessarily as a higher-dimensional form of \textit{time} rather than 
\textit{spacetime}.

  While the  symmetry group $\sltc$, as a subgroup of $\slthc$, acts on the subspace
  $\htwc$ embedded naturally within $\hthc$
  the action of 
 $\sltc$ on the full space $\hthc$ can also be considered. The $2 \times 2$ matrices 
$S\inn \sltc$ can be embedded in $3 \times 3$ matrices acting on $\bv_9 \inn \hthc$ 
as:
\begin{equation}
     \label{vinhinc}
          \bv_9 \;\, \to \;\,
		\left( \begin{array}{c|c} 
        \,\,\,\,\:\! S \;\!\!   \begin{array}{cc} &  \\  &  \end{array} \!\!\!   &
        \,     0  \begin{array}{cc} &  \\  &  \end{array} \!\!\!\!\!\!\!\!\!\! 
				                         \\  \hline
        \,\,\,\,\,\,  0  \!\! \begin{array}{cc}        &   \end{array}   &	  
		\,  1      \end{array}  \right) 
		\left( \begin{array}{c|c} 
        \,\,\,\, \bh_2     \!\!     \begin{array}{cc} &  \\  &  \end{array} \!\!\!   &
        \,  \psi  \begin{array}{cc} &  \\  &  \end{array} \!\!\!\!\!\!\!\!\!\! 
				                         \\  \hline
        \,\,\,\,\,\, \psi^{\dagger} \!\!\! \begin{array}{cc}    &   \end{array}   &	  
		\,  n      \end{array}  \right) 
		\left( \begin{array}{c|c} 
        \,\,\,\,\, S^{\dag} \!  \begin{array}{cc} &  \\  &  \end{array} \!\!\!   &
        \,  0  \begin{array}{cc} &  \\  &  \end{array} \!\!\!\!\!\!\!\!\!\! 
				                         \\  \hline
        \,\,\,\,\,\, 0 \!\! \begin{array}{cc}        &   \end{array}   &	  
		\,  1      \end{array}  \right) 
\end{equation}
   This combines the vector representation of $\sltc$ on $\bh_2 \inn \htwc$ and the 
spinor representation on $\psi \inn \ccc^2$, together with the scalar denoted $n \inn 
\rrr$ (in line with the notation used for the same expression in \cite{Novel} 
equation~19), in a single symmetry transformation that preserves  $L(\bv_9)  = 1$ of 
equation~\ref{lvnine}.
  Hence the choice of a preferred $\sltc$ implies a symmetry breaking pattern aligned 
with the isomorphism of vector spaces:
\begin{eqnarray}
  \hthc & \cong  & \htwc \, \oplus \, \ccc^2 \, \oplus \, \rrr  \label{symb1} \\   
	\bv_9   &  \to  &   (\bh_2 , \quad \;\, \psi, \quad \;\, n)  \label{symb2}    \\  
	  \mathbf{9}_{\mathrm{SL}(3,\ccc)}  &  \to  & 
   (\,\mathbf{4} \:\: + \:\: \mathbf{4} \:\: + \:\: 
\mathbf{1}\,)_{\mathrm{SL}(2,\ccc)}  \label{symb3}
\end{eqnarray} 
   where the first $\mathbf{4}$ denotes the real dimension of the vector 
representation, the second $\mathbf{4}$ that of the spinor and $\mathbf{1}$ is the 
trivial scalar, in each case as representations of the Lorentz group.
    Applying the translation symmetry of equation~\ref{rspill} in four dimensions 
only, corresponding to the subspace of external vectors ${\bv}_4 \equiv \bh_2 \inn 
\htwc \subset \hthc$, provides a natural mechanism for breaking the symmetry of the 
full group $\slthc$ through the necessary identification of the extended 4-dimensional 
background manifold $M_4$, with a spacetime metric structure, upon which the Lorentz 
group acts locally similarly as for the base space in figure~\ref{mtogmap}(b). 
 This symmetry breaking is depicted in figure~\ref{mtogmaph}
 (which is closely analogous to figure~1 of reference~\cite{Novel}, here for the full 
isochronal symmetry group $\hG = \slthc$ in place of $\sootn$).

\begin{figure}[htbp]  
\centering
\epsfxsize=12.5cm
\leavevmode
\epsffile[0 0 1765 918]{\gpath aKfig3e}
\vspace{-6pt}
\caption{\setb  (a) The gauge choice at each $x\in M_4$ depicted in 
figure~\ref{mtogmap}(a) is extended to $\hat{G}=\slthc$ as the full isochronal 
symmetry of the form $\lvni$, with (b) only the \textit{subgroup}  $\sltc \subset 
\slthc$ now acting on $\TM_4$ as the double cover of the Lorentz group $\soot$. A 
gauge field associated with the internal $\uo$ symmetry, together with the components 
$\psi$ and $n$ of $\bv_9 \inn \hthc$ in equation~\ref{vinhinc}, can be interpreted as 
`matter fields' over the external manifold $M_4$, augmenting the `vacuum' case of 
figure~\ref{mtogmap}(b).}
\label{mtogmaph}
\end{figure}

 The subgroup $\sltc$, as the double cover of $\soot$, is  distinguished in that it 
acts on  tangent space vectors $\bv_4 \inn \TM_4$ on the base manifold, as depicted in 
figure~\ref{mtogmaph}(b),  and is hence designated the \textit{external} symmetry.
 While collectively deriving from the components of temporal flow $\bv_9$,
the vector ${\bv}_4 \inn \TM_4$ is physically distinct from the spinor $\psi$ and 
scalar $n$ both in terms of their  properties under external symmetry transformations  
and in terms of their relation to the external spacetime manifold $M_4$ itself.
 That is, the symmetry breaking is not only felt in the identification of the 
components of the external  vector $\bv_4 \in \TM_4$ projected onto the physical 
spacetime $M_4$ but in the partitioning of  \textit{all} the components of $\bv_9$ 
into irreducible representations of the preferred $\sltc \subset \slthc$, including 
those with the physical properties of spinors and scalars.

   These latter objects, such as $\psi$ and $n$ identified from the `internal' 
components of $\bv_9$ in equations~\ref{vinhinc} and \ref{symb2}, 
 will be associated with and underlie physical `particle' states as deriving from the 
level of these elementary geometric structures of the theory through the symmetry 
breaking. 
  In turn the  \textit{internal} symmetry $G$
   is required to respect this partitioning into physical states and act collectively 
upon equivalent $\sltc$ representation objects as individual entities, for example 
upon the set of spinors, rather than act more generally upon the individual real 
components of $\bv_9$ as for the original full $\slthc$ symmetry.   
  For the general case this implies that the representations of the internal symmetry 
$G$ are aligned with those of the external symmetry $\sltc$, with $G$ consisting of 
remnant actions of the full isochronal symmetry $\hG$ of $\lvh$ which survive the 
partitioning of $\bvh$ under    
   the preferred $\sltc \subset \hG$.
 Particle states hence transform in representation multiplets under a symmetry 
breaking structure of the direct product form:
 \begin{equation} 
  \label{dirprod}
   \sltc \times G \subset \hG
 \end{equation}
   In the case of the $\hG = \slthc$ model the internal symmetry $G$ acts on the 
vector, spinor and scalar objects of equations~\ref{symb1}--\ref{symb3} as individual 
entities, without any mixing between the different types of Lorentz representations. 
Since for this model there is only one of each of a vector, spinor and scalar we have 
only one-dimensional representations for $G$, which in turn are only non-trivial for 
an Abelian internal symmetry as identified for a $\uo$ subgroup in the  breaking of 
the full   $\slthc$ symmetry  to:
\begin{equation}
 \label{suinsth}
   \sltc \times \uo \subset \slthc
\end{equation}
  Here the external $\sltc$ symmetry `locks on' to the tangent space $\TM_4$ leaving 
the residual internal $\uo$ symmetry as depicted in figure~\ref{mtogmaph}(b).
  This structure resembles the standard geometric picture for which a bundle of frames 
is `soldered' to the spacetime $M_4$, while a gauge symmetry bundle is arbitrarily 
attached, however here with the internal symmetry $\uo$ acting on the internal $\psi$ 
components.
 The internal symmetry $\uo$ and temporal components $\psi,n$, over and above those 
required to identify the base manifold $M_4$ as depicted in figure~\ref{mtogmaph}(b), 
give rise to gauge and further matter fields on the extended spacetime. Hence neither 
the extended spacetime $M_4$ nor the matter it contains are introduced independently 
of the flow of time itself, rather they derive collectively from the structure and 
symmetries of equation~\ref{lvnine}.

 The internal one-parameter group $\uo$, with elements represented by the $3 \times 3$ 
matrices $U = \mbox{diag}(e^{i\alpha/2},
  e^{i\alpha/2}, e^{-i\alpha})$ parametrised by $\alpha \inn \rrr$,
acts  non-trivially 
  on the spinor components  $\psi$ of equations~\ref{vinhinc},
  through $\bv_9 \to U \bv_9 U^{\dag}$, 
   while leaving
   the components of the vector $\bv_4 \inn \TM_4$ and the scalar $n$ invariant.
  In having a trivial action on the external vectors ${\bv}_4 \inn \TM_4$  the group 
$\uo$ belongs to the stability subgroup $\stab \subset \hG$, which could, 
  more generally, itself be considered a requirement for a  
  physical internal symmetry.   
 For the case of the model  considered only seven of the original sixteen generators 
of $\slthc$ survive the symmetry breaking to equation~\ref{suinsth}.
  The rank-4 Lie group $\slthc$ in fact contains a rank-4 subgroup decomposition, 
augmenting the rank-3 subgroup of equation~\ref{suinsth} as:
\begin{equation}
 \label{sudinsth}
   \sltc \times \uo \times \mbox{D}(1) \subset \slthc
\end{equation}
  with the dilation symmetry D(1) associated with a non-compact generator of $\slthc$. 
The D(1) group action $\bv_9 \to D \bv_9 D^{\dag}$,
 with $D = \mbox{diag}(e^{\lambda/2},
  e^{\lambda/2}, e^{-\lambda})$ parametrised by $\lambda \inn \rrr$, exhibits a 
complementary dilation effect on each of $\bv_4 \inn \TM_4$, $\psi$ and $n$ in 
equation~\ref{vinhinc} for this transformation.
 This dilation symmetry can be considered non-physical since it
 both relates different types of Lorentz representations and also
  does not respect the stability of the external spacetime with tangent space vectors 
 $\bv_4 \inn \TM_4 \equiv \htwc$ as a distinguished set of components of $\bv_9 \inn 
\hthc$, that is $\mbox{D}(1) \nsubst   \stab$.

  Identifying the  background manifold $M_4$ within the symmetry structures of the 
mathematical form $\lvf$ as depicted in figure~\ref{mtogmap} led to the 4-dimensional 
Minkowski spacetime of special relativity, that is with zero Riemannian curvature as 
implied by equations~\ref{maca3} and \ref{fdefaaa} for the $\hG=\soot$ case, as 
described in the previous subsection.
  Here breaking the full symmetry $\hG=\slthc$ of $\lvni$, in extracting the spacetime 
base manifold $M_4$
through a subset of four translational degrees of freedom
 of the higher-dimensional form, leaves a  space of symmetries associated with the 
product space $M_4 \times \uo$ underlying figure~\ref{mtogmaph}(b) and will result in 
a more flexible and dynamic 4-dimensional spacetime structure as employed for general 
relativity.

   Under the full form $\lvni$ we also have a
looser constraint on the four components $v^{a}(x)$ projected onto $\TM_4$ with: 
   \begin{equation}
 L(\v_4)=(v^0)^2- (v^1)^2-(v^2)^2-(v^3)^2= \eta_{ab} v^a v^b = \eta(\bv_4,\bv_4) = 
h^2.
 \label{lorform2}
\end{equation}
  with $h\inn \rrr$.
 With $M_4$ itself still originating locally out of a 4-dimensional translational 
symmetry of $\lvni$
   the Minkowski metric $\eta_{ab}$ implicit in the form $\lvfh$ in 
equation~\ref{lorform2} is sewn into the local tangent space structure everywhere on 
the base manifold.
  Since the manifold $M_4$ exists through the symmetries of $\lvni$ itself 
   there  are choices of local coordinates $\{x^a\}$ such that
    the vector field $\bv_4(x) = v^a e_a$  has the components $v^a=dx^a/ds$ with 
$\lvfh$ of equation~\ref{lorform2} expressed in this basis.
 Hence there exists a frame field $\{e_a(x)\}$ of local orthonormal basis vectors such 
that $g(e_a,e_b) = \eta_{ab}$ with respect to the metric $g(x)$. That is, we have a 
\textit{local} Lorentzian structure on $M_4$ as for the original case based on the 
full form $\lvf$ of equation~\ref{lorl} as described in the previous subsection. 
The manifold $M_4$ is drawn as a rectangular box in figure~\ref{mtogmaph}
  to represent the metric geometry of the base manifold $(M_4, g)$, as it was in 
figure~\ref{mtogmap}. 
 As will be reviewed in the opening of the following subsection,
 such a  metric structure $g_{\mu\nu}(x)$ on $M_4$ is associated in a one-to-one 
manner with the existence of an $\soot$ orthonormal frame bundle {\it OM}$_4$
 within the canonical $\glfrp$ general frame bundle $\FM_4$ over the base manifold 
$M_4$.

 The question then remains to identify a possible relation between the finite external 
Riemannian curvature of $M_4$ and the curvature of the internal gauge fields 
associated with the internal symmetry group $G$ identified in equation~\ref{dirprod}. 
With the aim of identifying such a relation in both a generally covariant and gauge 
invariant manner the geometry arising from the structure depicted in 
figure~\ref{mtogmaph}(b) will be considered in more detail
 for the general case. With the external symmetry acting on $\FM_4$ the broken 
symmetry structure of equation~\ref{dirprod} deriving from a full higher-dimensional 
form $\lvh$ can be accommodated on the product manifold \mbox{$M_4 \times G$}.
As a unifying framework for combining the external and internal symmetry this geometry 
is very similar to the principle fibre bundle structure employed in  Kaluza-Klein 
theories, which might be adapted for the present theory as will be described in 
section~\ref{one4}. In the following section we first review both the textbook 
geometrical setting and several relevant Kaluza-Klein models in the literature
 (summarising \cite{Unifi} chapters~3 and 4).

\section{Review of Kaluza-Klein Theories}
\label{one3}

\vspace{-3pt}

\subsection{Riemannian Geometry and General Relativity}

\label{one31}

  In the previous section we have identified a Lorentzian manifold $M_4$ with the 
local Minkowski metric $\eta$ deriving from the quadratic form $L(\bv_4)$ of 
equation~\ref{lorl} or \ref{lorform2}. In the latter case the identification of this 
extended metrical manifold breaks the full symmetry of the higher-dimensional form, 
such as the $\slthc$ symmetry of $\lvni$ as described for figure~\ref{mtogmaph}.

 More generally, any differentiable manifold $M$ is canonically associated with a  
linear frame bundle $\mbox{\it FM}$, with structure group $\mbox{GL}(n,\rrr)$ where 
$n$ is the dimension of the base manifold $M$, which is an example of a principle 
fibre bundle as described for the general case in the following subsection.
If $M$ is an $n$-dimensional Riemannian  manifold $(M,g)$, that is given a metric 
field with components $g_{\mu\nu}(x)$ on the manifold,  a subset of distinguished 
frames may be identified which are orthonormal with respect to the metric. This subset 
of frames over $M$ \textit{reduces} the total space of 
{\it FM} to a submanifold $\mbox{\it OM} \subset \mbox{\it FM}$ which is
 itself a principle fibre bundle  with structure group SO$^+(p,q)$ (or more generally 
O$(p,q)$) with $p+q = n$.
 There is a one-to-one correspondence between metric fields $g_{\mu\nu}(x)$ on $M$ and 
reductions of the structure group from $\mbox{GL}^+(n,\rrr)$ to SO$^+(p,q)$ on 
\textit{FM}, with each choice of field $g_{\mu\nu}(x)$ isolating one out of the many 
possible isomorphic copies of principle SO$^+(p,q)$-bundles.

  Given a Lorentzian metric in 4-dimensional spacetime,
 an $\soot$-bundle $\mbox{\it OM}_4$ can also be \textit{extended} to the frame bundle 
$\mbox{\it FM}_4$ with an $\soot$-valued Lorentz connection $A(x)$ (an example of 
which was introduced in equation~\ref{aegasth}) uniquely inducing a linear connection 
$\Gamma(x)$ for the extended bundle space. A $\glfrp$-valued linear connection 
$\Gamma$ identified in this way is compatible with the metric, that is $\nabla g = 0$.
 Here the kernel symbol $\nabla$ denotes the covariant derivative on  any 
differentiable manifold $(M,\Gamma)$ with a linear connection $\Gamma$.
 While a manifold with both a metric and a linear connection can be denoted 
$(M,g,\Gamma)$ in the general case $g(x)$ and
 $\Gamma(x)$  need not be related and the two objects may be introduced independently.

 With respect to a  frame field $\{e_a\}$ (here the indices $\{a,b,c\ldots \}$  denote 
the use of a general frame field on the manifold) the components of a linear 
connection $\Gamma^a_{\ph{a}bc}$ satisfy the relation $\nabla e_b = 
\Gamma^a_{\ph{a}bc} e^c \otimes e_a$, that is:
\begin{equation}
     \nabla_{\! c \,} e_b  =  \Gamma^a_{\ph{a}bc} e_a   
	 \label{gamene} 
\end{equation}
   with $\nabla_{\! c} \equiv \nabla_{\! e_c}$. In particular this equation 
establishes the convention for the order of the three indices for $\Gamma$. 
We also note here that the various possible sign conventions for  general relativity 
can be distilled down to the $\pm$ sign used for the right-hand side of just three 
expressions in the Riemannian geometry; in this paper we employ the following 
components: 
\begin{itemize}
 \item[1)] The spacetime metric tensor:  
   \begin{equation}
     \label{metcon}
      \eta_{ab} = \mbox{diag}(+1,-1,-1,-1)
   \end{equation}  
	   With `$+1$' for the time component this is a natural convention for the present 
theory based on forms of temporal flow.
 \item[2)] The Riemann curvature tensor:
 \begin{equation} 
{R}^a_{\ph{a}bcd} 
		  =   e_c {\Gamma}^a_{\ph{a}bd} - e_d {\Gamma}^a_{\ph{a}bc}
		      + {\Gamma}^a_{\ph{a}ec} {\Gamma}^e_{\ph{e}bd} - 
			    {\Gamma}^a_{\ph{a}ed} {\Gamma}^e_{\ph{e}bc} 
	    - c^{\:\! e \;\!\!}_{\ph{e}cd} {\Gamma}^a_{\ph{a}be}  \label{rabcd}
\end{equation} 
  
  Here $c^{\:\! e \;\!\!}_{\ph{e}cd}$  are the structure coefficients in the general 
frame field employed (which vanish if a coordinate basis is adopted). 
 \item[3)]  The Ricci tensor: 
 \begin{equation}
   \label{riccicon}
    R_{bc} = R^{d}_{\ph{d}bcd} \quad (=-R^{d}_{\ph{d}bdc}) 
  \end{equation}  
   This is equivalent to choosing the sign convention for 
the Einstein field equation as $G^{bc}=-\kappa T^{bc}$ with positive normalisation 
constant $\kappa$.
\end{itemize}
   
   The above three signs adopted here are `$(-+-)$'  relative to the original 
discussion of these conventions in~\cite{MTW}.
  Here a bold type $\bR$ will denote the Riemann tensor, with the components of 
equation~\ref{rabcd}, while  $R= g_{bc}R^{bc}$ is the  scalar curvature.
Further, for a general linear connection on the manifold $M$ the components of the 
torsion tensor $\bT$ can be written as:
\begin{equation}
     T^a_{\ph{a}bc} = - \Gamma^a_{\ph{a}bc} + \Gamma^a_{\ph{a}cb}
	                  - c^a_{\ph{a}bc} \label{t2gc}
\end{equation}   
  where again the final term is zero for a general coordinate frame.

  On a manifold with a metric 
  the unique linear connection $\Gamma$ that is both torsion-free ($\bT = 0$) and 
metric compatible ($\nabla g = 0$) is called the Levi-Civita connection.
 This linear connection is employed in  general relativity and can be written uniquely 
as a function of the metric tensor components $g_{\mu\nu}(x)$, expressed in a general 
coordinate frame (hence with $\{\mu,\nu,\rho\ldots\}$ indices) as:  
\begin{equation}
  \label{gtoGam}
    \Gamma^{\sigma}_{\ph{\sigma}\mu\nu} = \frac{1}{2} g^{\sigma\rho} 
(\partial_{\mu}g_{\rho\nu} + \partial_{\nu}g_{\mu\rho} - \partial_{\rho}g_{\mu\nu}) 
\end{equation}

For a spacetime manifold $M_4$ with a metric $g_{\mu\nu}(x)$ 
 equation~\ref{guneen} implies that in general
 a tetrad field $\teta$ can also be interpreted as the gravitational field. 
  If $A(x)$ is chosen to be the unique torsion-free Lorentz connection expressed in 
terms of a given tetrad field, then the associated linear connection $\Gamma(x)$ is 
the unique Levi-Civita connection. 
  While gravitation in Einstein's original theory of 1915 is described through the 
freedom of the metric field $g_{\mu\nu}(x)$, together with its relation to the 
Levi-Civita connection $\Gamma(x)$ of equation~\ref{gtoGam} on the spacetime manifold 
$M_4$,
 an equivalent formulation of general relativity can be given in terms of a tetrad 
field $\teta$ together with a Lorentz connection $A(x)$.  This latter approach was 
introduced in 1956 by Utiyama~\cite{Uti} in which general relativity is considered as 
a type of gauge theory invariant under local Lorentz transformations. 
 (This is also the interpretation of the Lorentz connection $A(x)$ in
   equations~\ref{aegasth}--\ref{fdefaaa} for the flat geometry of 
subsection~\ref{one22}).

  As well as tensor representations the Lorentz group also has spinor representations 
via the group $\sltc$ as the double cover of $\soot$. Hence spinor fields can be 
introduced on a spacetime manifold with an arbitrary metric $g_{\mu\nu}(x)$ via the  
tetrad field $\teta$.
 (For the $\lvni$ model described  for the present theory in subsection~\ref{one23}  
such a spinor field  derives from the components $\psi \inn \ccc^2$ of the space 
$\hthc$ as introduced in equation~\ref{vinhinc} and pictured in 
figure~\ref{mtogmaph}(b)).
 This structure also permits gravitation to be expressed in terms of  an 
$\sltca$-valued  connection,  accommodating a description of both vector and spinor 
objects in spacetime.
 However the dynamics of such an $\sltc$ `gauge theory' of gravitation (see for 
example \cite{Carm2}) are different to those of a standard Yang-Mills gauge theory. 
Such an approach is also in
 contrast with  Kaluza-Klein theories for which an internal gauge theory itself 
derives from a structure of general relativity with extra spatial dimensions.

    Finally in this subsection we review the standard use of the Lagrangian formalism 
to derive physical equations of motion as applied to general relativity and gauge 
theories. 
 In the 4-dimensional spacetime of general relativity the scalar curvature
  $R = g_{\mu\nu}R^{\mu\nu}$ is adopted as the principle geometric contribution to the 
total scalar Lagrangian function,  with 
 the field equations  determined from the Einstein-Hilbert action integral~(see for 
example \cite{HawkEl} page~75):
\begin{equation}
   I = \int (\alpha(R - 2\Lambda) + \lag)   \sqrt{\vert g \vert}\, d^4 x  
\label{einhil}   
\end{equation}
  Here   $\Lambda$ is the cosmological constant, $\lag$ is the Lagrangian function for 
matter fields and $\alpha$ is a normalisation constant.
 The magnitude of the metric determinant 
$\vert g \vert$ is employed in the 4-dimensional invariant volume element $\sqrt{\vert 
g \vert}\, d^4 x$.
  The vacuum equations for general relativity, that is with $\lag=0$ and $\Lambda = 
0$, are obtained by requiring stationarity $\delta I = 0$ for the action  in 
equation~\ref{einhil}  under any variation  $\delta g_{\mu\nu}$ of the metric 
components, leading  to the Einstein vacuum equation with the Einstein tensor:
\begin{equation}
    \label{lagtoein}
    G^{\mu\nu} := R^{\mu\nu} - \frac{1}{2}R \, g^{\mu\nu} = 0
\end{equation}

  For the non-vacuum case 
   the energy momentum tensor $T^{\mu\nu}$ for a general matter Lagrangian $\lag \neq 
0$ can be defined under variation of the metric $\delta g_{\mu\nu}$ through:
\begin{equation}
 \label{enmom}
   \delta I \; = \; \delta \!\! \int \lag \, \sqrt{\vert g \vert } \, d^4x \; = \;
	   \int \fhs \, T^{\mu\nu} \, \delta g_{\mu\nu} \, \sqrt{\vert g \vert } \, d^4x 
\end{equation}
 Hence for the full action integral of equation~\ref{einhil} stationarity  under the 
metric variation gives Einstein's field equation for the general case, with $\kappa = 
\frac{-1}{2\alpha}$  adopted as the normalisation constant
 (the $\Lambda = 0$ case was quoted in item `3)' above):
\begin{equation}
   \label{einlamt}
    G^{\mu\nu} + \Lambda g^{\mu\nu} = - \kappa T^{\mu\nu}
\end{equation}

  The Maxwell Lagrangian for the electromagnetic field is constructed as:
\begin{equation}
  \label{lagem}
   \lag_{\mathrm{em}} \, = \, - \frac{1}{4}F_{\mu\nu}F^{\mu\nu}
\end{equation}
 in terms of the electromagnetic field strength tensor components $F_{\mu\nu} = 
\pal_{\mu} A_{\nu} - \pal_{\nu} A_{\mu}$.
 Under variation $\delta A_{\mu}(x)$ of the electromagnetic gauge field $A_{\mu}(x)$ 
the Euler-Lagrange equation for $\lag_{\mathrm{em}}$ yields Maxwell's equation for the 
source-free  case, that is $\pal_{\mu}F^{\mu\nu}_{\ph{\mu\nu}} = 0$ (which can be 
written $\nabla_{\!\!\mu}F^{\mu\nu}_{\ph{\mu\nu}} = 0$ in a curved spacetime).
  The form of the Lagrangian for a non-Abelian gauge theory is  guided by the Abelian 
case of electromagnetism, motivating the   Lorentz and gauge invariant  Yang-Mills  
Lagrangian:
\begin{equation}
 \label{lagym}
 \lag_{\mathrm{YM}} \, = \, - \frac{1}{4} F^{\alpha}_{\ph{\alpha}\mu\nu}
   F_{\alpha}^{\ph{\alpha}\!\mu\nu}
\end{equation}
  as a direct generalisation of equation~\ref{lagem}.
   For the non-Abelian case there is a further contraction over the index $\alpha = 
1\ldots n_G$, with $n_G = \dim (G)$ for the generators of the group $G$, between the 
adjoint and coadjoint representations, which are related by the Killing metric 
$K_{\alpha \beta}$ (which in a suitable basis  is simply $-\delta_{\alpha\beta}$ for 
the compact simple Lie groups relevant for the internal gauge symmetries in particle 
physics).  In this case the Euler-Lagrange equation for $\lag_{\mathrm{YM}}$ under 
variation of the gauge field components $A^{\alpha}_{\ph{\alpha}\mu}(x)$ yields the 
non-linear second order differential equation:
\begin{equation}
  \label{ymol}
  D_{\mu}F^{\alpha \;\!\mu\nu} = \partial_{\mu} F^{\alpha \;\!\mu\nu}
     + \cstr A^{\beta}_{\ph{\beta}\mu} F^{\gamma \;\!\mu\nu}  = 0 
\end{equation}
 Here $D_{\mu}$ is the gauge covariant derivative and again $\pal_{\mu}$ is replaced 
by $\nabla_{\!\;\!\!\mu}$ for a curved spacetime.
 
  In general relativity the
  energy-momentum tensor $T^{\mu\nu}$ can be derived directly from  the matter 
Lagrangian $\lag$  as described for equation~\ref{enmom}.  Substituting the Yang-Mills 
Lagrangian $\lag_{\mathrm{YM}}$ of equation~\ref{lagym} into equation~\ref{enmom} 
results in:
\begin{equation}
  \label{temkk}
      T^{\mu\nu} =  F^{\alpha\:\! \mu}_{\ph{\alpha \mu}\rho}  \:\! 
F_{\alpha}^{\ph{\alpha}\rho\nu}
	                +\frac{1}{4} g^{\mu\nu} \;\! F^{\alpha}_{\ph{\alpha}\rho 
\sigma}\:\! F_{\alpha}^{\ph{\alpha}\rho\sigma} 
\end{equation}
   This  yields an energy-momentum tensor $T^{\mu\nu}$ which is symmetric, gauge 
invariant and complies necessarily with the Einstein equation~\ref{einlamt} since it 
derives from the Einstein-Hilbert action of equation~\ref{einhil} via 
equation~\ref{enmom}. In equation~\ref{temkk} $\alpha$ is a Lie algebra index (which 
is absent for the Abelian case of electromagnetism) while all other indices relate to  
spacetime coordinates on $M_4$.

\subsection{Principle Fibre Bundle Structure}
\label{one32}

   In subsection~\ref{one22} we introduced \textit{two} independent differentiable 
manifolds, the base space $M_4$ and Lie group $\hG=\soot$ with points labelled by $x 
\inn M_4$ and $h \inn \hG$,
   associated with the 4-dimensional form of temporal flow $\lvf$ of 
equation~\ref{lorl} through the respective `translational' $\{r^a\}$  and `isochronal' 
$\sigma_h$   symmetries as described for equation~\ref{lv2syms}.
    The map between the manifolds $h:M_4 \to \hG$, mapping $x \to h(x)$ as depicted in 
figure~\ref{mtogmap}(a), represents a local choice of gauge, or orthonormal frame, in 
which to express the tangent vector $\bv_4(x)$ on $M_4$.  This association between 
$M_4$ and $\hG$ may be examined more concisely through the structure  of a 
\textit{single} differentiable manifold, namely a principle fibre bundle $P$, which 
combines the geometric properties of a base space $M$ and a Lie group $G$ together 
with their mutual relation.

   In the general case the structure group $G$ of a principle fibre bundle $P=(M,G)$ 
does not need to be related to a symmetry on the tangent space to the base manifold 
$M$, as it is for the above case of figure~\ref{mtogmap}. Indeed for the augmented 
case of figure~\ref{mtogmaph}(a) the full symmetry group $\hG=\slthc$  acts only 
partially  on the tangent space of $M_4$ while  the internal symmetry group $G=\uo$ 
does not act on the external tangent space at all, as pictured in 
figure~\ref{mtogmaph}(b). This latter figure  contains information about both the 
external and internal symmetry, the geometries of which we are ultimately aiming to 
relate.
 Hence it is the generalisation in which $M$ and $G$ are initially  introduced 
independently, and then related through the principle bundle space $P=(M,G)$, that we 
shall review here for the benefit of the subsequent application to the case of a 
higher-dimensional form $\lvh$ such as presented in subsection~\ref{one23}.
 (For more details on principle bundle structures generally see for example 
\cite{kob,amp,fecko}).

 For the present theory it will be assumed that the structure of principle bundles 
with a trivial global topology
  will be sufficient. In this case the bundle $P=(M,G)$ is diffeomorphic to the 
product space $M \times G$, which can be expressed as
   $P \equiv U \times G$ where a single `subset' $U \subset M$  may be identified with 
the entire base manifold $M$. This triviality is implied in deriving the bundle 
structure through the symmetries of $\lvh$ as described for figures~\ref{spillout}(c), 
\ref{mtogmap} and \ref{mtogmaph}.
 In this case a  change of trivialisation, or gauge transformation, may apply over the 
entire volume of the base space $M$.

  The Lie algebra $L(G)$ of a Lie group $G$ can be represented by a basis of 
left-invariant vector fields $\{X_{\alpha}\}$, for $\alpha = 1\ldots n_G$, on the 
group manifold,  that is with $L_{h\ast}X_{\alpha}(h') = X_{\alpha}(hh')$ where 
$L_{h\ast}$ is the differential of the left action $L_h$ of $G$ on itself and with 
$h,h' \inn G$.
 While the \textit{right} action of $G$ on the group manifold itself induces 
left-invariant vector fields, including any  basis vector $X_{\alpha}$, the right 
action $R_{\exp(tA)} : p \to p \:\! \exp(tA)$ of $G$  on points of a principle bundle 
manifold $p \inn P\equiv M \times G$ induces `vertical' vector fields in the tangent 
space $\mbox{\it TP}$ with:  
\begin{equation}
   \label{atova}
    V_p^A(f(p)) = \frac{d}{dt} \: f(p \:\! \exp(tA)) \, \vert_{t=0}
\end{equation}
   where $f(p)$ is any smooth real-valued function on the bundle space,
    $A\inn L(G)$ and $V_p^A$ is a tangent vector to the fibre of $P$ at the point
	 $p\inn P$. The map $A \to V_p^A$ described in equation~\ref{atova} represents an 
\textit{isomorphism} of the Lie algebra $L(G)$ into the space of vector fields 
residing in the vertical tangent space $\mbox{\it VP}\subset \mbox{\it TP}$.

  While this structure relates to $\mbox{\it VP}$, the space of vectors tangent to the 
individual fibres of $P$, different fibres may be related by an additional structure 
called a \textit{connection} on the principle bundle which, conceptually, is a smooth 
assignment of a `horizontal' subspace $H_pP$ of the full tangent space $T_pP$ at each 
point $p\inn P$ such that:
\begin{eqnarray}
              T_pP &  =  &  V_pP \oplus H_pP   \label{vplush}  \\
   R_{h\ast} H_pP  &  =  &  H_{ph}P         \label{racthp}     
\end{eqnarray}
  Compatibility of the horizontal subspaces on $P$ with the right action $R_h$ of $G$ 
on the bundle space is described by the latter requirement, where $R_{h\ast}$ is the 
differential of this map for any $h\inn G$. 
The tangent space decomposition of equation~\ref{vplush} 
is sketched in figure~\ref{pbunbasis}.
\begin{figure}[htbp]  
\centering
\epsfxsize=10cm
\leavevmode
\epsffile[0 0 1201 680]{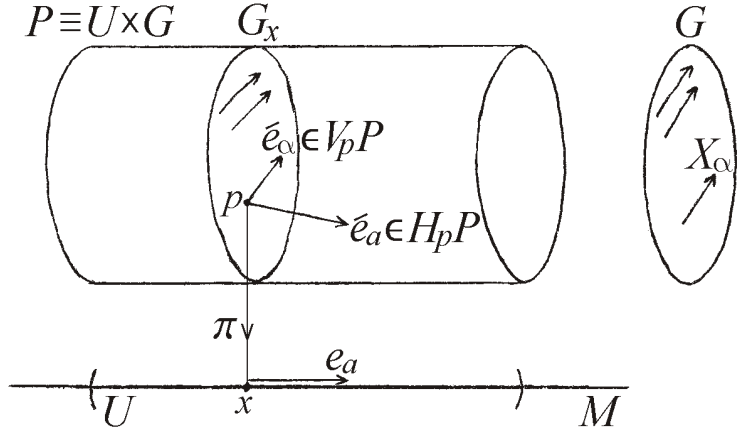}
\caption{\setb Vertical and horizontal basis vectors in a local trivialisation $U 
\times G$ of a principle bundle  $P$, together with their associated basis vectors on 
the group space $G$ and the base manifold $M$ respectively.  (Since the bundle is  
trivial  we take $U=M$.)}
\label{pbunbasis}
\end{figure} 

At every point $p \inn P$ a basis for the tangent space of the principle bundle can be 
expressed in terms of these complementary subspaces. Such a basis $\{ {\acute e}_i \} 
= \{{\acute e}_{\alpha}, \, {\acute e}_a\}$ consists of the subset $\{{\acute 
e}_{\alpha}\} \inn \mbox{\it VP}$, with $\alpha = 1\ldots n_G$, (that is vectors of 
the form $V_p^A$ in equation~\ref{atova}, tangent to the fibres $G_x$ over each point 
$x\inn M$) and the  subset
$\{{\acute e}_a \}\inn \mbox{\it HP}$, with $a = 1\ldots n$,  (where ${\acute e}_a $ 
is the `horizontal lift' of the basis vector $e_a \inn T_xM$ to the point $p \inn P$ 
such that $\pi_{\ast} {\acute e}_a = e_a$, where $\pi$ is the projection between 
manifolds $\pi: P \to M$).

 An `acute' mark above a kernel symbol, such as for ${\acute e}$, denotes an object 
defined on a principle bundle space in the horizontal lift basis. In all cases 
 the indices $\{i,j,k\ldots \}$ denote elements defined in the tangent space   
$\mbox{\it TP}$ on the principle bundle $P$ itself; $\{ \alpha,\beta,\gamma\ldots \}$ 
in the vertical subspace on $P$ or on the manifold $G$; and $\{ a,b,c\ldots \}$ in a 
complementary subspace on $P$ or on the base space $M$. 
  There is a one-to-one correspondence between  basis vector fields
  ${\acute e}_{\alpha}$ on $P$ and basis vector fields $X_{\alpha}$ on $G$, and
  similarly between basis vector fields ${\acute e}_{a}$ on $P$ and basis vector 
fields $e_a$ on $M$. The relations between these vector fields are implied in 
figure~\ref{pbunbasis}.

    The defining structure for  the horizontal subspace $\mbox{\it HP}$ of 
equation~\ref{vplush} and \ref{racthp} can be specified  via a smooth Lie 
algebra-valued 1-form $\omega \inn L(G) \otimes T^{\ast}\!P$,  mapping vectors $X\inn 
\mbox{\it TP}$ into elements of $L(G)$. This connection 1-form $\omega$
  on $P$  has the properties (the first of which is essentially the reverse of 
equation~\ref{atova}):
\begin{eqnarray}
  & \mbox{(i)}  &   \omega (V^A) = A \qquad \mbox{with} \; A\inn L(G)  \label{wvaalg} 
\\
   & \mbox{(ii)}  &  R^{\ast}_h \, \omega = \mbox{Ad}(h^{-1}) \omega \quad \mbox{i.e.} 
\;
       R^{\ast}_h \, \omega_{ph}(X) = h^{-1} \omega_p(X)h   \quad \mbox{with} \; X\inn 
T_pP  \qquad \label{rwgog}  \\
  & \mbox{where} & H_pP \equiv \{ X \inn T_pP \mid \omega(X) = 0 \} \label{hpom}
\end{eqnarray}
is the horizontal subspace. Here $R^{\ast}_h$ is the pull-back map associated with the 
right action $R_h$ of the group on $P$.

 The specification of a connection $\omega$ and corresponding horizontal subspace 
allows `parallel transport' between the fibres to be defined.
  As depicted within figure~\ref{paraCD} in subsection~\ref{one41} given a point 
$p_1\inn P$ with $\pi(p_1)=x_1$ and a curve $C$ on the base space from $x_1$ to $x_2$ 
a connection  on a principle bundle $P$ specifies a unique horizontal lift of the 
curve $C$ to the curve $C'$ on $P$, by advancing locally within the horizontal 
subspace $\mbox{\it HP} \subset \mbox{\it TP}$.
 That is, the tangent vector to the curve $C'$ at any $p\inn P$ always lies within the 
horizontal subspace $H_pP$ and projects under $\pi_{\ast}$ to a tangent vector to $C$ 
at $x= \pi (p)$.
 The path $C'$ then represents the parallel transport of $p_1$ mapped between fibres 
to the unique point $p_{2C'} \inn P$, with $\pi(p_{2C'})=x_2$. In 
subsection~\ref{one41}
 this geometric structure of the gauge connection $\omega$ on the bundle space will be  
modelled by a parallel transport according to a linear connection $\acute{\Gamma}(p)$ 
defined on $P$ itself.

Since $P$ itself is a differentiable manifold  real-valued structure coefficients 
$\check{c}^{\:\!i}_{\ph{i}jk}(p)$ can be defined for any frame field $\{\check{e}_i\}$ 
on $P$ through the relation (with a `check' on an object as for $\check{e}$ denoting a 
field on the bundle space generally):
\begin{equation}
   [\check{e}_j, \check{e}_k] = \check{c}^{\:\!i}_{\ph{i}jk}(p) \check{e}_i
       \label{eecefull}
\end{equation}
   The horizontal lift basis $\{{\acute e}_i\} =
     \{{\acute e}_{\alpha}, {\acute e}_a\}$
for the tangent space $\mbox{\it TP}$, introduced above, is adapted to a given 
connection $\omega$ such that ${\acute e}_{\alpha} \inn V_pP$ and  ${\acute e}_a \inn 
H_pP$ at any $p \inn P$, as was depicted in figure~\ref{pbunbasis}, with 
$\omega(\acute{e}_{\alpha}) = X_{\alpha}$ and $\omega(\acute{e}_{a}) = 0$, by the 
definition of the horizontal lift basis.
 In this basis the full set of structure coefficients on $P$ are given by: 
\begin{eqnarray}
  \lbrack \acute{e}_{\alpha}, \acute{e}_{\beta} \rbrack & = &
                   c^{\gamma}_{\ph{\gamma}\;\!\!\alpha\beta}\acute{e}_{\gamma} 
\label{hatb1}  \\
  \lbrack \acute{e}_{\alpha}, \acute{e}_{b} \rbrack & = &  0   \label{hatb2} \\
  \lbrack \acute{e}_{a}, \acute{e}_{b} \rbrack & = &
           \acute{c}^{\:\! \alpha}_{\ph{\alpha}ab}\acute{e}_{\alpha}
		   =-F^{\alpha}_{\ph{\alpha}ab}(p)\acute{e}_{\alpha}  \label{hatb3}
\end{eqnarray}

 In equation~\ref{hatb1} the $c^{\gamma}_{\ph{\gamma}\;\!\!\alpha\beta}$ are the 
structure constants of the Lie algebra of $G$, expressing the Lie algebra isomorphism 
described after equation~\ref{atova}.
  Since right actions induce the basis vectors $\acute{e}_{\alpha}$ of the subspace 
$\mbox{\it VP}$, via equation~\ref{atova}, each $\acute{e}_{\alpha}$ generates a right 
translation and hence
 equation~\ref{hatb2} expresses the right-invariance of the fields \mbox{$\acute{e}_b 
\inn \mbox{\it HP}$}, consistent with equation~\ref{racthp}.
    For the third equation the structure coefficients 
	$\acute{c}^{\;\! d}_{\ph{d}ab}$ are  zero since here a coordinate basis is taken 
for $\{e_a\}$ on the base manifold $M$ in order to simplify the expressions.
 In fact with $\mbox{d}e^a = 0$ for the dual coframe basis $\{e^a\}$ on  $M$ and
  each $\acute{e}^{\:\! a} = \pi^{\ast} e^a$ on $P$ we have
   $\mbox{d}\acute{e}^{\:\! a} = -\fh \acute{c}^{\;\! a}_{\ph{a}jk}
    \acute{e}^{\:\! j} \wedge \acute{e}^{\:\! k} = 0$ and hence each 
	$\acute{c}^{\;\! a}_{\ph{a}jk} = 0$. 
 Equation~\ref{hatb3} demonstrates the intimate relationship between the horizontal 
lift basis and the physical manifestation of the gauge connection in terms of the 
internal curvature components
  $F^{\alpha}_{\ph{\alpha}ab}(p)$  on the principle bundle.

 Here the components of the Lie algebra-valued curvature 2-form $\Omega(p)$ on a 
principle bundle $P$  are denoted $F^{\alpha}_{\ph{\alpha}ab}(p)$ in order to match 
the notation of the references for the following two subsections.
 The curvature 2-form itself on $P$ is defined by:
 \begin{equation}
 \label{omdef}
  \Omega = \mbox{d}\omega \,\scirc\, \mbox{hor}
 \end{equation}
 where $\mbox{d}\omega$ is the exterior derivative of the connection 1-from $\omega$ 
 and `hor' maps vectors $X$ in the tangent space {\it TP}  to their horizontal 
components in the decomposition of equation~\ref{vplush}, with the
  vertical component mapped to zero (that is, $\mbox{hor}:  X \to X_H$, such that $X_H 
\subset  HP$  with $\omega(X_H)=0$ by equation~\ref{hpom}).

   For a particular trivialisation $\psi: P \to M \times G$ on the principle bundle, 
with corresponding section $\sigma(x): M \to P$ mapping $x \inn M$ to $\psi^{-1}(x,e)$ 
with $e \inn G$ the group identity element,  a 
  direct product basis  $\{ {\ddot e}_i \} = \{{\ddot e}_{\alpha}, \, {\ddot e}_a\}$ 
for the tangent space 
   consists of the subset $\{{\ddot e}_{\alpha}\} \inn \mbox{\it VP}$, tangent to the 
fibres $G_x$ over each point $x\inn M$, and the  subset
$\{{\ddot e}_a \}$  with ${\ddot e}_a = \sigma_{\ast} e_a $ for each basis vector $e_a 
\inn T_xM$ (hence with $\pi_{\ast} {\ddot e}_a = e_a$).
 Each vector ${\ddot e}_a$ defined on the section $\sigma(x)$ is Lie transported via 
the right action of $G$ on $P$ such that the basis covers the entire principle bundle.
 The `double dot' mark above the kernel symbol, such as for ${\ddot e}$, denotes an 
object defined on a principle bundle space in the direct product basis.
 Relative to the horizontal lift basis $\{\acute{e}_i\}$  we have:
\begin{equation} 
 \label{ddandhl}
  \ddot{e}_{\alpha} = \acute{e}_{\alpha} \quad \mbox{and} \quad
     \ddot{e}_a = \acute{e}_a + \omega^{\alpha}_{\ph{\alpha}a}\acute{e}_{\alpha}
\end{equation}
 where $\omega^{\alpha}_{\ph{\alpha}a}(p)$ are the components of the connection on $P$ 
for this section. With $\Omega = \Omega^{\alpha} X_{\alpha}$ the curvature components 
at any $p \inn P$:
\begin{equation}
    F^{\alpha}_{\ph{\alpha}ab} = \Omega^{\alpha}(\acute{e}_a,\acute{e}_b)
	                 = \Omega^{\alpha}(\ddot{e}_a,\ddot{e}_b)
\end{equation}
  are identical in the horizontal lift basis and a direct product basis since 
 the curvature vanishes on vertical components, equation~\ref{omdef}, and
  these bases differ only by a vertical vector, equation~\ref{ddandhl}. 
 However, while the horizontal lift basis represents a physical geometrical structure 
a direct product basis represents a passive choice of gauge.

 Given a  section $\sigma(x)$ on $P$ the  representative of the curvature $\Omega$ on 
the base space is defined by the pull-back map as the 2-form 
 $F(x)=\sigma^{\ast}\Omega(p)$ while the representative of the connection $\omega$ is 
the gauge field $A(x) = \sigma^{\ast}\omega(p)$ on $M$.
 Equation~\ref{omdef} above can be written as the structure equation
  $\Omega = \mbox{d} \omega + \fh [ \omega, \omega ]$ on $P$ which pulls-back to 
equation~\ref{fdefaaa} on the base manifold $M$.
 The curvature on the base manifold $F(x) = F^{\alpha}(x) X_{\alpha}$ also 
 takes values in the Lie algebra. 
  With a coordinate basis employed on $M$ the components of the curvature $F$ are 
written $F^{\alpha}_{\ph{\alpha}\mu\nu}(x)$ on the base manifold, as is the case for 
equations~\ref{lagem}--\ref{temkk}. 
  In general for a particular section $\sigma : M \to P$, and corresponding 
trivialisation $P \equiv M \times G$, the components of the curvature and gauge fields 
on $M$ can be expressed respectively as:
 \begin{eqnarray}
     F^{\alpha}_{\ph{\alpha}\mu\nu}(x) & = &
	    \sigma^{\ast}\Omega^{\alpha}(e_{\mu}, e_{\nu})\;\;	=\;\;
	    \Omega^{\alpha}(\sigma_{\ast}e_{\mu}, \sigma_{\ast}e_{\nu})\;\;	=\;\;
		\delta^a_{\ph{a}\mu}\delta^b_{\ph{b}\nu}F^{\alpha}_{\ph{\alpha}ab}(x,e)
		\label{fonbase} \\
	A^{\alpha}_{\ph{\alpha}\mu}(x) & = &
	    \sigma^{\ast}\omega^{\alpha}(e_{\mu}) \qquad	=\;\;
	    \omega^{\alpha}(\sigma_{\ast}e_{\mu}) \qquad\quad	=\;\;
		\delta^a_{\ph{a}\mu} \omega^{\alpha}_{\ph{\alpha}a}(x,e)
		\label{aonbase}
 \end{eqnarray}
 
  where $\{e_{\mu}\}$ is a coordinate basis on $M$.
 The fibre dependence of the curvature components $F^{\alpha}_{\ph{\alpha}ab}(p)$ on 
$P$ may be deduced by application of the Jacobi identity in the horizontal lift basis
 and use of equations~\ref{hatb2} and \ref{hatb3}:
\begin{eqnarray}
 & &  \lbrack \acute{e}_{\alpha}, \lbrack \acute{e}_{a}, \acute{e}_{b} \rbrack \rbrack  
+
 \lbrack \acute{e}_{a}, \lbrack \acute{e}_{b}, \acute{e}_{\alpha} \rbrack \rbrack  +
 \lbrack \acute{e}_{b}, \lbrack \acute{e}_{\alpha}, \acute{e}_{a} \rbrack \rbrack  = 0   
   \nonumber  \\
 & \Rightarrow & \lbrack \acute{e}_{\alpha},\,
  F^{\beta}_{\ph{\beta}ab}\acute{e}_{\beta} \rbrack +
     \qquad\, 0 \qquad + \qquad\, 0 \qquad = 0   \nonumber \\
 & \Rightarrow & (\acute{e}_{\alpha} \, 
    F^{\beta}_{\ph{\beta}ab})\acute{e}_{\beta} +
       F^{\gamma}_{\ph{\beta}ab}
	      c^{\beta}_{\ph{o}\alpha\gamma}  \acute{e}_{\beta} = 0  
	                                          \nonumber \\
 & \Rightarrow &   \acute{e}_{\alpha} \, F^{\beta}_{\ph{\beta}ab} = 
         - c^{\beta}_{\ph{o}\alpha\gamma} \,    
		 F^{\gamma}_{\ph{\beta}ab}  \label{ecderv}
\end{eqnarray}
  The final expression describes the directional derivative of the coefficients 
$F^{\beta}_{\ph{\beta}ab}$ with respect to the vector field  $\acute{e}_{\alpha}$,
 which generates right translations by the gauge group $G$, 
 and hence expresses gauge transformations of the curvature $\Omega$.
 This is  the expected transformation property  for the components of the Lie 
algebra-valued curvature 2-form  under infinitesimal gauge transformations, since the 
curvature transforms under the adjoint representation for finite gauge 
transformations, as expressed immediately after equation~\ref{fdefaaa} for example.

\subsection{General Relativity with Extra Dimensions}  
\label{one33}

  The unifying framework for gravitation and gauge theories reviewed here is 
constructed in the mathematical setting of a principle fibre bundle. Keeping within 
the spirit of Einstein's original 4-dimensional spacetime theory of gravitation and 
the incorporation of electromagnetism in the extension to a 5-dimensional arena by 
Kaluza and Klein~\mbox{\cite{Kaluza,Klein}}, the generalisation for geometric 
unification with non-Abelian gauge theory is founded upon a metric tensor $\check{g}$, 
now defined upon the manifold of the principle bundle $P=(M_4,G)$ itself
 (\cite{Cho}, see also \cite{Kerner}, \cite{ChMaMa} sections I--V and 
\cite{Orzalesi}).

  We note that conventions vary in the literature -- in particular with respect to the 
assignment of index labels such as $\{a,b,\ldots\}$, $\{\alpha,\beta,\ldots\}$, and 
$\{i,j,\ldots\}$ which in this paper are associated with objects on the  manifolds 
$M_4$, $G$ and $P$ respectively, in the manner described alongside and in  
figure~\ref{pbunbasis}. The conventional order of the indices for the linear 
connection coefficients $\Gamma^a_{\ph{a}bc}$ also varies, with the convention of 
equation~\ref{gamene} adopted here,
while the sign of the Ricci tensor $R_{bc} = R^{d}_{\ph{d}bcd}$  of 
equation~\ref{riccicon} also differs in some of the references. Hence in turn a number 
of derived expressions here will have signs differing to those in  the literature.

   In addition to the metric $g_{ab}(x)$ on the base manifold $M_4$ the 
   $\mbox{Ad}(G)$-invariant Killing form $K$ 
 defines a natural bi-invariant metric on the group manifold $G$. That is,  both the 
left $L_h$ and right $R_h$ group actions, for any $h\inn G$, are isometries on $G$, 
with for example $(R^{\ast}_h \, K_{h'h})(X,Y) = K_{h'}(X,Y)$ for all 
$X,Y\inn T_{h'}G$  for the Killing metric $K$ at any point $h'\inn G$.
 As a matrix of components $K$ is invertible provided $G$ is a semi-simple Lie group 
and negative definite if $G$ is compact. In the latter case   a basis for the Lie 
algebra can be chosen such that the Killing form has components $K_{\alpha\beta} = 
-\delta_{\alpha\beta}$, as noted after equation~\ref{lagym}.
 Here we choose metric components 
 $g_{\alpha\beta} = +K_{\alpha\beta}$ in order to match the signature convention of 
equation~\ref{metcon}, with spacelike components having a negative norm.
  In terms of the group structure constants $\cstr$ in a left-invariant basis 
$\{X_{\alpha}\}$ on the group manifold the components of the Killing metric are:  
\begin{equation}
  \label{killmet}
g_{\alpha\beta} = K_{\alpha\beta} =  
c^{\rho}_{\ph{\rho}\alpha\sigma}c^{\sigma}_{\ph{\sigma}\beta\rho} 
\end{equation}

 A gauge connection 1-form $\omega$ on a principle bundle $P$ specifies a 
right-invariant horizontal subspace {\it HP} of the tangent space {\it TP}, as 
described for equations~\ref{wvaalg}--\ref{hpom}. A unique metric $\check{g}$ may be 
defined on such a principle bundle space, aligned with the gauge connection structure 
with:
 \begin{equation}   
  \check{g} (X,Y) = g (\pi_{\ast} X, \pi_{\ast} Y)  +  K (\omega(X), \omega(Y))  
\label{ggxyk}
\end{equation}
  where $X,Y \inn \mbox{\it TP}$, while here $g$ and $K$ are the metrics on the base 
space $M_4$ and group space $G$ respectively. This construction yields an intuitively 
natural metric on the bundle space in the sense that the vertical $\mbox{\it VP}$ and 
horizontal $\mbox{\it HP}$ subspaces of the tangent space of $P$, as depicted in 
figure~\ref{pbunbasis}, are then \textit{orthogonal} with respect to $\check{g}$, with 
$\check{g}(X,Y)=0$ if $X \inn \mbox{\it VP}$ and $Y\inn \mbox{\it HP}$ for example.

   Alternatively, and perhaps more in the spirit of the original Kaluza-Klein theory, 
a metric $\check {g}$ rather than a connection $\omega$ can be considered as the 
fundamental entity on $P$. That is, the bundle is initially endowed with a 
pseudo-Riemannian metric $\check{g}(p)$ with \textit{certain restrictions} -- namely 
compatibility with a Lorentzian metric $g_{ab}(x)$ for vectors projected onto $M_4$ 
and with the Killing metric $g_{\alpha\beta}$ for vectors tangent to the fibres $G_x$ 
and the requirement of invariance under the right action of $G$ on $P$:
\begin{equation}
 \label{rtrang}
  (R^{\ast}_h \, \check{g}_{ph})(X,Y) =   \check{g}_p(X,Y) = \check{g}_{ph}(R_{\ast}X, 
R_{\ast}Y)
\end{equation} 
 for any $p\inn P$, $h\inn G$ and $X,Y \inn \mbox{\it TP}$.
  This latter property then \textit{implies} the existence of a subspace {\it HP}, 
orthogonal to {\it VP}, which is right-invariant and hence is equivalent to the 
existence of a connection 1-form $\omega$ on the bundle $P$, which is related to 
$\check{g}$ as described in equation~\ref{ggxyk}.

  From either perspective from the relation of $\check{g}$ to $\omega$ in 
equation~\ref{ggxyk} in the horizontal lift basis $\{\acute{e}_i\} = 
\{\acute{e}_{\alpha}, \acute{e}_a\}$,  with  $\acute{e}_{\alpha} \inn \mbox{\it VP}$ 
and $\acute{e}_a \inn \mbox{\it HP}$, for the tangent space on $P$ the metric 
$\acute{g}$, and its inverse, take respectively the simple forms:
\begin{equation}
  \acute{g}_{ij} =  \left( \begin{array}{c|c}
                 g_{ab} &  0  \\
				          \hline
				    0    & g_{\alpha\beta} 
          \end{array}  \right)    
		       \qquad \mbox{and} \qquad 
  \acute{g}^{ij} = \left( \begin{array}{c|c}
                 g^{ab} &  0  \\
	            		  \hline
				    0    & g^{\alpha\beta}   
          \end{array}  \right)     \label{ggghlb}
\end{equation}
   That is with the components of the metric on the base space $M_4$ identified as 
$g_{ab} = \acute{g}(\acute{e}_a, \acute{e}_b)$ and those of the Killing metric on the 
group space identified as $g_{\alpha\beta} = \acute{g}(\acute{e}_{\alpha}, 
\acute{e}_{\beta})$. The off-diagonal block components in equation~\ref{ggghlb} are 
all zero, with for example $\acute{g}_{a\beta} = \acute{g}(\acute{e}_a, 
\acute{e}_{\beta}) = 0$ describing the orthogonality of any
 $X = X^a \acute{e}_a\inn H_pP$ to any $Y = Y^{\beta}\acute{e}_{\beta}\inn V_pP$ 
  at any $p\inn P$ with respect to this right-invariant metric $\acute{g}$.

   Under a change of frame from a horizontal lift basis to a direct product basis 
$\{\acute{e}_i\} \to \{\ddot{e}_i\}$, aligned with a particular choice of  
trivialisation $\psi: P \to M_4 \times G$ as described for equation~\ref{ddandhl},
 equation~\ref{ggghlb} is transformed to: 
\begin{equation}
  \ddot{g}_{ij} =  \left( \begin{array}{c|c}
                 g_{ab} + 
g_{\alpha\beta}\omega^{\alpha}_{\ph{\alpha}a}\omega^{\beta}_{\ph{\beta}b}  & 
				                   \omega^{\alpha}_{\ph{\alpha}a} g_{\alpha\beta}    
\\
				          \hline
	 g_{\alpha\beta}\omega^{\beta}_{\ph{\beta}b}  & g_{\alpha\beta} 
          \end{array}  \right)    
		       \quad \mbox{and} \quad 
  \ddot{g}^{ij} = \left( \begin{array}{c|c}
                 g^{ab} &  -g^{ab}\omega^{\beta}_{\ph{\beta}b}  \\
	            		  \hline
				 -\omega^{\alpha}_{\ph{\alpha}a}g^{ab}      & g^{\alpha\beta}  +
				      g^{ab}\omega^{\alpha}_{\ph{\alpha}a}\omega^{\beta}_{\ph{\beta}b} 
          \end{array}  \right)      \label{gmetug}
\end{equation}
   In this latter basis the trivialisation dependent components 
$\omega^{\alpha}_{\ph{\alpha}a}(p)$ of the connection 1-form  on $P$  for the 
non-Abelian internal symmetry are found alongside
 the Killing metric components $g_{\alpha\beta}$ and
 the external spacetime metric components $g_{ab}$,  framed within the elements of the 
full metric $\ddot{g}_{ij}$ on the bundle space. 
 The form of equation~\ref{gmetug} is preserved under any coordinate transformation on 
$P$ corresponding to a different choice of gauge section $\sigma: M_4 \to P$ (i.e. 
choice of trivialisation diffeomorphism $\psi: P \to M_4 \times G$) as well as under 
general coordinate transformations on $M_4$, but is not generally covariant under 
arbitrary changes of  coordinates on $P$.

  As described in the opening of subsection~\ref{one31} a principle bundle of linear 
frames can be constructed over any differentiable manifold, including the case for 
which the `base manifold' is actually the space of a given principle fibre bundle $P$ 
itself. 
  While the metrics $g$ and $K$ on the manifolds $M_4$ and $G$ can be naturally 
extended to the metric $\check{g}$ of equation~\ref{ggxyk} on the principle bundle $P 
= (M_4,G)$ with a gauge connection $\omega$, a  
  linear connection $\Gamma(x)$ on the manifold $M_4$  can also be generalised to the 
domain of the larger manifold $P$. As described for equation~\ref{gamene}
such a linear connection $\check{\Gamma}(p)$ will define covariant differentiation 
with $\check{\nabla} \check{e}_j = 
\check{\Gamma}^i_{\ph{i}j} \otimes \check{e}_i$ in a general tangent space basis 
$\{\check{e}_i\}$ for {\it TP} with dual basis $\{\check{e}^i\}$ for $T^\ast \! P$, 
where
\begin{equation}
 \label{gamform}
  \check{\Gamma}^i_{\ph{i}j} =  \check{\Gamma}^i_{\ph{i}jk} \check{e}^k 
\end{equation}
  are a set of linear connection 1-forms on $P$ as the base space of the frame bundle 
{\it FP}.  
The identification of the smooth symmetric gauge covariant rank-2 tensor field 
$\check{g}$ on $P$, equation~\ref{ggxyk},   endows the principle bundle itself with 
the structure of a pseudo-Riemannian manifold $(P,\check{g})$. In  
 turn a connection $\check{\Gamma}$ compatible with the metric $\check{g}$, and hence 
with the geometric structure of the underlying manifold $P$, may be extended from the 
notion of a metric connection on $M_4$, defining a  structure denoted 
$(P,\check{g},\check{\Gamma})$.

 Further guided by Einstein's general theory of relativity in 4-dimensional spacetime, 
the unique linear connection  which is torsion-free, $\check{\bT}=0$, and compatible 
 with the metric, $\check{\nabla} \check{g} = 0$, that is the Levi-Civita connection, 
may be defined on the bundle space $P$. The corresponding  connection coefficients can 
be written, defining $\Gamma_{ijk} = g_{il}\Gamma^l_{\ph{i}jk}$ and  $c_{ijk} = 
g_{il}c^{\;\! l}_{\ph{l}jk}$, as:
\begin{equation}
    \check{\Gamma}_{ijk} = \frac{1}{2} (\che_j(\chg_{ik}) + \che_k(\chg_{ij}) - 
\che_i(\chg_{jk}))
	                    -\frac{1}{2} (\chc_{ijk} + \chc_{kji} + \chc_{jki})   
\label{hgamijk}
\end{equation}
  which expresses equation~\ref{gtoGam} in a general frame. These coefficients take a 
relatively simple form in the  horizontal lift basis on $P$, as employed for the 
metric in equation~\ref{ggghlb} and the structure coefficients of 
equations~\ref{hatb1}--\ref{hatb3}, with a coordinate basis  adopted on the base space 
$M_4$. 
 In this basis the connection coefficients $\Gamma^a_{\ph{a}bc}(x)$ on the base space 
$M_4$ contribute to the set in equation~\ref{hgamijk} for $x=\pi(p)$, with:
\begin{equation}
  \acute{\Gamma}^a_{\ph{a}bc}(p) = \Gamma^a_{\ph{a}bc}(x) = \frac{1}{2} 
g^{ad}(e_b(g_{cd}) + e_c(g_{bd}) - e_d(g_{bc}))   \label{hgamext}
\end{equation}
which is simply equation~\ref{gtoGam}, since the structure coefficients
  $\acute{c}^{\;\! a }_{\ph{a}bc}(p)$
 on $P$  vanish in this basis.
 The connection coefficients $\check{\Gamma}^i_{\ph{i}jk}(p)$ are also related to the 
internal curvature through equation~\ref{hgamijk} since in the horizontal lift basis, 
by equation~\ref{hatb3},  we have 
 $\acute{c}^{\:\! \alpha}_{\ph{\alpha}ab}(p) = -F^{\alpha}_{\ph{\alpha}ab}(p)$.
 In fact via equation~\ref{hgamijk} we find in the horizontal lift basis on the bundle 
$P$ 
 terms such as (see \cite{Cho} equation~22): 
\begin{equation}
 \acute{\Gamma}^{\alpha}_{\ph{\alpha}ab} = +\frac{1}{2} {F}^{\alpha}_{\ph{\alpha}ab}
   \qquad \mbox{and} \qquad
 \acute{\Gamma}^a_{\ph{a}b\gamma} = \acute{\Gamma}^a_{\ph{a}\gamma b}  = 
      +\frac{1}{2}g^{ac} g_{\gamma\beta} {F}^{\beta}_{\ph{\beta}bc}   \label{hgamint}
\end{equation} 
  The complete set of coefficients for the Levi-Civita connection on $P$ is listed 
  under `Cho~\cite{Cho}' as the first case in table~\ref{Gamsetsr} in the following 
subsection.

 In turn the components of the Riemann curvature tensor $\check{R}^i_{\ph{i}jkl}$ can 
be calculated for this Levi-Civita connection on $P$ via equation~\ref{rabcd}. 
Hence this Riemann curvature on the total bundle space $P$ is intimately related to
\textit{both} the \textit{external} curvature on $M_4$ via equation~\ref{hgamext} and 
the \textit{internal} curvature, associated with gauge group $G$, which is drawn into 
the Riemannian geometry through equation~\ref{hgamint}.

 Having the metric $\chg_{ij}$ on $P$ the Ricci tensor $\check{R}_{jk} = \chg^{il} 
\check{R}_{ijkl}$ (equation~\ref{riccicon}) and scalar curvature $\check{R} = 
\chg^{ij} \check{R}_{ij}$ may also be computed, where here the latter is found in the 
horizontal lift basis to be (with differing sign convention to \cite{Cho}):   
\begin{equation}
  \acute{R}(p) = R_M + R_G + \frac{1}{4}F^2    \label{hrscal}
\end{equation}
 Here $R_M$ is the usual scalar curvature on the base manifold (which varies with the 
point $x = \pi(p) \inn M_4$ under $p\inn P$) and $R_G$ is the constant scalar 
curvature on the group manifold $G$.
 The term $F^2 = {F}^{\alpha}_{\ph{\alpha}ab}(p){F}_{\alpha}^{\ph{a}ab}(p)$, 
constructed from the connection on $P$  
 can be expressed as $F^2 = 
{F}^{\alpha}_{\ph{\alpha}\mu\nu}(x){F}_{\alpha}^{\ph{a}\mu\nu}(x)$ in terms of the 
non-Abelian gauge fields on $M_4$, with the
gauge covariant curvature components $F^{\alpha}_{\ph{\alpha}\mu\nu}(x)$ on the base 
space $M_4$   deriving from $F(x) = \sigma^{\ast} \Omega(p)$ as described
 for equation~\ref{fonbase} towards the end of subsection~\ref{one32}. 
 (In this sense the ${F}^{\alpha}_{\ph{\alpha}ab}$ entries in table~\ref{Gamsetsr} can 
be interpreted as curvature components directly on $M_4$).
 Hence each term in equation~\ref{hrscal} is gauge invariant.

  As a scalar $\acute{R}(p)$ in equation~\ref{hrscal} is a quantity which is also 
independent of the basis $\{\che_i\}$ in which it is determined. The equations of 
motion for the theory are then derived by adopting $\check{R}(p)$ as the scalar 
Lagrangian function together with $\sqrt{\vert \chg \vert} \, d^4 x \, d^{n_G} G$ as 
the invariant volume element, where $\vert \chg \vert$ is the magnitude of the 
determinant of the metric $\chg_{ij}$ on $P$, in the Einstein-Hilbert action integral 
on the bundle space:
\begin{equation}
  I_{m} = \int \check{R}\sqrt{\vert \chg \vert} \; d^4 x \; d^{n_G} G  \label{einhilf}
\end{equation}
 with $m=4+n_G$.
 The integration over the group manifold $G$, with volume $V_G$, is trivial and the 
above expression reduces to the 4-dimensional action integral:
\begin{equation}
   I_4 = V_G \int  \check{R}\sqrt{\vert g \vert} \; d^4 x    \label{einhilk}
\end{equation}
  where $\vert g \vert$ is here the determinant of the metric $g_{ab}$ on $M_4$. The 
variational principle is then applied under the constraint $\delta I_{m} = 0$, and 
hence $\delta I_{4} = 0$, with respect to restricted variations of the metric 
$\delta\chg$ on the bundle space, consistent with equation~\ref{rtrang}, as will be 
discussed further before equation~\ref{einfiek} in the following subsection. Within 
this restriction 
 this again follows the prescription for the original theory of general relativity on 
a \mbox{4-dimensional} spacetime manifold $M_4$ with scalar curvature $R \equiv R_M$ 
for which the field equations can be determined from the Einstein-Hilbert action 
integral of equation~\ref{einhil}.

 By comparison of equations~\ref{hrscal} and \ref{einhilk} with \ref{einhil} the 
constant $R_G$ in this version of  Kaluza-Klein theory appears as a cosmological 
constant term which, however, is problematically too large by a factor of $\sim \!\! 
10^{120}$ if a natural normalisation  is used with the length scale of the group space 
$G$ taken to be of order the Planck length~\cite{Cho}. 
On the other hand 
  the $F^2$ term in equation~\ref{hrscal}  effectively contributes the content for the 
matter Lagrangian $\lag$ in equation~\ref{einhil} in the form of equation~\ref{lagem} 
or \ref{lagym}. Hence, as a particularly elegant feature of Kaluza-Klein theory, the 
external \textit{geometry} of the 4-dimensional spacetime manifold along with a 
\textit{matter} contribution from the internal gauge fields is identified within a 
single \textit{geometrical} object in the form of the scalar curvature $\check{R}$ on 
the principle bundle space.

\subsection{Theories with Torsion on the Bundle}
\label{one34}

  The problematic cosmological term $R_G$ in equation~\ref{hrscal} can be addressed by
  exploiting the flexibility within the Kaluza-Klein approach on a principle fibre 
bundle that opens up if the metric $\chg_{ij}$ is \textit{not} treated as the 
fundamental field of the theory, as it was from the perspective of the paragraph 
leading to equation~\ref{rtrang}.
 While the same natural metric $\chg_{ij}(p)$ of equation~\ref{ggghlb} can be 
constructed on a bundle with a gauge \mbox{connection $\omega$},
 a linear connection $\check{\Gamma}^i_{\ph{i}jk}(p)$ on $P$ may be defined with  some 
independence from $\chg_{ij}(p)$, unlike the Levi-Civita connection of 
equation~\ref{hgamijk}.  In this case it is possible to derive a curvature scalar 
$\check{R}$ on $P$ such that the cosmological term vanishes, that is with $R_G=0$ in 
equation~\ref{hrscal} (see for example  \cite{Kopcz,OrzPau,Kalin,Katan}). 

   One way to achieve this is to require the linear connection 
$\check{\Gamma}^i_{\ph{i}jk}$ on $P$ to incorporate a description of absolute 
parallelism on the bundle fibres $G_x$ of figure~\ref{pbunbasis}.
 On the group manifold $G$ itself the list of canonical geometric objects includes a 
basis of left-invariant vector fields $\{X_{\alpha}\}$ and the Maurer-Cartan 1-form 
$\theta = X_{\alpha}\theta^{\alpha}$, which satisfies equation~\ref{maca2}, as well as 
the structure constants $\cstr$ and the Killing form metric $g_{\alpha\beta}$ of 
equation~\ref{killmet}.
 Employing the derivative action of the left-invariant basis vectors $\{X_{\alpha}\}$ 
the right-invariance of the Killing metric implies $X_{\alpha} g_{\beta\gamma} = 0$.  
In turn the covariant derivative, defined in terms of linear connection coefficients 
$\Gamma^{\alpha}_{\ph{\alpha}\beta\gamma}$ on $G$, of the Killing metric vanishes:
\begin{eqnarray}
    \nabla_{\!\alpha} g_{\beta\gamma}  & = & 
	       X_{\alpha} g_{\beta\gamma} - 
\Gamma^{\delta}_{\ph{\delta}\beta\alpha}g_{\delta\gamma}
	                               - 
\Gamma^{\delta}_{\ph{\delta}\gamma\alpha}g_{\beta\delta}
								    \;  = \; 0   \\
  \mbox{provided} \quad	\Gamalp & = & -\rho\, \cstr 
 \quad \mbox{for any} \quad \rho \inn \rrr		\label{gamlamc}					 
\end{eqnarray}
  by the antisymmetry in the indices of  $c_{\alpha\beta\gamma}$. 
  Hence for any value of $\rho$ this linear connection is metric compatible, with 
  $\nabla g = 0$ on $G$. 
  However  the torsion is zero only for $\rho = \fh$ which hence represents the unique 
Levi-Civita connection on $G$ defined in terms of the Killing metric  on the group 
manifold. For this case with the 
 linear connection defined with components $\Gamma^{\alpha}_{\ph{a}\beta\gamma} = -\fh 
c^{\alpha}_{\ph{a}\beta\gamma}$, while the torsion vanishes by equation~\ref{t2gc},
 the Riemann curvature is finite as can be seen from equation~\ref{rabcd}. 
  In general the curvature and torsion on any manifold are independent geometric 
concepts where either one may be non-zero while the other is zero.

  The Riemann curvature  is zero on $G$ only for the case of $\rho = 0$ or 
\mbox{$\rho=1$} in equation~\ref{gamlamc}, for which the torsion is finite, in 
contrast with the above Levi-Civita connection.
 	The choice of linear connection coefficients $\Gamalp = 0$ is equivalent to 
inducing parallel transport on the group manifold via the left action $L_h$  of $G$ on 
itself, for any $h\inn G$, that is with a complete parallelism on $G$  defined
 in terms of the  self-parallel frame composed 
 of left-invariant basis vector fields $\{X_{\alpha}\}$ on $G$.
   The other case with vanishing curvature for $\Gamalp = -\cstr$ in this 
left-invariant basis corresponds to a parallelism described by a right-invariant frame 
field under the action $R_h$. In either case  
 the resulting Riemann curvature vanishes with 
$R^{\alpha}_{\ph{\alpha}\beta\gamma\delta} = 0$, as can be shown using 
equation~\ref{rabcd} together with the Jacobi identity expressed in terms of the 
structure constants. 
  These latter two cases, while having finite torsion $ 
T^\alpha_{\ph{\alpha}\beta\gamma} =  \mp \cstr $ by equation~\ref{t2gc}, in describing  
an absolute parallelism on $G$ can be considered as geometrically natural metric 
connections on the group manifold.

  For a linear connection with components $\Gamalp = 0$ or $\Gamalp = -\cstr$   
employed on the bundle fibres $G_x$ a subset of the torsion components on $P$ are also 
necessarily non-zero,  with  $\check{T}^{\alpha}_{\ph{\alpha}\beta\gamma}(p) \neq 0$. 
Hence with the torsion allowed to be non-zero on the bundle space $P$ this version of 
Kaluza-Klein theory resembles the Einstein-Cartan theory on 4-dimensional spacetime 
for which $\Gamma$ and $g$ are treated as independent geometric objects. Here we 
briefly review four such approaches in the literature.

  In Kopczy\'{n}ski \cite{Kopcz} a  linear connection $\check{\Gamma}$ with finite 
torsion is constructed in terms of the structure on a principle bundle
with a gauge connection $\omega$ without reference to any metric. The `gravitational 
field' on $P$ is described by the combination of both the metric $\chg$ of 
equation~\ref{ggxyk} and the components of $\check{\Gamma}$ as listed in the 
corresponding column under `Kop \cite{Kopcz}' in table~\ref{Gamsetsr}. With these 
components the scalar curvature on $P$ is found to be $\check{R} = R_M + (\alpha 
-\alpha^2)K^2$, with $K^2 = K^{\alpha\beta}K_{\alpha\beta}$ (in \cite{Kopcz} the 
metric on $G$ is $g_{\alpha\beta} = \lambda K_{\alpha\beta}$, here we take $\lambda = 
1$ consistent with equation~\ref{killmet}). 
 For  Einstein-Cartan theory
  the connection is compatible with the metric, which is achieved by setting $\beta = 
0$ in row `5)' of the `Kop~\cite{Kopcz}' column of table~\ref{Gamsetsr}. 
 While this reference shows that the connection coefficients can be greatly simplified 
compared with the Levi-Civita case, as listed under `Cho \cite{Cho}' in the first 
column of table~\ref{Gamsetsr}, in order to achieve the correct dynamics a more 
complicated Lagrangian function is postulated with the scalar 
 $\check{R} + \frac{\mu}{2} \, \check{T}^i_{\ph{i}jk}\check{T}_i^{\ph{i}jk}$, 
including a quadratic torsion term, employed in place of $\check{R}$  alone in 
equation~\ref{einhilf}.
The cosmological constant $\Lambda$ obtained in this approach is arbitrary, and may be 
set to be zero or very small by a suitable choice of the parameters $\alpha$ and 
$\mu$.

 In Orzalesi and Pauri~\cite{OrzPau} the main motivation is to describe a linear 
connection $\check{\Gamma}$ on the principle bundle which is gauge covariant.
 In particular requiring the Ricci curvature on the fibre space to be gauge invariant 
implies the adoption of zero curvature on the group manifold, that is the case $\rho = 
0$ or $\rho = 1$ as described above after equation~\ref{gamlamc}. 
 This construction requires a relatively minimal modification of
  the Levi-Civita connection, as can be seen by comparing the entries of column 
`O$+$P~\cite{OrzPau}' with column `Cho~\cite{Cho}' in table~\ref{Gamsetsr}. Here the 
simple scalar Lagrangian $\check{R}$ on the bundle space is again adopted, 
 with the resulting vanishing of the  $\Lambda \equiv R_G $ term (as seen in the 
bottom line of table~\ref{Rsetsr})  interpreted as a consequence of the underlying 
gauge $G$-symmetry of the Riemannian geometry on $P$. Without  a finite $R_G$ term the 
vacuum solution corresponds to a zero Einstein tensor $G_{\mu\nu}(x) = 0$ together 
with zero internal curvature components $F^{\alpha}_{\ph{\alpha}\mu\nu}(x)=0$ on the 
base space $M_4$.

  In Kalinowski \cite{Kalin} the linear connection 1-forms 
   $\check{\Gamma}^{i}_{\ph{i}j} = \check{\Gamma}^{i}_{\ph{i}jk} \che^k$ of 
equation~\ref{gamform} on $P$ are defined as
     the horizontal part of the  Levi-Civita connection 1-forms, the latter here 
denoted  $\check{\Gamma}^{i}_{\!\!\! \mbox{\tiny L} \; j}$  with the components of 
equation~\ref{hgamijk} in general, as constructed in the horizontal lift basis. That 
is
  $\acute{\Gamma}^{i}_{\ph{i}j} = \acute{\Gamma}^{i}_{\!\!\! \mbox{\tiny L} \; j}
  \,\scirc\, \mbox{hor}$  in the notation of equation~\ref{omdef} (although here for 
the linear connection $\acute{\Gamma}(p)$ the manifold $P$ is considered as the base 
space for the Riemannian geometry), with
 the components of this linear connection $\acute{\Gamma}^{i}_{\ph{i}jk}$  listed in  
column `Kal~\cite{Kalin}' of table~\ref{Gamsetsr}. The factors of $\lambda$ arise as 
here the metric on $G$ is taken to be $g_{\alpha\beta} = \lambda^2 K_{\alpha\beta}$.
 This linear connection $\check{\Gamma}^{i}_{\ph{i}jk}$ is metrical, invariant under 
the $G$-action, again with non-zero torsion and, while motivated in the context of 
gauge derivatives of spinor fields, again leads to a vanishing cosmological constant 
as seen in the bottom row of table~\ref{Rsetsr}.

 In Katanaev \cite{Katan} an initially completely general 
$\check{\Gamma}^i_{\ph{i}jk}(p)$ on the principle bundle manifold is considered. Four 
conditions are postulated for $\check{\Gamma}$ in 
a geometrically meaningful way related to
 the structure group $G$ over $P$  and, as for the previous reference, with emphasis 
on  horizontal propagation. In particular for column `Kat~\cite{Katan}' of 
table~\ref{Gamsetsr}
 on taking $c=1$ for entry `5)' $\acute{\Gamma}^{\alpha}_{\ph{\alpha}ab}= c 
F^{\alpha}_{\ph{\alpha}ab}$   the change in a tangent vector to $P$ under parallel 
transport using these linear connection coefficients equals the change in the vector 
due to the basis transformation under parallel transport of the fibres using the gauge 
connection. Entry `4)' in this column is included for compatibility with the natural 
metric of equation~\ref{ggxyk}.
The coefficients listed represent the case presented in  \cite{Katan} with finite 
torsion and the absence of a cosmological constant term, although a different choice 
of $\check{\Gamma}$ consistent with the postulates is possible. A small modification   
within this framework would be to set  entries `1)' and `2)' equal to zero under 
column `Kat~\cite{Katan}' in table~\ref{Gamsetsr}, corresponding to taking $\rho = 0$ 
rather than $\rho=1$ in  equation~\ref{gamlamc}. This reference is of significance for 
the present paper in that it highlights the possibility of a natural geometric origin 
for $\check{\Gamma}$ on $P$ without any appeal to the Levi-Civita connection.

   The complete set of linear connection coefficients for reference \cite{Cho}, 
augmenting equations~\ref{hgamext} and \ref{hgamint}, are collected in the first 
column of table~\ref{Gamsetsr}. These are listed alongside the linear connection 
coefficients  $\acute{\Gamma}^{i}_{\ph{i}jk}$ in the horizontal lift basis on the 
bundle space $P$ for the above four cases with non-zero torsion. Where necessary signs 
have been aligned to the conventions used here, with the linear connection index order 
in    
    $\acute{\Gamma}^{i}_{\ph{i}jk}$
    as described for equation~\ref{gamene} together with the conventions listed for 
items 1) -- 3) in subsection~\ref{one31}.

\begin{table}[htbp]
\centering
\begin{tabular}{|l||c|c|c|c|c|}
 \hline
    $\quad \acute{\Gamma}^i_{\ph{i}jk}$ & Cho~\cite{Cho} 
  &	Kop~\cite{Kopcz}  &   O$+$P~\cite{OrzPau} 
    & Kal~\cite{Kalin}  & Kat~\cite{Katan}   \\  
 \hline				
1)  $\acute{\Gamma}^{\alpha}_{\ph{\alpha}\beta\gamma}$  &
 $-\fhs \cstr$ & $\,\:\;-\alpha \cstr\,\:\;$ & $-\cstr$ or $0$ & 0 & $-\cstr$ 
   \\ 
 \hline
2)  $\acute{\Gamma}^{\alpha}_{\ph{\alpha}\gamma a}$  &
 0 & 0 & 0 & 0 & $-\omega^{\beta}_{\ph{\beta}a}
                      \acute{\Gamma}^{\alpha}_{\ph{\alpha}\beta\gamma}  $ 
  \\ 
  \hline
3)  $\acute{\Gamma}^{a}_{\ph{a}b\gamma}$  & 
   $\!\fhs g^{ac}g_{\gamma\beta}F^{\beta}_{\ph{\beta}bc}\!$ & 0 &
      $\!\fhs g^{ac}g_{\gamma\beta}F^{\beta}_{\ph{\beta}bc}\!$ & 0 & 0 
	   \\
 \hline
4)   $\acute{\Gamma}^{a}_{\ph{a}\gamma b}$  & 
   $\!\fhs g^{ac}g_{\gamma\beta}F^{\beta}_{\ph{\beta}bc}\!$ & 0 &
   $\!\fhs  g^{ac}g_{\gamma\beta}F^{\beta}_{\ph{\beta}bc}\!$ &      
    $\!\frac{\lambda}{2}g^{ac}g_{\gamma\beta}F^{\beta}_{\ph{\beta}bc}\!$
	   &  $\! c g^{ac}g_{\gamma\beta}F^{\beta}_{\ph{\beta}bc}\!$
	    \\
\hline
5)   $\acute{\Gamma}^{\alpha}_{\ph{\alpha}ab}$  & 
   $ \fhs F^{\alpha}_{\ph{\alpha}ab}$ & $\beta F^{\alpha}_{\ph{\alpha}ab}$  &
   $\fhs F^{\alpha}_{\ph{\alpha}ab}$ & $\frac{\lambda}{2}F^{\alpha}_{\ph{\alpha}ab}$ &
   $c F^{\alpha}_{\ph{\alpha}ab}$ 
      \\
 \hline
6)   $\acute{\Gamma}^a_{\ph{a}bc}$   & $\Gamabc$ & $\Gamabc$ & $\Gamabc$ 
       & $\Gamabc$ & $\Gamabc$  
	     \\
	   \hline 
  \end{tabular}
  \caption{\setb Linear connection components $\acute{\Gamma}^i_{\; jk}$ on a 
principle bundle extracted from \protect\cite{Cho} equation~22, \protect\cite{Kopcz} 
page~367, \protect\cite{OrzPau} equation~19, \protect\cite{Kalin} equation~29, and the 
case in \protect\cite{Katan} with non-zero torsion on $G$.
 The $\{\alpha,\beta,\gamma\ldots\}$ and $\{a,b,c\ldots\}$  index convention is 
explained near the opening of subsection~\ref{one33} with reference to 
figure~\ref{pbunbasis}. 
 All components are expressed in the horizontal lift basis and 
$\acute{\Gamma}^{a}_{\;\beta\gamma}= \acute{\Gamma}^{\alpha}_{\; a\gamma}= 0$ in all 
five cases. Each of
 $\lambda > 0$, $\alpha$, $\beta$ and $c$, where used as coefficients, are real 
constant parameters.}
\label{Gamsetsr}
\end{table} 

   Only the first case in table~\ref{Gamsetsr} describes a torsion-free linear 
connection, yet each of the five cases is a Kaluza-Klein theory providing a unifying 
structure for general relativity together with gauge field theory. In part the purpose 
of collecting together this range of linear connection coefficients on $P$ together 
with their motivating arguments is to demonstrate that a significant degree of 
flexibility is possible within Kaluza-Klein theory while still maintaining this 
unified framework.

 For any linear connection on the bundle space $P$, such as defined by any of the five 
sets of connection coefficients $\acute{\Gamma}^i_{\ph{i}jk}$ listed in 
table~\ref{Gamsetsr}, the Riemann curvature tensor can be determined according to 
equation~\ref{rabcd} applied  in the horizontal lift basis on $P$. The corresponding
 Ricci  curvature components $\acute{R}_{\alpha\beta}$ and $\acute{R}_{ab}$ are listed 
here in the first and fourth rows of table~\ref{Rsetsr} for these five theories.
In all cases the entries in this table calculated here agree with the corresponding 
equations of the respective references within the sign conventions adopted.

\begin{table}[htbp]
\centering
 \hspace*{-19pt}
\begin{tabular}{|r||c|c|c|c|c|}
 \hline
     & Cho~\cite{Cho} 
  &	Kop~\cite{Kopcz}  &   O$+$P~\cite{OrzPau} 
    & Kal~\cite{Kalin}  & $\!\!$Kat \cite{Katan}$\!\!$ 
	  \\  
 \hline				
 $\acute{R}_{\alpha\beta} = R_{(G)\alpha\beta} \, + \!$  &
 $-\frac{1}{4}F_{\alpha bd}F_{\beta}^{\ph{\beta}bd}$      &
  0 & $-\frac{1}{4}F_{\alpha bd}F_{\beta}^{\ph{\beta}bd}$  & 0 & 0 \\ 
 \hline
  $ R_{(G)\alpha\beta} = \!$  &
   $\frac{1}{4}K_{\alpha\beta}$ &  $\!(\alpha - \alpha^2) K_{\alpha\beta}\!$
    & 0 & 0 & 0 \\ 
  \hline
  $K^{\alpha\beta}\acute{R}_{\alpha\beta} = \!$  & 
   $R_G - \frac{1}{4}F^2$  &  $R_G  $ &
      $- \frac{1}{4}F^2$ & 0 & 0 \\
 \hline
  $\acute{R}_{ab} = R_{(M)ab} \, + \!$  & 
   $\fh F^{\beta}_{\ph{\beta}ad}F_{\beta b}^{\ph{\beta b}d}$ & 0 &
      $\fh F^{\beta}_{\ph{\beta}ad}F_{\beta b}^{\ph{\beta b}d}$  &
	  $\frac{\lambda^2}{4} F^{\beta}_{\ph{\beta}ad}F_{\beta b}^{\ph{\beta b}d}$
	  & $c^2 F^{\beta}_{\ph{\beta}ad}F_{\beta b}^{\ph{\beta b}d}$
     \\
\hline
   $g^{ab}\acute{R}_{ab} = R_M\, + \!$   & 
   $\fh F^2$ & 0  &
   $\fh F^2$ & $\frac{\lambda^2}{4}F^2$ &
   $c^2 F^2$  \\
 \hline
  $\acute{R} = \acute{g}^{ij}\acute{R}_{ij} = \!$   &
     $\! R_{\! M} \! + \! R_{\! G} \! + \! \frac{1}{4}F^2 \!$  & $R_M + R_G$   &
	  $R_M+\frac{1}{4}F^2$ &  $R_M+\frac{\lambda^2}{4}F^2$  &
	    $R_M+c^2F^2$   
  \\
	   \hline 
  \end{tabular}
  \caption{\setb Composition of the scalar curvature $\acute{R}$ on the bundle space 
for the five cases of table~\ref{Gamsetsr}. Contributions to the components of the 
Ricci curvature on the bundle include 
 $R_{(G)\alpha\beta}$ and $R_{(M)ab}$ from the group manifold and base space 
respectively,
 with  
 $R_G = K^{\alpha\beta}R_{(G)\alpha\beta}$ and $R_M = g^{ab}R_{(M)ab}$ being the 
respective scalar curvatures.}
\label{Rsetsr}
\end{table} 

   The scalar curvature constructed in the horizontal lift basis on the principle  
bundle space can be written as:
\begin{equation}
   \acute{R}\, = \, \acute{g}^{ij}\acute{R}_{ij} \,  = \,
   g^{ab}\acute{R}_{ab} \, + \, K^{\alpha\beta}\acute{R}_{\alpha\beta} 
    \label{rtwobits}
\end{equation}
  owing to the simple form of the metric $\acute{g}$ in this basis as expressed in 
equation~\ref{ggghlb}. Hence the Ricci curvature components $\acute{R}_{a\beta}$ and 
$\acute{R}_{\alpha b}$ are not required in order to determine the scalar curvature on 
the bundle.
  
  If  the four factors of $\fh$ in the `Cho~\cite{Cho}' column in 
table~\ref{Gamsetsr}, 
   for the case of the Levi-Civita connection coefficients 
$\acute{\Gamma}^i_{\ph{i}jk}$ on the bundle, listed in rows `1)', `3)', `4)' and `5)' 
are replaced by the real factors $f_1$, $f_3$, $f_4$ and $f_5$ respectively then the 
scalar curvature in the horizontal lift basis is found to be:
\begin{eqnarray}
  \label{rrrffff}
   \acute{R} & = & R_M \, + \, R_G \, + \, \chi F^2  \\
   \mbox{with} \quad \chi & = &  f_3 - f_3f_4 + f_4f_5 - f_3f_5   \label{rrrffff2}
\end{eqnarray}
  These equations reproduce the scalar curvature for the Levi-Civita case, with each 
$f_i = \fh$, as quoted originally in equation~\ref{hrscal}, and also apply to each 
subsequent case of table~\ref{Gamsetsr} as quoted in the final row of 
table~\ref{Rsetsr}.

Equations~\ref{rrrffff} 
 and \ref{rrrffff2} show that $f_3$ is the only coefficient which is sufficient in 
itself to introduce a non-trivial $F^2$ term, alongside $R_M$, into the scalar 
curvature $\acute{R}$, and hence including  
 $\acute{\Gamma}^{a}_{\ph{a} bc}  =   {\Gamma}^{a}_{\ph{a} bc}$ and
   $\acute{\Gamma}^{a}_{\ph{a}b\gamma} = 
         f_3 \,  g^{ac}g_{\gamma\beta}F^{\beta}_{\ph{\beta}bc}$
 as the only non-zero  $\acute{\Gamma}^i_{\ph{i}jk}$ coefficients  
 might be considered as a further, `minimal', Kaluza-Klein model. While  not developed 
here as a serious physical proposal this minimal model further demonstrates the 
flexibility within the Kaluza-Klein framework, obtaining the appropriate link between 
the external geometry and internal curvature with a seemingly much simpler linear 
connection on the bundle compared with the Levi-Civita case.  More generally, 
equations~\ref{rrrffff} and \ref{rrrffff2} display the mutual consequences of the 
non-zero $\acute{\Gamma}^i_{\ph{i}jk}$ terms for the models listed in 
table~\ref{Gamsetsr}.

  The derivation of Einstein's equations in 4-dimensional spacetime from the 
Einstein-Hilbert action of equation~\ref{einhil} was described in 
subsection~\ref{one31}. In the vacuum case with $\lag=0$ and $\Lambda = 0$ variation 
of the metric $\delta g_{\mu\nu}$ on $M_4$ leads to the equation of motion $G^{\mu\nu} 
=0$ of equation~\ref{lagtoein}.
  For the Kaluza-Klein extension to a scalar curvature $\check{R}$  on a principle 
bundle space the same steps lead
 to the requirement of the stationarity of 
 the action integral over the full bundle space in equation~\ref{einhilf}, that is 
$\delta I_{m}=0$, under variation of the extended metric $\delta \chg_{ij}$ on $P$. 
   However for the  Kaluza-Klein theories described here the variations in the metric 
$\chg_{ij}$ on the bundle space
are  \textit{not} arbitrary since 
 the right-invariance  of equation~\ref{rtrang} and the general form of the metric in 
equation~\ref{gmetug} should be preserved on $P$.
 This limits the metric variations to the components $\delta g_{ab}$ and $\delta 
\omega^{\alpha}_{\ph{\alpha}a}$ in equation~\ref{gmetug} and leads to two equations of 
motion on the base manifold $M_4$. Applying the variation $\delta g_{ab}$ under 
$\delta I_{m} = 0$ for the action in equation~\ref{einhilf}, with the curvature 
$\acute{R}$ of equation~\ref{rrrffff},  leads to the generally non-zero solution 
expressed in a general coordinate basis on $M_4$ 
(see for example \cite{Kalin}~page~394):
\begin{equation}
  G^{\mu\nu} + R_G g^{\mu\nu}  \: = \:
  2\chi( - F^{\alpha \mu}_{\ph{\alpha \mu}\rho}F_{\alpha}^{\ph{\alpha}\rho\nu}
	                -\frac{1}{4} g^{\mu\nu} \, F^{\alpha}_{\ph{\alpha}\rho 
\sigma}F_{\alpha}^{\ph{\alpha}\rho\sigma} )  \: = \:
   -\kappa T^{\mu\nu}
      \label{einfiek}      
\end{equation}
  A necessarily finite cosmological term with $R_G \neq 0$ arises only for 
    the case of the Levi-Civita connection in the `Cho~\cite{Cho}' columns of 
tables~\ref{Gamsetsr} and \ref{Rsetsr}, while the factor of $\chi$ is determined by 
equation~\ref{rrrffff2}.  On the other hand the variation $\delta 
\omega^{\alpha}_{\ph{\alpha}a}$ on the bundle leads, on the spacetime manifold $M_4$, 
to:
\begin{equation}
	D_{\mu}F^{\alpha \;\! \mu\nu} \: = \:  0    \label{yangmk}
\end{equation}
   Equation~\ref{einfiek} with $R_G = 0$ is the Einstein field equation with an 
energy-momentum tensor $T^{\mu\nu}$, reproducing equation~\ref{temkk}, composed purely 
from the gauge fields $A^{\alpha}_{\ph{\alpha}\mu}(x)$, with the latter being subject 
to equation~\ref{yangmk} which is the Yang-Mills field equation (or Maxwell's equation 
$\nabla_{\! \mu}F^{\mu\nu}_{\ph{\mu\nu}}=0$ in the case of the Abelian internal 
symmetry group $G=\uo$).  Hence the source-free Yang-Mills field equation~\ref{ymol} 
has been derived \textit{without} the explicit introduction of the Yang-Mills 
Lagrangian of equation~\ref{lagym}. Rather 
such a `Lagrangian term'
  $F^2 = F^{\alpha}_{\ph{\alpha}ab}(p) F_{\alpha}^{\ph{\alpha}ab}(p)
    \equiv F^{\alpha}_{\ph{\alpha}\mu\nu}(x) F_{\alpha}^{\ph{\alpha}\mu\nu}(x)$ has 
been incorporated  within an Einstein-Hilbert action deriving from the geometry of the 
bundle space.
 In this way  non-Abelian Kaluza-Klein theory provides a unified framework for the 
combined Einstein-Yang-Mills field equations.

\section{Geometric Unification through One Dimension}
\label{one4}


\subsection{Construction of a Linear Connection on $P \equiv M_4 \times G$}
\label{one41}

   In this section,  guided by the framework of Kaluza-Klein theories described in the 
previous section, the aim is to determine a relation between the external and internal 
geometry  arising out of the symmetries of the full multi-dimensional form of temporal 
flow $\lvh$, as originally derived from the one dimension of time in 
subsection~\ref{one21} and building upon the structures described in 
section~\ref{one2} generally.

 Initially in subsection~\ref{one22} we considered the form $L(\bv_4) = 
\eta_{ab}v^av^b = 1$ of equation~\ref{lorl},
  with Minkowski metric
    $\eta = \mbox{diag}(+1,-1,-1,-1)$ and  Lorentz $\soot$ symmetry. These structures 
are projected locally onto a 4-dimensional base space $M_4$, which itself derives from 
the translation symmetry of $\lvf$, as depicted in figure~\ref{mtogmap}.
  An extension from equation~\ref{lorl} to equation~\ref{lvnine} for a full 
9-dimensional form 
  $L(\bvh)= L(\bv_9) = \det(\bv_9) = 1$ with $\bv_9 \inn \hthc$ and full symmetry 
group  $\hat{G}=\slthc$, via $\sltc$ as the double cover of the Lorentz group, was 
then described in subsection~\ref{one23}.
   In this case the extended base manifold $M_4$ arises out of four of the nine 
translational degrees of freedom of $\lvni$ leading directly to the symmetry breaking 
structure described for figure~\ref{mtogmaph}.

 The breaking of the full symmetry through 
 the extraction of a preferred subgroup  $\sltc \subset \slthc$
 acting on the tangent space $\TM_4$ of the external spacetime $M_4$, as described for 
equations~\ref{vinhinc}--\ref{symb3},  leaves a residual internal $\uo$ symmetry, as 
described for equation~\ref{suinsth}.
 Given that both $\sltc$ and $\uo$ are contained within the initial unbroken full 
symmetry $\slthc $
 of figure~\ref{mtogmaph}(a)  a correlation between the external curvature ${\bR}$ and 
internal curvature ${F}$ is implied in the symmetry breaking to the structure of 
figure~\ref{mtogmaph}(b), in particular with the case of both 
${\bR}=0$ and ${F}=0$ simultaneously possible.

 A principle bundle structure 
$P \equiv M_4 \times \uo$ underlying figure~\ref{mtogmaph}(b), representing the space 
of broken symmetries of the full form of time $\lvni$, 
   emerges in the identification of the base manifold $M_4$. 
  The base space $M_4$ is also naturally associated with the frame bundle $\FM_4$, 
which is itself a particular type of principle fibre bundle as described in the 
opening of subsection~\ref{one31}.
  In the symmetry breaking the degrees of freedom of the $\sltc \subset \slthc$ 
subgroup part of the original full \textit{gauge} connection are converted into the 
freedom of a \textit{linear} connection on $M_4$. That is,
  an $\sltca$-valued connection on ${P}$ can be extended to a
   $\mbox{gl}(4,\rrr)$-valued connection on the frame bundle, together with the 
associated  tetrad $e^{\mu}_{\ph{\mu}a}(x)$ and metric $g_{\mu\nu}(x)$ fields on 
$M_4$, as related in equation~\ref{guneen} and familiar from the theory of general 
relativity, as also described in subsection~\ref{one31}.

  For the generalisation with $\hG$ as the full isochronal symmetry of the full 
multi-dimensional form of time $\lvh$ the same symmetry breaking mechanism results in 
the bundle structure  $P \equiv M_4 \times G$, where $G$ is the internal symmetry 
identified as described for equation~\ref{dirprod}. This principle bundle again
 implicitly combines the geometry of the frame bundle on $M_4$,
  described in terms of a metric $g(x)$ and Riemann curvature $\bR(x)$ with an 
underlying local $\soot$ symmetry, and the geometry of $P$ itself, expressed in terms 
of a gauge field $A(x)$ and gauge curvature $F(x)$ on $M_4$ with an internal local $G$ 
symmetry.  
  Hence, in accommodating both the structure of the external geometry on $M_4$ and 
that of the internal gauge group $G$  in the bundle space, in principle all the 
necessary geometric structures
 for relating the external and internal curvature  can be identified on the bundle 
space $P \equiv M_4 \times G$.

 For the original case with a full form of simply $\lvf$ a preferred  globally defined 
orthonormal basis can be identified 
 on  the extended Riemannian manifold $M_4$, as described 
 in subsection~\ref{one22}, 
 supporting  a linear connection with all coefficients $\Gamma^a_{\ph{a}bc}(x)=0$.
  Such a global frame is 
 adapted to  the natural absolute parallelism and 
  canonical zero curvature  described for figure~\ref{mtogmap}.
 A Lie group manifold ${G}$ also supports a natural absolute parallelism with all 
linear connection coefficients $\Gamma^{\alpha}_{\ph{\alpha}\beta\gamma}(h)=0$ for all 
$h\inn G$ in a left-invariant basis, also with zero Riemann curvature, as described 
following equation~\ref{gamlamc} in subsection~\ref{one34}.

   The question can then be asked concerning the possible generalisation to a natural 
parallelism and corresponding linear connection $\check{\Gamma}^i_{\ph{i}jk}(p)$ on 
the larger space of symmetries $P \equiv M_4 \times G$, and the manner in which this 
may perturb the external curvature on the base manifold $M_4$.  
 In particular we seek  a linear connection on $P$, taking precedence over any 
possible metric structure,  for which  parallel transport in the  horizontal and 
vertical directions on $P$  closely reflects the geometry of the base manifold $M_4$.
 While the principle bundle space $P \equiv M_4 \times G$ is \textit{not} itself 
considered as a physical space, it opens up an `internal' freedom in this larger space 
of symmetries for the broken form $\lvh$ that might in principle perturb the parallel 
transport on the external physical spacetime $M_4$ component.

  Given a gauge connection $\omega$ on $P \equiv M_4 \times G$ associated with the 
internal  symmetry $G$, which in general is \textit{not} flat, a distinguished 
horizontal subspace in the tangent space of $P$ is  identified.
 In turn
  any tetrad basis $\{e_a\}$ on $M_4$ can be mapped from any point $x \inn M_4$ into 
the natural horizontal components of the tangent space at any point $p \inn P$
 with $\pi(p) = x$, that is 
 to the corresponding horizontal lift basis vectors $\{\acute{e}_a\}$, as related in 
figure~\ref{pbunbasis}. The horizontal lift basis is of direct physical significance 
for the internal geometry as described for equation~\ref{hatb3}, with 
$F^{\alpha}_{\ph{\alpha}ab}(p)$ being the internal curvature components. In connecting 
the fibres of $P$ over $M_4$ the geometry described by the horizontal subspace in 
principle provides a means of perturbing the geometry of the base manifold itself. In 
order to study this structure we introduce a linear connection $\acute{\Gamma}(p)$ in 
the horizontal lift basis on $P$ which defines a parallel transport in the tangent 
space {\it TP}  describing the internal relation between the fibres over $M_4$ 
associated with the connection $\omega$.

  That is, the geometry on $P$ is interpreted  \textit{both} in terms of a gauge 
connection $\omega$ with curvature $\Omega(p)$ on the principle bundle space and at 
the same time in terms of a linear connection $\acute{\Gamma}(p)$ with $P$ considered 
as the base space of {\it FP}. 
 Extending the structure of $\Gamma(x)$ on $M_4$ to $\acute{\Gamma}(p)$ on $P \equiv 
M_4 \times G$ mirrors the extra-dimensional extension of general relativity in 
Kaluza-Klein theories, as described in section~\ref{one3}, relating the gauge symmetry 
structure to a structure of Riemannian geometry on $P$ itself.

 For completeness
 a full set of linear connection coefficients 
   $\acute{\Gamma}^i_{\ph{i}jk}(p)$ on the bundle space $P \equiv M_4 \times G$ 
   is constructed in the horizontal lift basis  $\{{\acute e}_i\} = \{{\acute 
e}_{\alpha}, {\acute e}_a\}$, which was depicted in figure~\ref{pbunbasis} after which 
we also described the index convention used here.
  With emphasis on parallel transport in the horizontal directions  we take parallel 
propagation in the vertical directions to be trivial in this basis,  with:
\begin{equation}
\label{gamvert}
  \acute{\Gamma}^a_{\ph{a}b\gamma} \; = \;
  \acute{\Gamma}^{\alpha}_{\ph{\alpha}b\gamma} \; = \;
  \acute{\Gamma}^a_{\ph{a}\beta\gamma} \; = \;  
  \acute{\Gamma}^{\alpha}_{\ph{\alpha}\beta\gamma} \; = \; 0
\end{equation}

  That is, with vanishing coefficients for propagation along the fibres of the bundle 
as indicated by the Greek $\gamma$ for the third index.
  Here the final set $\acute{\Gamma}^{\alpha}_{\ph{\alpha}\beta\gamma}(p) =  0$ is 
essentially imported from the parallelism 
${\Gamma}^{\alpha}_{\ph{\alpha}\beta\gamma}(h) =  0$ on the manifold $G$ and applied 
to vertical transport generally on $P$.
 This vertical structure is similar to that for the linear connection on the bundle as 
described for the  `Kal~\cite{Kalin}' column of table~\ref{Gamsetsr}  in 
subsection~\ref{one34}, which was constructed with 1-forms 
 $\acute{\Gamma}^{i}_{\ph{i}j} \equiv \acute{\Gamma}^{i}_{\ph{i}j} \, \scirc \, 
  \mbox{hor}$   
 in the horizontal lift basis and hence with each 
   $\acute{\Gamma}^{i}_{\ph{i}j\gamma} =  0$
  as for equation~\ref{gamvert}.

  With the focus upon parallel propagation in the horizontal directions on $P$, 
conceived from a geometrical point of view as perturbing the geometry of the base 
manifold $M_4$, which is of primary physical interest as the external spacetime,
   the relevant  structures are sketched in figure~\ref{paraCD}.
In general, for finite external Riemann curvature $\bR \neq 0$ on $M_4$ the parallel 
transport of a given tangent vector $\bu_1 \inn T_{x_1}\!M_4$ from $x_1 \inn M_4$ 
along two different paths $C$ and $D$ to a point $x_2 \inn M_4$ will result in two  
different vectors $\bu_{2C},\bu_{2D} \inn T_{x_2}M_4$. Similarly for non-trivial 
internal curvature $\Omega \neq 0$ on $P$ the horizontal lifts of the curves $C$ to 
$C'$ and $D$ to  $D'$  will generally lead from any given point $p_1 \inn P$, with 
$\pi(p_1) = x_1$,  to two different points $p_{2C'}$ and $p_{2D'}$ on the fibre over 
$x_2 \inn M_4$, as also depicted in figure~\ref{paraCD}. The aim is then to understand 
how the external and internal curvature might be mutually dependent for the space of 
symmetries and geometric structure $P \equiv M_4 \times G$  arising from the symmetry 
of the full form $\lvh$ broken over $M_4$. 
\begin{figure}[htb]  
\centering
\epsfxsize=10.5cm
\leavevmode
\epsffile[0 0 1282 1041]{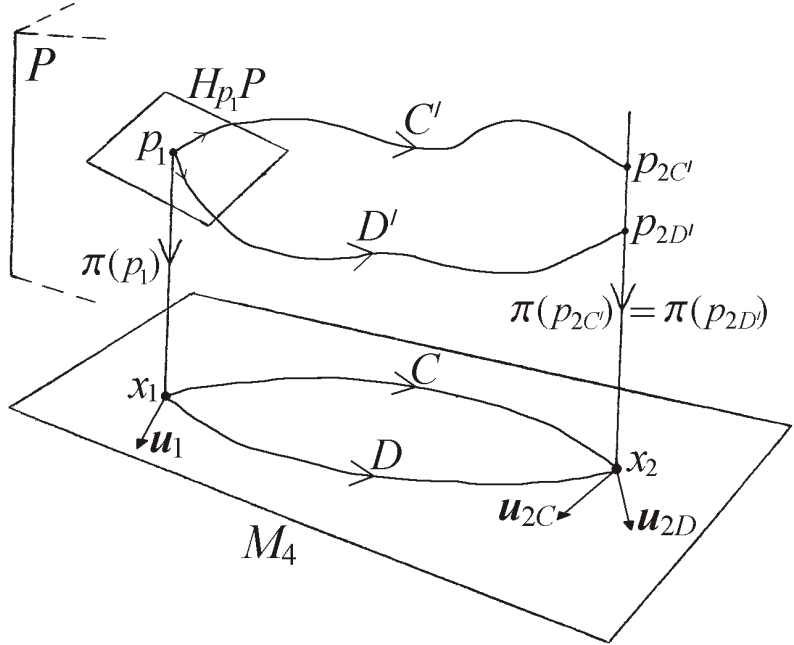}
\vspace{-5pt}
\caption{\setb Parallel transport of $\bu_1 \inn T_{x_1}\!M_4$ over two paths $C$ and 
$D$ between the same two points $x_1,x_2$ on the base space $M_4$ and the respective 
horizontal lifts $C'$ and $D'$ between the corresponding fibres in the principle 
bundle space $P$. While the geometric structure is similar to that reviewed in 
subsection~\ref{one32} here the bundle space $P \equiv M_4 \times G$ derives from the 
broken symmetries of the full temporal form $\lvh$, as described for the example of 
figure~\ref{mtogmaph}(b).}
\label{paraCD}
\end{figure}

  We note here that the diagram in figure~\ref{paraCD} might be interpreted quite 
literally for a toy model with $P \equiv M_2 \times \uo$, with for example a local 
SO$^+$(1,1) symmetry on $M_2$ and  fibres with a single parameter $\alpha \inn \rrr$ 
for $e^{i\alpha} \inn \uo = G$ as the internal gauge symmetry group. However the 
figure also holds metaphorically both for the bundle $P \equiv M_4 \times \uo$ arising 
from the $\lvni$ model of figure~\ref{mtogmaph}(b) and for the more realistic case $P 
\equiv M_4 \times G$ arising from a yet higher-dimensional form $\lvh$, where in 
general $G$ is  a non-Abelian internal symmetry group, and it is this more general 
case that we describe here.

  Here  the physical significance of the linear connection $\acute{\Gamma}(p)$ on $P$ 
derives from its relation to a linear connection $\Gamma(x)$ on the base manifold 
$M_4$, and the associated Riemannian geometry, when generalised for the larger 
 space of symmetries of the full form $\lvh$ broken over the base manifold.
 With emphasis on horizontal propagation, generalised from the manifold $M_4$ into the 
space $P \equiv M_4 \times G$,
  a relation between the \textit{local} structure of parallel transport on the bundle 
described by a linear connection $\acute{\Gamma}(p)$ and the geometry of an internal 
gauge connection relating the fibres in the horizontal directions will be established.
 The horizontal propagation on $P$
  is described by the four sets of linear connection coefficients with a Latin $c$ 
replacing the $\gamma$ index
  for each set in equation~\ref{gamvert}.
  We consider each of these four cases in turn.
  
 The parallel transport of $\acute{e}_b$ along $\acute{e}_a$ on the bundle $P$ when 
projected onto $M_4$ with $e_b = \pi_{\ast}(\acute{e}_b)$ and  $e_a = 
\pi_{\ast}(\acute{e}_a)$ is taken to be equivalent to the parallel transport of $e_b$ 
along $e_a$ directly on the base space $M_4$. This is essentially the same motivation 
for these coefficients as for all of the references summarised in table~\ref{Gamsetsr} 
and, with $\pi(p) = x$, implies that:
\begin{equation}
 \label{gamabcg}
  \acute{\Gamma}^a_{\ph{a}bc}(p) \; = \; {\Gamma}^a_{\ph{a}bc}(x)
\end{equation}

  However the parallel transport of $\acute{e}_b$ along $\acute{e}_a$ also has freedom 
in the vertical directions of the bundle space, as parametrised  by the connection 
coefficients $\acute{\Gamma}^{\alpha}_{\ph{\alpha}bc}(p)$. That is, the covariant 
derivative for the vector field $\acute{e}_b$ with respect to $\acute{e}_a$ can, from 
equation~\ref{gamene}, be written generally as:
\begin{equation}
\label{covhor}
  \acute{\nabla}_{\!\acute{e}_a} \acute{e}_b  \; = \;
   \acute{\Gamma}^i_{\ph{i}ba} \acute{e}_i \; = \;
   \acute{\Gamma}^c_{\ph{c}ba} \acute{e}_c \; + \;
   \acute{\Gamma}^{\alpha}_{\ph{\alpha}ba} \acute{e}_{\alpha}
\end{equation}

  In order to determine a natural expression for the coefficients 
$\acute{\Gamma}^{\alpha}_{\ph{\alpha}ba}(p)$, corresponding to a natural parallelism 
on the bundle space, we first briefly review another standard structure of 
differentiable geometry. In general on any differentiable manifold $M$ there is a 
one-to-one correspondence between a vector field $V$ and its flow $\Phi_t$, associated 
with the integral curves of $V$. The flow $\Phi_t$ is a one-parameter group of 
transformations on $M$, and a diffeomorphism for each value of the parameter $t\inn 
\rrr$. As a diffeomorphism $\Phi_t$ induces a pull-back map of tensor fields of any 
type in the space $T^r_s\!M$ on $M$, that is:
\begin{equation}
   \Phi^{\ast}_t \; : \; T^r_s\!M \to T^r_s\!M
\end{equation}
  which is known as Lie transport or Lie dragging.
   While the geometric origin  differs Lie transport in general shares a number of 
features in common with parallel transport via a linear connection. Both forms of 
transport preserve the $(r,s)$ degree of a tensor field and commute with tensor 
products and contractions. In addition both Lie transport and parallel transport give 
rise to derivatives that are derivations of the tensor algebra. The Lie derivative of 
a tensor field $X \inn T^r_s\!M$ with respect to a vector field $V$ is defined by:
\begin{equation}
\label{lieder}
  \lag_V X \; := \; \frac{d}{dt} \Big\vert_{t=0}  \Phi^{\ast}_t X
           \; = \; \lim_{t \to 0} \,
  \frac{\Phi^{\ast}_{t} X  \; - \; X}{t}
\end{equation}
 The covariant derivative is defined by
 \begin{equation}
  \nabla_{\!\!\: V } X \; := \; \frac{d}{dt} \Big\vert_{t=0}  {\mathcal T}^{\lambda}_t 
X
           \; = \; \lim_{t \to 0} \,
  \frac{{\mathcal T}^{\lambda}_{t} X  \; - \; X}{t}
\end{equation}
 where ${\mathcal T}^{\lambda}_t$ is the operation of parallel transport backwards by  
parameter distance $t$ along the integral curve $\lambda$ of the vector field $V$. 

 Parallel transport is generally an independent structure on a manifold $M$ which can 
be introduced by endowing the manifold with a linear connection $(M,\Gamma)$. For the 
case of the bundle manifold $P \equiv M_4 \times G$ with a gauge connection $\omega$ 
the distinguished horizontal subspace has a geometric origin and physical meaning and 
hence Lie dragging along vector fields \mbox{$\acute{e}_a \inn \mbox{\it HP}$} 
describes a natural rule for transport in the bundle space. This structure can in turn 
be modelled in terms of a parallel transport of vector fields
 by introducing appropriate linear connection coefficients  on the bundle space. This 
linear connection uniquely specifies a covariant derivative on the bundle via 
equation~\ref{gamene}, as for example in equation~\ref{covhor} above in the horizontal 
lift basis on $P$.

  On the other hand the Lie derivative of equation~\ref{lieder} of a vector field $X 
\inn \TM = T^1_0\!M$ with respect to the vector field $V$ is given by the simple Lie 
bracket  expression \mbox{$\lag_{V}X = [V,X]$}. Applying this to horizontal lift basis 
vector fields on the bundle $P \equiv M_4 \times G$ we have from equation~\ref{hatb3} 
(hence adopting a coordinate basis on $M_4$):
\begin{equation}
\label{liehor}
  \lag_{\acute{e}_a} \acute{e}_b \; = \; \lbrack \acute{e}_{a}, \acute{e}_{b} \rbrack 
		  \; = \; -F^{\alpha}_{\ph{\alpha}ab}\acute{e}_{\alpha} 
\end{equation}

 We require that the vertical component in the parallel transport of a horizontal 
basis vector $\acute{e}_b$ along the field $\acute{e}_a$, as described by the 
coefficients 
 $\acute{\Gamma}^{\alpha}_{\ph{\alpha}ba}$ in equation~\ref{covhor}, should be 
directly associated with the natural Lie transport according to the local coefficients 
$-F^{\alpha}_{\ph{\alpha}ab}$ of equation~\ref{liehor}. Hence, taking into account the 
asymmetry in the $ab$ indices of the components $F^{\alpha}_{\ph{\alpha}ab}$ of the 
gauge curvature 2-form $\Omega$, we have simply:
\begin{equation}
 \label{gamome}
  \acute{\Gamma}^{\alpha}_{\ph{\alpha}ab}(p) \; = \; F^{\alpha}_{\ph{\alpha}ab}(p)
\end{equation}
  These linear connection coefficients  derive from a similar argument as the 
 corresponding components listed in  `row 5)' under `Kat~\cite{Katan}' in 
table~\ref{Gamsetsr} for the case of the parameter $c=1$ 
  as described in subsection~\ref{one34}.
   Here however these coefficients are considered as inducing a perturbation of the 
original flat geometry of the external spacetime $M_4$, associated with the symmetries 
of the original form $\lvf$,
 arising through the extra degrees of freedom of the internal gauge structure for the 
larger space $P \equiv M_4 \times G$ associated with the symmetries of the broken full 
form $\lvh$.

 Unlike the Levi-Civita connection coefficients described in subsection~\ref{one33}, 
following Cho~\cite{Cho} and as listed in the first column of table~\ref{Gamsetsr}, 
here the connection components are motivated directly in terms of a natural 
parallelism on the manifold $P \equiv M_4 \times G$ rather than via a metric structure 
on the bundle. However, since there is an internal gauge connection $\omega$ it is 
also the case here that the structure of the natural metric $\acute{g}(p)$ with the 
components $\acute{g}_{ij}$ of equation~\ref{ggghlb} can be identified in the 
horizontal lift basis on principle bundle $P$. In particular this metric  encapsulates 
the decomposition of the tangent space {\it TP} into vertical and horizontal 
subspaces, equation~\ref{vplush}, in terms of the orthogonality of any two vectors -- 
one in each of {\it VP} and {\it HP}.
  Containing the components of the Killing form $g_{\alpha\beta} = K_{\alpha\beta}$
 of equation~\ref{killmet}  for the gauge symmetry group $G$ the metric $\acute{g}$ of 
equation~\ref{ggghlb} on $P$ is also right-invariant under the action of the gauge 
group, with the scalar product of any two right-invariant fields via $\acute{g}$ being 
independent of the fibre coordinates, as implied in equation~\ref{rtrang}.

 Since $\acute{g}(p)$ is a non-degenerate symmetric rank-2 tensor field on  $P$ the 
principle bundle space itself \textit{can} be considered as a Riemannian manifold. 
Hence here we are considering the geometry of $(P,\acute{g},\acute{\Gamma})$ with a 
pair of structures, both a metric and a linear connection, defined on $P$. After 
having determined a link between $\acute{\Gamma}$ and the gauge curvature $\Omega$ on 
the bundle via equation~\ref{gamome} we next establish the relationship between 
$\acute{\Gamma}$ and $\acute{g}$ in order to specify  the remaining connection 
coefficients.

   In particular we require the linear connection $\acute{\Gamma}$ to be compatible 
with the metric $\acute{g}$ of equation~\ref{ggghlb} to the extent at least that 
parallel transport in the horizontal subspace of $P$, with local tangent space basis  
$\{\acute{e}_a\}$, should preserve the partition of {\it TP} into vertical and 
horizontal subspaces as described by their orthogonality according to the metric 
components
 $\acute{g}_{a\beta} = \acute{g}_{\alpha b} =  0$. This is  reasonable since although 
$\acute{g}(p)$ is not considered a physical metric on $P$ (unlike  $g_{ab}(x)$ which 
determines spacetime intervals in the appropriate units on $M_4$) the partition of 
{\it TP} according to equation~\ref{vplush} does have a distinct physical meaning. 
Hence we require $\acute{\nabla}_{\!\acute{e}_a \,} \acute{g}_{\alpha b} = 0$, that 
is:
\begin{eqnarray}
   \acute{\nabla}_{\! a \,} \acute{g}_{\alpha b} & = & 
   \acute{e}_a \acute{g}_{\alpha b}
     -  \acute{\Gamma}^{c}_{\ph{c}\alpha a} \acute{g}_{cb}
     - \acute{\Gamma}^{\gamma}_{\ph{\gamma}ba} \acute{g}_{\alpha \gamma}
	   \; = \; 0  \label{nabnon} \\
	& \Rightarrow &  \acute{\Gamma}^{c}_{\ph{c}\alpha a} \acute{g}_{cb}
	  = - \acute{\Gamma}^{\gamma}_{\ph{\gamma}ba} \acute{g}_{\alpha \gamma}
	     \nonumber  \\
	& \Rightarrow &  \acute{\Gamma}^{d}_{\ph{d}\alpha a} =
	   - \acute{g}^{bd} 
	\acute{\Gamma}^{\gamma}_{\ph{\gamma}ba} \acute{g}_{\alpha\gamma} 
	     \nonumber  \\
	& & \qquad  = g^{bd}g_{\alpha\gamma}F^{\gamma}_{\ph{\gamma}ab}
	 \label{gamggome}
\end{eqnarray}
    via equation~\ref{gamome}. The determination of these coefficients 
$\acute{\Gamma}^{d}_{\ph{d}\alpha a}(p)$ again follows by a similar argument that led 
to the entry of row `4)' for Katanaev~\cite{Katan} in table~\ref{Gamsetsr} as 
described in subsection~\ref{one34}. The generalisation of equation~\ref{nabnon} for 
compatibility of the full metric $\acute{g}(p)$ with  the covariant derivative in 
directions within the horizontal subspace  can be written:
\begin{equation}
\label{nabgen}
  \acute{\nabla}_{\! a \,} \acute{g}_{ij}  =  
   \acute{e}_a \acute{g}_{ij}
     -  \acute{\Gamma}^{k}_{\ph{k}ia} \acute{g}_{kj}
     - \acute{\Gamma}^{k}_{\ph{k}ja} \acute{g}_{ik}
	   \; = \; 0  
\end{equation}

  Since $\acute{g}$ is right-invariant on $P$, and given that 
  $\acute{\Gamma}^a_{\ph{a}bc} = {\Gamma}^a_{\ph{a}bc}$ from equation~\ref{gamabcg} 
and $\acute{g}_{bc} = g_{bc}$ from equation~\ref{ggghlb}, the 
$\acute{\nabla}_{\! a \,} \acute{g}_{bc}  = 0$ components of equation~\ref{nabgen} on 
the bundle space
 reduce to simply $\nabla_{\! a \,} {g}_{bc}  = 0$ on the base space $M_4$.
   This is just the statement of metric compatibility for the Levi-Civita connection 
   ${\Gamma}^a_{\ph{a}bc}(x)$ employed on the base space, which is adopted as 
consistent with general relativity in 4-dimensional spacetime.
  
  On the other hand the remaining components of equation~\ref{nabgen} describe the 
preservation of Killing metric $\acute{g}_{\alpha\beta} = g_{\alpha\beta} = 
K_{\alpha\beta}$ under parallel transport in the horizontal subspace on $P$. This in 
turn constrains the remaining linear connection components on the bundle space: 
 \begin{eqnarray}
   \acute{\nabla}_{\! a \,} \acute{g}_{\alpha \beta} & = & 
   \acute{e}_a \acute{g}_{\alpha \beta}
     -  \acute{\Gamma}^{\gamma}_{\ph{\gamma}\alpha a} \acute{g}_{\gamma\beta}
     - \acute{\Gamma}^{\gamma}_{\ph{\gamma}\beta a} \acute{g}_{\alpha \gamma}
	   \; = \; 0  \nonumber \\
	& \Rightarrow &  \acute{\Gamma}^{\gamma}_{\ph{\gamma}\alpha a} 
\acute{g}_{\gamma\beta} + 
      \acute{\Gamma}^{\gamma}_{\ph{\gamma}\beta a} \acute{g}_{\alpha \gamma} = 0
	     \nonumber  \\
	& \Rightarrow &  \acute{\Gamma}^{\gamma}_{\ph{\gamma}\alpha a} = 0
	 \label{gamfin}
\end{eqnarray}

  The adoption of $\acute{\Gamma}^{\gamma}_{\ph{\gamma}\alpha a} = 0$ in the final 
line as the simplest solution for the second line above for these metric compatible 
connection coefficients in the horizontal lift basis on $P$ is analogous to setting 
$\rho = 0$ in equation~\ref{gamlamc} for the simplest metric compatible connection in 
a left-invariant  basis on a Lie group manifold $G$.  

 The full set of linear connection coefficients  $\acute{\Gamma}^i_{\ph{i}jk}(p)$ on 
the bundle space deduced above is in fact compatible with the metric $\acute{g}$ of 
equation~\ref{ggghlb} generally, that is the metric is covariantly constant with 
$\acute{\nabla}\acute{g} = 0$ and the covariant derivative of $\acute{g}(p)$ is zero 
in any direction on the bundle space, including along the vertical fibre components.
  This full set, that we have arrived at in the horizontal lift basis in 
equations~\ref{gamvert}, \ref{gamabcg}, \ref{gamome}, \ref{gamggome} and \ref{gamfin},  
is summarised here:
\begin{equation}
 \label{gamsetkk}
   4)\; \acute{\Gamma}^{a}_{\ph{a}\gamma b} = 
g^{ac}g_{\gamma\beta}F^{\beta}_{\ph{\beta}bc},
   \qquad
   5)\; \acute{\Gamma}^{\alpha}_{\ph{\alpha}ab} = F^{\alpha}_{\ph{\alpha}ab},
   \qquad
   6)\; \acute{\Gamma}^a_{\ph{a}bc} = \Gamabc
\end{equation}
   with all other $\acute{\Gamma}^i_{\ph{i}jk} = 0$, and may be compared with the 
models of  table~\ref{Gamsetsr}.  
 Although the motivation differs significantly, this set is equivalent to that listed 
under `Kal~\cite{Kalin}' in the penultimate column of table~\ref{Gamsetsr} for 
$\lambda = 2$, with the motivation for employing this latter value here related to the 
geometrical argument in \cite{Katan}. This latter argument also has the benefit of 
fixing the geometry of $\acute{\Gamma}$, which has finite torsion as for the models of 
subsection~\ref{one34}, without any reference to the Levi-Civita connection on $P$.

  Since we have introduced the metric $\acute{g}$ on the bundle space
   it is \textit{possible} to define the unique metric compatible and torsion-free 
Levi-Civita connection on $P$, as described in subsection~\ref{one33}, with the 
components of equation~\ref{hgamijk} as listed for the horizontal lift basis in the 
first column  of table~\ref{Gamsetsr} under `Cho \cite{Cho}'. However in the present 
theory at no stage is ${P}$ considered to be an extended \textit{physical} space or 
spacetime structure, hence neither the metric $\acute{g}_{ij}$ nor a linear connection 
$\acute{\Gamma}^i_{\ph{i}jk}$ on ${P}$ have a physical meaning of the kind that such 
objects have on the external base space $M_4$. Hence the  Levi-Civita connection is 
not here considered to be a natural structure on the bundle space as it is for the 
base manifold, and an alternative argument for the form of $\acute{\Gamma}$ on ${P}$ 
has been presented.

  Observations regarding the structure of the theory presented here, together with
  the broader discussion of Kaluza-Klein theories in section~\ref{one3}, have led to 
the conjectured linear connection components on $P$ as summarised in 
equation~\ref{gamsetkk}.
 This proposal for the properties of  $\acute{\Gamma}$, constructed in the 
distinguished horizontal lift basis on $P$, appropriate for the present theory has 
been influenced by  consideration of all cases collected in table~\ref{Gamsetsr}, as 
we summarise here:

\begin{itemize}
 
 \item[a)]  The bundle $P \equiv M_4 \times G$ serves as an arena to relate the 
external and internal symmetry structures in a geometric framework closely analogous 
to that of \cite{Cho},   deriving from a generalisation of figure~\ref{mtogmaph}(b) 
for the present theory.

\item[b)] The linear connection $\acute{\Gamma}$ on $P$ may be compatible with the 
natural metric $\acute{g}$ of equation~\ref{ggghlb}, while the torsion can be finite 
as initially emphasised in~\cite{Kopcz}, and here $(P,\acute{g},\acute{\Gamma})$ does 
\textit{not} represent a physical spacetime.

\item[c)]  In deriving physical equations on the base space $M_4$
  compatibility with gauge covariance should be observed, as emphasised in   
\cite{OrzPau}, as well as general covariance and the simultaneous possibility of 
${\bR} = 0$ and ${F} = 0$ on $M_4$.
 
 \item[d)]  Parallel propagation via $\acute{\Gamma}$ in the vertical subspace
  of equation~\ref{vplush} 
  is taken to be trivial,
   with $\acute{\Gamma}^{i}_{\ph{i}j} \equiv \acute{\Gamma}^{i}_{\ph{i}j} \, \scirc \, 
  \mbox{hor}$ similarly as
   for  \cite{Kalin}, that is  any vector  $X(p) \inn V_pP$ (i.e. with $\pi_{\ast} X = 
0$) is mapped to zero by the 1-forms
 $\acute{\Gamma}^{i}_{\ph{i}j}$ on $P$.

 \item[e)] Parallel propagation via $\acute{\Gamma}$ in the horizontal directions on 
$P$ is determined by the geometry of the gauge curvature over the base space $M_4$, 
following \cite{Katan} in the fifth column of table~\ref{Gamsetsr} for the case $c=1$.

\end{itemize}

  On the bundle space $(P, \acute{g}, \acute{\Gamma})$ both the metric $\acute{g}$ of 
equation~\ref{ggghlb} and the components of the metric compatible linear connection 
$\acute{\Gamma}$ of equation~\ref{gamsetkk} transform in a gauge covariant manner on 
$P$. This follows since both $\acute{g}$ and $\acute{\Gamma}$ are closely associated 
with the horizontal subspace {\it HP} of the tangent space on $P$ which itself is 
right-invariant as described for equation~\ref{racthp}, by the definition of a gauge 
connection on $P$. 
  That is, $\acute{\Gamma}$ is defined in part in terms of a parallel transport on the 
bundle space that follows the contours of the horizontal lift subspace as described 
for figure~\ref{paraCD}, with the non-zero coefficients of equation~\ref{gamsetkk}  
 representing a gauge covariant perturbation from an initial structure, consistent 
with all
 $\Gamma^a_{\ph{a}bc}(x) = 0$ throughout $M_4$ under $\lvf$, in the extension onto $P 
\equiv M_4 \times G$ for the symmetries of the broken full form $\lvh$. The analysis 
of this perturbation led to equations~\ref{gamome} and \ref{gamggome}, which via 
equation~\ref{ecderv} imply for the respective coefficients of 
equation~\ref{gamsetkk}:
\begin{equation}
 \label{gamtall}
   4)\;\; \acute{e}_{\alpha} \, \acute{\Gamma}^{a}_{\ph{a}\beta b} \, = \,
          c^{\gamma}_{\ph{o}\alpha\beta} \,    
		 \acute{\Gamma}^{a}_{\ph{a}\gamma b},
   \qquad
   5)\;\; \acute{e}_{\alpha} \, \acute{\Gamma}^{\beta}_{\ph{\beta}ab} \, = \,
         - c^{\beta}_{\ph{o}\alpha\gamma} \,    
		 \acute{\Gamma}^{\gamma}_{\ph{\beta}ab},
   \qquad
   6)\;\; \acute{e}_{\alpha} \, \acute{\Gamma}^{a}_{\ph{a}b c} \, = \, 0
\end{equation}
   These transformation properties are a direct consequence of this construction of a 
linear connection on $P$, inheriting this structure from the gauge curvature $\Omega$ 
on the bundle.
 An alternative interpretation would be to \textit{first} extend $\Gamma$ on $M_4$ to
  a linear connection $\acute{\Gamma}$ on $P \equiv M_4 \times G$ with trivial 
vertical propagation and parallel transport in the horizontal directions defined in 
gauge covariant manner, since $G$ does not represent an extension into a physical 
space. The gauge covariant parallelism of $\acute{\Gamma}$ itself then determines a 
local relation between the horizontal subspaces on $P$ which is equivalent to the 
structure  of a  gauge curvature $\Omega$. 
(This approach is analogous to taking the metric $\acute{g}$ as the primary entity on
 $P$ in conventional Kaluza-Klein theories, as described leading to 
equation~\ref{rtrang}).  
  From either perspective,
  this gauge covariant parallelism via the set $\acute{\Gamma}^i_{\ph{i}jk}$ of 
equation~\ref{gamsetkk} on $P$  provides an appropriate  description  of the geometry 
on the bundle as a means of deriving a gauge invariant relation between the external 
and internal curvature on the base space $M_4$.

\subsection{Perturbation to the Einstein-Hilbert Action on $M_4$}
\label{one42}

 The main purpose of constructing a linear connection $\acute{\Gamma}(p)$ on $P$, as 
described in the previous subsection, is to provide a means through which a 
correlation between the external and internal curvature may be explicitly described. 
On the spacetime manifold $M_4$ any relationship between the external geometry, 
expressed in terms of the Einstein tensor with components $G_{\mu\nu}(x)$, and the 
internal geometry, expressed in terms of the gauge curvature with components 
${F}^{\alpha}_{\ph{\alpha}\mu\nu}(x)$, must transform covariantly both under general 
coordinate transformations and under gauge transformations. One technique for 
obtaining such a relation is to first identify a scalar `Lagrangian' function which 
has these invariance properties.

 Towards this end a Riemann curvature tensor $\acute{\bR}$ can be defined on the 
bundle space $P$ itself in terms of the linear connection components 
$\acute{\Gamma}^i_{\ph{i}jk}(p)$ of equation~\ref{gamsetkk}, with the components 
 $\acute{R}^i_{\ph{i}jkl}$ determined by equation~\ref{rabcd}. In turn the Ricci 
curvature $\acute{R}_{jk} = \acute{R}^l_{\ph{l}jkl}$ and scalar curvature
 $\acute{R} = \acute{g}^{jk}\acute{R}_{jk}$ can also be constructed. 
 In these expressions the Killing form  $\acute{g}_{\alpha\beta} = g_{\alpha\beta} = 
K_{\alpha\beta}$ subcomponents  of $\acute{g}_{ij}$ relate the Lie algebra adjoint and 
coadjoint representations as usual, with for example $F_{\alpha \;\! ab} = g_{\alpha 
\beta} F^{\beta}_{\ph{\beta}ab}$, and  may be employed as a mathematical structure on 
$P$ in the derivation of scalar quantities through the contraction of indices 
associated with a basis for the Lie algebra of $G$, 
 as for example in the second term of equation~\ref{rtwobits}. That is,
  unlike the case of Kaluza-Klein theories for which ${g}_{\alpha\beta} \propto 
K_{\alpha\beta}$ is interpreted as a \textit{physical} metric, here   
 the components ${g}_{\alpha\beta} = K_{\alpha\beta}$  are employed in terms of the 
Killing form in the usual sense, both on the group manifold $G$ and the bundle space 
$P$. 
 Further, the 
 $\acute{g}_{a \beta} = \acute{g}_{\alpha b} = 0$ subcomponents 
 of $\acute{g}_{ij}$ on $P$  simply express the physical distinction between the 
vertical and horizontal subspaces in the decomposition of equation~\ref{vplush}. Only 
the remaining components $\acute{g}_{ab}(p)$ implicitly represent a physical metric in 
that they are inherited directly from the  metric $g_{ab}(x)$ on the base space 
manifold $M_4$ as described for equation~\ref{ggghlb}. This physical metric
 on the  spacetime manifold $M_4$ itself derives
  from the local Minkowski metric $\eta_{ab}$ of equation~\ref{lorform2} for the 
present theory.

  The resulting scalar $\acute{R}(p)$ can be determined directly from the more general 
case of equation~\ref{rrrffff}, here with $R_G = 0$ and with $\chi = 1$ from 
equation~\ref{rrrffff2} via the non-zero coefficients $f_4 = f_5 = 1$ implied 
equation~\ref{gamsetkk}. As noted in the previous subsection, following 
equation~\ref{gamsetkk}, this set of $\acute{\Gamma}^i_{\ph{i}jk}(p)$ is equivalent to 
setting $\lambda =2$ in the `Kal~\cite{Kalin}' column of table~\ref{Gamsetsr} and
  we have simply  $\acute{R}(p)  = R_M +  F^2$, as  can be read off directly from the 
bottom line of the `Kal~\cite{Kalin}' column of table~\ref{Rsetsr}.
 It can also be seen from that column that the finite scalar curvature $\acute{R}$ on 
$P$ arises entirely from the $g^{ab}\acute{R}_{ab}$ contribution in 
equation~\ref{rtwobits}, consistent with an augmentation to the geometry of the base 
space $M_4$.

  This scalar $\acute{R}(p)$ on $P$ is independent of the location on the fibre over 
any $x\inn M_4$ and as a scalar field for any given $p\inn P$ it takes the same value  
in any tangent space basis for {\it FP}. 
Hence  in a direct product basis associated with a section $\sigma : M_4 \to P$
 this field is simply $\ddot{R}(p) = \acute{R}(p)$ and we can unambiguously define: 
\begin{equation}
 \label{rtilde}
   \tilde{R}(x) = \sigma^{\ast}\ddot{R}(p)  =  R_M + F^2
\end{equation}
   as a gauge invariant scalar field on the base space $M_4$. 
  The value of this scalar field $\tilde{R}(x)$  is  equivalent to $\acute{R}(p)$ for 
any $p\inn P$ such that $\pi(p)=x \inn M_4$.
 Hence $\tilde{R}(x)$ is a real scalar function on $M_4$ which contains information 
about both the external and internal geometry, is invariant under both coordinate and 
gauge transformations on the base space, and therefore makes a suitable `Lagrangian' 
candidate on $M_4$.
 This expression derives from the geometry on the bundle $P \equiv M_4 \times G$ in a 
physically meaningful way that can be expressed in terms of entities defined on the 
base space $M_4$.

   The means of constructing  the scalar field of equation~\ref{rtilde} can be 
considered as a perturbation to general relativity on the base space deriving from the 
need to take into account the symmetries and internal space of the full form $\lvh$ 
when broken over $M_4$ and the geometric structures entailed.
 This perturbation carries with it  consequences for the Riemannian geometry on $M_4$ 
that follow from the embedding of the 4-dimensional spacetime manifold  in the 
structures of  the higher-dimensional form of temporal flow $\lvh$.
 We hence conjecture  that a relation between the external and internal curvature can 
be determined  via the scalar function $\tilde{R}(x)$ of equation~\ref{rtilde}, 
interpreted as a geometric perturbation to the vacuum case for the Einstein-Hilbert 
action on the base space $M_4$. 
 
 This approach is justified in part since the Einstein-Hilbert vacuum solution, 
$G^{\mu\nu}=0$ of equation~\ref{lagtoein}, correlates with the zero Riemann curvature, 
implied in equations~\ref{maca3} and \ref{fdefaaa}, on $M_4$ for the original vacuum 
case of figure~\ref{mtogmap}(b) deriving from the symmetries of $\lvf$ alone, as
 associated with the canonical flat Lorentz connection $A(x) = h^{\ast}\theta(h)$ of 
equation~\ref{aegasth}.
 On generalising from that case, with a linear connection $\Gamma(x) = 0$ in suitable 
coordinates on $M_4$, to a linear connection $\acute{\Gamma}(p)$ defined on the space  
$P \equiv M_4 \times G$ deriving from the broken symmetries of the higher-dimensional 
form $\lvh$, as exemplified in figure~\ref{mtogmaph}(b), the geometry of the  base 
manifold $M_4$ is taken to be determined by a perturbation to the Einstein-Hilbert 
action with the scalar  $R_M(x)$ augmented to $\tilde{R}(x)$ in equation~\ref{rtilde}.

For Kaluza-Klein theory, reviewed in section~\ref{one3}, originating as a physical 
higher-dimensional spacetime extension of general relativity, to be interpreted as a 
unified theory of gravitation and gauge fields in a 4-dimensional spacetime the 
symmetry  of general coordinate transformations in the extended spacetime has to be 
\textit{broken} down to  4-dimensional general covariance together with the local 
gauge symmetry.
 This is equivalent to placing restrictions on the metric of the extended space,
  in conformity with equation~\ref{rtrang},
  which then possesses a set of isometries described by Killing vector fields which 
have a one-to-one relationship with the left-invariant vector fields on the manifold 
of an apparent gauge group $G$. In this way a principle fibre bundle structure 
emerges on the extended space, exhibiting symmetries such that the freedom in 
variation of the metric $\ddot{g}_{ij}$, as expressed in a direct product basis in 
equation~\ref{gmetug}, is effectively reduced to the components $g_{ab}$ and 
$\omega^{\alpha}_{\ph{\alpha}a}$.
  The construction of an Einstein-Hilbert action integral on this bundle space 
$P=(M_4,G)$ then leads to corresponding  equations of motion as described for 
equations~\ref{einfiek} and \ref{yangmk}.
 A dynamical mechanism for this process in which an extended 4-dimensional base 
manifold $M_4$ of general relativity survives while the \textit{extra} dimensions
 apparently  lose any sense of external spatial significance, sometimes called 
`dimensional reduction' or  `spontaneous compactification', then remains to be 
specified. That is, the origin of the above restrictions on the metric for the full 
space remains to be accounted for.

 The origin of the bundle structure in Kaluza-Klein theories hence contrasts sharply 
with that for the present theory. While for Kaluza-Klein theory the degrees of freedom 
of the manifold of the internal symmetry group $G$ can be interpreted as extra 
dimensions of space over the spacetime manifold $M_4$,
here the full geometric structure $P \equiv M_4 \times G$ arises out of the broken 
symmetries of the full form of temporal flow $\lvh$ as described in 
subsection~\ref{one23}, however with only the  degrees of freedom of $M_4$ ever 
interpreted as an extended physical space. For example for the 9-dimensional form 
$\lvni$, as employed in figure~\ref{mtogmaph}, the base manifold $M_4$ arises out of a 
parametrisation of a 4-dimensional subset of the degrees of freedom of the 
`translational' symmetry of the components $\bv_{9}$ under $\lvni$, with an internal 
gauge field   over the resulting base space deriving from  the internal `isochronal' 
$G=\uo$ symmetry of the same temporal form.
 With the only \textit{physical} space being the manifold $M_4$, providing the arena 
for general relativity in a 4-dimensional spacetime,  no `compactification' from a 
higher-dimensional extended spacetime is required. The spacetime geometry on $M_4$ 
derives from the local Minkowski  metric $\eta_{ab}$ implicit in the 4-dimensional 
temporal form of equation~\ref{lorform2} in the projection out of a higher-dimensional 
form such as $\lvni$, breaking the symmetry of this full temporal form.

   However, while the interpretation differs there is significant overlap between the 
geometric structure employed for the present theory and that of the Kaluza-Klein 
theories reviewed in section~\ref{one3}. Indeed,
 the argument here leading to the set of linear connection coefficients 
$\acute{\Gamma}^i_{\ph{i}jk}(p)$ of equation~\ref{gamsetkk}  is heavily influenced by 
the range of models studied as summarised 
 in points `a) -- e)' towards the end of the previous subsection. This argument  
focuses on the horizontal transport in $P$ skirting over the base manifold $M_4$ 
through the internal degrees of freedom, and in appealing in particular to 
references~\cite{Kalin} and \cite{Katan} meets halfway with Kaluza-Klein theory.

 In standard Kaluza-Klein theory the action $I_m$ for the scalar curvature
   $\check{R}(p)$ defined on the bundle space $P$ in equation~\ref{einhilf} reduces to 
the 4-dimensional action integral $I_4$ of equation~\ref{einhilk} owing to the trivial 
integration over the fibre degrees of freedom.  The point of view adopted here is that 
all fields in the expression
  $\tilde{R}(x) = R_M + F^2$ of equation~\ref{rtilde}  are defined \textit{directly} 
on the base space $M_4$ itself, with the components of the gauge curvature on the base 
space related to those on the bundle space as described for equation~\ref{fonbase}. 
Being  invariant under both gauge and coordinate transformations on $M_4$ this scalar 
$\tilde{R}(x)$   can be employed as a Lagrangian function in the simple action 
integral:
\begin{equation}
   \tilde{I} = \int \tilde{R}   \sqrt{\vert g \vert}\; d^4 x 
    = \int (R_M + F^2) \sqrt{\vert g \vert}\; d^4 x  \label{einhilp}   
\end{equation}
 defined directly on the base space $M_4$. 
  This is the action on the base manifold $M_4$ that  arises through the external and 
internal symmetry degrees of freedom described by the bundle space
 $P \equiv M_4 \times G$.
 As denoted by the `tilde' on $\tilde{I}$ this function is considered as a 
perturbation to the Einstein-Hilbert action for the vacuum case, namely  
equation~\ref{einhil} with  $\Lambda = 0$ and $\lag = 0$ as described in 
subsection~\ref{one31}.
   Equation~\ref{einhilp} incorporates the perturbation  to the effective scalar 
Lagrangian $R_M(x) \to \tilde{R}(x)$ on the base space $M_4$ reflecting the 
augmentation from the vacuum case of $\lvf$, as pictured in figure~\ref{mtogmap}, with 
zero external curvature as concluded in subsection~\ref{one22}, to the symmetries of 
the broken full form of $\lvh$, as exemplified in figure~\ref{mtogmaph} and described 
in subsection~\ref{one23}.
 The full Einstein-Hilbert action of equation~\ref{einhil} can be written:
\begin{equation}
   I = \int (\alpha R_M  + \lag)   \sqrt{\vert g \vert}\; d^4 x  \label{einhilnol}   
\end{equation}
  where the cosmological constant $\Lambda$ has been dropped in correspondence with 
the lack of a finite $R_G$ term in equations~\ref{rtilde} and \ref{einhilp}.  Further 
comparison between the above two equations shows that equation~\ref{einhilp} describes 
a perturbation to general relativity equivalent to the introduction of a Lagrangian 
term $\lag = +\alpha  F^2$ in the original Einstein-Hilbert action of 
equation~\ref{einhilnol}.
  While the mathematical conclusion is identical to Kaluza-Klein theory, here the 
interpretation involves a more minimal impact on the arena of general relativity in 
4-dimensional spacetime, namely without a \textit{physical} augmentation
  into a higher-dimensional extended spacetime.

  The equation of motion obtained by requiring 
  $\delta \tilde{I}=0$ for equation~\ref{einhilp}, under variations $\delta 
g_{\mu\nu}(x)$ of the metric on $M_4$, follows the derivation of 
equation~\ref{einfiek} and can be written here as:  
\begin{equation}
 \label{gchift}
  G^{\mu\nu} \: = \: 
  2\chi(  - F^{\alpha \mu}_{\ph{\alpha\mu} \rho}F_{\alpha}^{\ph{\alpha}\rho\nu}
	                -\frac{1}{4} g^{\mu\nu} \, F^{\alpha}_{\ph{\alpha}\rho 
\sigma}F_{\alpha}^{\ph{\alpha}\rho\sigma})   \: =: \:  -\kappa T^{\mu\nu}   
\end{equation}
 with $\chi=1$.
   At the purely theoretical level the factor of $\chi=1$ in this relation between the  
external and internal geometry in the breaking of the full form $\lvh$ arises directly 
from equations~\ref{rrrffff} and \ref{rrrffff2} and the relation of the linear 
connection $\acute{\Gamma}^i_{\ph{i}jk}(p)$ on the bundle 
 to the gauge curvature components $F^{\alpha}_{\ph{\alpha}ab}(p)$ as
 summarised in equation~\ref{gamsetkk}. 
 These geometric arguments establish a direct relationship between the external and 
the internal curvature, with energy-momentum here \textit{defined} through 
 $-\kappa T^{\mu\nu} := G^{\mu\nu}$  on the right-hand side of this expression.

 Through equation~\ref{gchift} for the internal symmetry $G=\uo$ case direct contact 
is made between gravitation in the form of the geometric curvature of spacetime and 
the familiar laboratory phenomena of the electromagnetic field. In practice powerful 
electromagnetic effects are generally observed for which the associated gravitational 
field is immeasurably small and  the 
  appropriate units and a normalisation factor  connecting the left-hand side and 
central expression of equation~\ref{gchift} are a matter for empirical convention, as 
for the factor of $\kappa= \frac{8\pi G_{\! N}}{c^4}$ on the right-hand side of this 
equation. 
  On the other hand the bare mathematical relations are needed to understand the 
theoretical basis of the unification.

   By further considering the stationarity $\delta \tilde{I}=0$ for the action in 
equation~\ref{einhilp}, now with respect to variation in the gauge field components
 $A^{\alpha}_{\ph{\alpha}\mu}(x)$, which are related to the connection components on 
the bundle space as explained for equation~\ref{aonbase}, leads, as described earlier 
for equation~\ref{yangmk}, to the Yang-Mills vacuum equation:
\begin{equation}
	D_{\mu}F^{\alpha \;\!\mu\nu} \: = \:  0    \label{delyym}
\end{equation}  
  that is equation~\ref{ymol},  with $D_{\mu}$ the gauge covariant derivative in 
curved spacetime.
  For the case of an Abelian internal $\uo$ symmetry this relation expresses Maxwell's 
equation for a source-free electromagnetic field. Such a structure arises for example 
for the internal $\uo$ symmetry of equation~\ref{suinsth} from the breaking of the 
full $\slthc$ symmetry for the $\lvni$ model depicted in figure~\ref{mtogmaph}.

  Given the  relation of equation~\ref{gchift} itself, 
 a number of further consequences for the  equations of motion
 may be deduced without the need for the Lagrangian formalism.
 A fundamental difference between the Einstein equation
    \mbox{$G^{\mu\nu}  = - \kappa T^{\mu\nu}$} (with or without the cosmological term 
of equation~\ref{einlamt}) and other equations of motion
 is that, assuming that all fields are associated with energy-momentum $T^{\mu\nu}$,   
\textit{all} fields are subject to the constraint of the  contracted Bianchi  identity 
$\nabla_{\!\mu}G^{\mu\nu} = 0$ and in principle `no physical entity escapes this 
surveillance' (\cite{MTW} page~475).
    For example the Einstein equation in the form of equation~\ref{gchift}, for case 
of the $\uo$ internal symmetry, mutually constrains the evolution of both the 
gravitational and  electromagnetic field.
 In this way
 it can be shown  (\cite{MTW} page~472) that the
 source-free form of the Maxwell equation $\nabla_{\!\mu}F^{\mu\nu} = 0$ (that is, 
equation~\ref{delyym} for the $\uo$ case)  does not need to be derived independently,  
rather it may instead be deduced from $\nabla_{\!\mu}G^{\mu\nu} = 0$ applied to the 
Einstein equation for the electromagnetic field.  This observation, together with 
further consequences for the equations of motion, is 
 reviewed in more detail in (\cite{Unifi} section~5.2).

 A further aim for the present theory  would be to derive equation~\ref{gchift}, or an 
equivalent expression,  by a direct  geometrical means  without reference to a scalar 
Lagrangian function and the Einstein-Hilbert action.
  Here equation~\ref{gchift} has been considered to arise as a perturbation to the 
Einstein vacuum equations, derived for equation~\ref{lagtoein} in terms of the 
stationarity of the Einstein-Hilbert action under variations of the metric $\delta 
g_{\mu\nu}(x)$. 
 With equation~\ref{gchift} itself conjectured to arise inevitably out of the 
geometric constraints implied in the breaking of the full $\lvh$ symmetry over $M_4$ 
any explicit reference to the Lagrangian formalism might in principle be avoided 
entirely.
 
   In conclusion, the construction presented in this section  leading to the relation 
between the external and internal curvature of equation~\ref{gchift} is based on the 
broken symmetry structure $P \equiv M_4 \times G$, which itself derives from the 
identification of the base manifold $M_4$ from a parametrisation of the translation 
symmetry of the full multi-dimensional form of time $\lvh$ as described in 
section~\ref{one2}. The mathematical framework for this geometric unification closely 
resembles that of the Kaluza-Klein theories reviewed in section~\ref{one3}, and which 
have provided an essential guide for the present theory.  
Attempting to justify all the steps along the way, via the linear connection on the 
bundle space of equation~\ref{gamsetkk}, scalar function on the base space of 
equation~\ref{rtilde} and action integral of equation~\ref{einhilp}, the aim has been 
to arrive at the relation of equation~\ref{gchift} with minimal assumptions.
 This equation shows how  a relation between the external and internal curvature  can 
be achieved for the present theory through a multi-dimensional temporal form as an 
expression of the unifying structure inherent in basing the theory on the underlying  
one-dimensional flow of time.

\section{Summary and Discussion}
\label{one5}

\subsection{Conceptual Picture}

The fundamental role of time in physical theories as well as the universal nature of 
time in infusing both experiments and our interactions with the world more generally 
is widely recognised. Physical theories themselves are typically constructed in terms 
of postulated particles, fields or other entities evolving in time according to 
equations of motion or consider time as an essential component of a 4-dimensional 
spacetime manifold, which may itself be embedded within a larger structure possessing 
`extra dimensions'.

  The underlying idea described in this paper, developed in sections~\ref{one2} and 
\ref{one4},
  contrasts with the Kaluza-Klein models, reviewed in section~\ref{one3},   since here  
`extra dimensions' are \textit{not} required to satisfy an explicitly geometric, or 
spacetime, symmetry.
For the theory presented here we study a general higher-dimensional `symmetry of time' 
and describe how a large scale  extended 4-dimensional spacetime geometry  supporting 
physical structures  can derive from the underlying one-dimensional temporal flow 
itself.
The means of obtaining multiple spatial, extended and also extra dimensions from a 
one-dimensional element of time has been described here in subsections~\ref{one21}, 
\ref{one22} and \ref{one23} respectively, elaborating upon 
 (\cite{Unifi} chapter~2) and 
(\cite{Novel} sections~2 and 3). 
 These structures exploit the symmetries of the general multi-dimensional flow of 
time, expressed as $\lvn$ as derived for equation~\ref{lvn}.

 We note that the starting point for the theory is the mathematical interval
  \mbox{$\delta s \inn \rrr$} as a \textit{purely} one-dimensional entity, and hence 
any typical `picture' of this element of time as seemingly  embedded \textit{within} a 
given two or three-dimensional space, as for figure~\ref{spillout}(a) necessarily 
drawn on a two-dimensional page, is inevitably somewhat misleading. An accurate 
representation of the one-dimensional interval of time \textit{is} provided
 however, mathematically and directly, by the real element $\delta s \inn \rrr$, as 
for the left-hand side of equations~\ref{sfind} and \ref{seqxxx}. The mathematical 
structure of multi-dimensional space itself is then derived \textit{from} the interval 
of time through the implicit arithmetic structure and symmetries of this real 
interval, as for example for equations~\ref{sfind} and \ref{lvthree} as pictured in 
figures~\ref{spillout}(b) and (c) respectively.
  That is, time is not considered here to flow through an independent geometric 
manifold or expand into a pre-existing space or spacetime, rather these 
multi-dimensional structures are implicit within the arithmetic forms and symmetries 
of one-dimensional time itself.

One of the attractions of using a symmetry of extra \textit{spatial} dimensions in 
Kaluza-Klein theories, as well as its intuitive appeal as an extension to a 
4-dimensional spacetime geometry, is that it limits the set of possible higher 
symmetries and 
mathematical structures
 to consider. 
  For the present theory the construction of  symmetries of multi-dimensional forms of  
\textit{time}  also greatly limits the choice of symmetry groups and their 
representations, based on the form of equation~\ref{lvn}. However, while considering 
natural extensions to the full higher-dimensional form of  time $\lvh$  we retain the 
significance and necessity of a $(1+3)$-dimensional metrical manifold as a background 
arena for observations. 
Indeed, while possessing different geometrical properties compared with other higher 
symmetries of $\lvh$,
 the symmetry of the \textit{space} part of spacetime,  that is the SO(3) subgroup of 
the Lorentz symmetry $\soot$ acting approximately globally over extended regions of 
the base manifold, 
  describes the symmetry of an extended 3-dimensional approximately
 Euclidean arena through which the physical world is actually observed.

Developing the theory as described for the three stages of figure~\ref{spillout} and  
given the symmetries of an appropriate full multi-dimensional form $\lvh$ 
    the possibility of identifying an  extended 4-dimensional space with a
	 local Minkowski metric $\eta_{ab}(x)$ as well as a local $\soota$-valued 
connection 1-form gives a \textit{geometric} meaning to $M_4$ as being not just a 
numerical parameter space for translational degrees of freedom, from 
equation~\ref{rspill}, but rather implicitly possessing a Riemannian structure as an 
arena for the observation of physical objects in \textit{time} and \textit{space}, as 
described for figures~\ref{mtogmap} and \ref{mtogmaph}.
 The identification of such an extended base space is possible given a `spacetime' 
symmetry as a subgroup of the full symmetry of $\lvh$ which acts on the local tangent 
space of $M_4$.
 The interplay between the necessary geometrical form of spacetime and the general 
mathematical form of temporal flow breaks the symmetry of the latter structure, in 
turn shaping the physical form of the observed world.  
 Both the extended manifold $M_4$, essential as the backdrop for observations in the 
world, and the local external symmetry group
 are identified through symmetries of the full form $\lvh$, leaving residual temporal 
components together with an internal symmetry $G$.
In this way both the external 4-dimensional spacetime manifold and matter fields on 
the manifold arise together in the breaking of a higher-dimensional form of temporal 
flow without the need to introduce either the spacetime background or a postulated 
material substratum \textit{independently} of time itself. 

 At the elementary level
      the equality in equation~\ref{seqxxx} signifies two different ways of expressing 
the \textit{same} interval of time $\delta s$ by applying simple arithmetic rules, 
implicit in the structure of the real line $\rrr$ itself.  In turn the flow of time 
can be considered to be intrinsically  
  accompanied by  structures which represent the geometric form of spacetime through 
the appropriate quadratic form, in the simplest case with $p=2$ in 
equation~\ref{seqxxx} and  $\alpha_{ab}$ identified with the Minkowski metric.
  Such a structure can be algebraically contained within a more general 
higher-dimensional form for equation~\ref{seqxxx}, with for example $p>2$ describing a 
cubic or higher polynomial form.
 In subsection~\ref{one23} we  presented these ideas through the example of an 
$\slthc$ symmetry of the cubic form $\lvni$ projected over $M_4$ as depicted in 
figure~\ref{mtogmaph} with an internal $G=\uo$ symmetry identified.
 The symmetry breaking structure exemplified in figure~\ref{mtogmaph}(b) 
 motivates the employment of the geometric framework of the principle bundle space 
  $P \equiv M_4 \times G$ for the general case, as analysed in detail in 
section~\ref{one4} (and building upon  \cite{Unifi} section~5.1).

 As described in section~\ref{one4}  the principle bundle ${P} \equiv M_4 \times {G}$ 
is \textit{not} considered here to represent a physical space or spacetime.
  While in Kaluza-Klein theory the parameter space of the internal symmetry group $G$ 
augments the physical spacetime arena, here the physical spacetime $M_4$ itself 
derives from a 4-dimensional component of the translational symmetry of $\lvh$ that 
exhibits the appropriate metrical structure. 
 For this theory the base space $M_4$, as the \textit{only} physically extended 
manifold,
  arises spontaneously as a background arena for observations out of the symmetries 
inherent in the form $\lvh$  with no mechanism of `compactification' from a larger 
manifold required. The necessary identification of the geometry of $M_4$ provides a 
 natural mechanism for breaking the full symmetry of the form $\lvh$.

 In subsection~\ref{one41} we described the relation between the internal curvature
  $\Omega(p)$ on $P$ as a principle bundle and a natural linear connection 
$\acute{\Gamma}(p)$ on $P$ as a base space for a Riemannian geometry. From this latter 
perspective the notion of full `general covariance' is not useful owing to the 
physical significance of the product space structure $P \equiv M_4 \times G$  and with 
a distinguished reference frame provided by the horizontal lift basis associated with 
the gauge connection $\omega$ as a further physically meaningful structure on the 
bundle space. 
With horizontal basis vector fields $\acute{e}_a(p)$ on $P$ having a one-to-one 
correspondence with basis vector fields $e_a(x)$ on $M_4$, as described for 
figure~\ref{pbunbasis}, the horizontal lift basis reflects and augments the structure 
of the spacetime manifold $M_4$ which is the primary physical arena and upon which the 
full machinery of general relativity, including the properties of general covariance, 
can be fully deployed.
 Further, in linking the coefficients of the linear connection $\acute{\Gamma}(p)$
  with the gauge curvature structure on $P$, guided by figure~\ref{paraCD} and leading 
to equation~\ref{gamsetkk}, the corresponding gauge covariant transformation 
properties of equation~\ref{gamtall} lead to gauge invariant expressions on the base 
manifold $M_4$ culminating in equation~\ref{gchift}, relating the external and 
internal curvature, as described in subsection~\ref{one42}.

\subsection{Higher Symmetries}

In the absence of a structure of extra \textit{spatial} dimensions in general the full 
form of temporal flow $\lvh$  is not required to be associated with a metric geometry. 
 It may be that higher symmetries, such as SO$^+(1,n-1)$ with $n > 4$ acting on a 
quadratic form $\lvn$ as an  $n$-dimensional extension of equation~\ref{lorl}, 
\textit{could} be interpreted in a geometrical spatial manner, but this feature is 
relatively incidental in comparison with the fundamental requirement that it must 
describe a symmetry of a general homogeneous polynomial form of time as derived for 
equation~\ref{lvn}.   
  In fact through investigating possible symmetries of time a significant example
    \textit{is} identified for the symmetry group $\sltwoo$ acting on the 
10-dimensional space $\htwo$, constructed in terms of the octonion algebra 
  as described in (\cite{Unifi} section~6.3), 
  as the two-to-one
    covering group of the 10-dimensional Lorentzian symmetry $\sootn$.

The question concerning the natural mathematical augmentation to
  higher-dimensional forms  $\lvh$ and their connection with 
  empirical observations of the physical world
  is considered in (\cite{Unifi} chapters~6--9) and summarised in \cite{Novel}.
   In (\cite{Novel} section~4) a
   27-dimensional  space of  elements $\bv_{27} \equiv \mcX \inn \htho$ with cubic 
temporal form $L(\bv_{27}) = \det(\mcX) = 1$ and the 
   exceptional Lie group $\hG= \esi \equiv \sltho$ as the full isochronal symmetry
    is described.
  Given the extra dimensions of the full  vector object $\bv_{27} \inn \htho$ the  
identification of a  Riemannian geometry parametrised over  an extended  4-dimensional 
spacetime manifold $M_4$ breaks the full $\esi$ symmetry (\cite{Novel} section~5).

  The form  $\lvt$ itself can be motivated as an augmentation from the 4-dimensional 
form $\lvf$ of equation~\ref{lorl}, building on $\bv_4 \equiv \bh_2 \inn \htwc$ in 
equation~\ref{vtoh},  via the action of $\slthc$ on the form $\lvni$ with $\bv_9 \inn 
\hthc$, hence with a $2\times 2 \to 3 \times 3$ matrix structure as described here in 
subsection~\ref{one23}, or via the action of $\sltwoo$ on the form $\lvte$ with 
$\bv_{10} \inn \htwo$, that is with $\ccc \to \ooo$ where the octonions are the 
largest normed division algebra as presented in~(\cite{Unifi} chapter 6); that is via 
either of two possible natural extensions:
\begin{eqnarray}
 \sltc \;\, \mbox{on} \;\, \htwc &  \rightarrow &
                              \slthc \;\, \mbox{on} \;\, \hthc \nonumber \\
  \downarrow \qquad\quad  &   &  \qquad\quad \downarrow  \label{esiext}  \\
 \sltwoo \;\, \mbox{on} \;\, \htwo & \rightarrow  &
                    \quad\, \esi  \;\, \mbox{on} \;\, \htho  \nonumber
\end{eqnarray}

 In all cases the  elements $\bv_n$ of the \textit{representation space} (such as 
$\htho$) for the  $n$-dimensional flow of time, subject to the form $\lvn$ of 
equation~\ref{lvn},
  are the starting point for the present theory. Certain forms have rich symmetry 
properties which then motivates the study of a \textit{group structure} with actions 
defined on this space, as is the case for the $\hG = \esi$ isochronal symmetry of 
$\lvt$. The construction of an extended spacetime manifold $M_4$ from a 4-dimensional 
subset of the translation symmetries of $\lvt$, similarly as described for 
figure~\ref{mtogmaph}, is a prerequisite for the identification of physical entities 
on the base manifold.   
  The subgroup $\sltc \subset \esi$ acting on the projected $\bv_4 \in \TM_4$ 
components of the physical manifold $M_4$ partitions \textit{all} the components of 
$\bv_{27}$ into vector, spinor or scalar objects as irreducible elements of the 
structure of the physics on $M_4$. 
 Interest then turns from the mathematical symmetry of $\esi$ on $\lvt$ to the local 
symmetry of this physical structure, with the components of $\bv_{27}$ partitioned 
into representation multiplets under the broken symmetry in the form of a direct 
product of the external symmetry $\sltc$ and an internal symmetry $G$ as described for 
equation~\ref{dirprod}.

  Through the augmentation of the full symmetry $\hat{G}$ acting on $\lvh$ from 
$\slthc$ on the vectors $\bv_9 \inn \hthc$ to  $\esi$ acting on $\bv_{27} \inn \htho$ 
   the decomposition of equations~\ref{vinhinc}--\ref{symb3} is augmented to a vector, 
four spinor and seven scalar pieces in total (\cite{Unifi} table 8.2 and \cite{Novel} 
table 2) under the external $\sltc \subset \esi$ symmetry component of 
 $\lvt$. The internal symmetry group is found to be $G = \suth_c \times \uo_Q \subset  
\esi$ (\cite{Novel}  equation~42), as an extension from the internal $\uo$ identified 
in equation~\ref{suinsth} in this paper.
 While acting trivially on the vector $\bv_4 \equiv \bh_2 \in \htwc \subset \htho$ 
this internal
   $\suth_c \times \uo_Q$ symmetry is \textit{found} to act upon the set of four Weyl 
spinors in the manner of a colour singlet charged lepton together with a colour 
triplet of fractionally charged $d$-quarks. That is, the particle states associated 
with these components of $\bv_{27}$  transform in representation multiplets under the 
broken symmetry structure in a manner resembling known elements of the Standard Model 
of particle physics, which itself is based upon empirical observations.

   For the augmented  case the $G = \uo_Q \subset \esi$ symmetry can still be 
considered as a component of the internal symmetry in itself, with the results of this 
paper applying  for equation~\ref{suinsth} here as well as for the embedding within 
$\esi$ as the full symmetry group, and also for non-Abelian internal symmetry groups 
such as $G=\suth_c$.
 While the geometry of the simpler case with $\slthc$ as the full symmetry 
 was introduced in subsection~\ref{one23}, in section~\ref{one4} we determined the 
relationship between the external Riemannian curvature and internal gauge curvature in 
equation~\ref{gchift} for the full theory 
   bearing in mind the need to  incorporate also non-Abelian and product internal 
symmetries such as $G = \suth_c \times \uo_Q$.

  Further generalisation first to an $\ese$ symmetry of a 56-dimensional quartic form 
\mbox{$\lvfs$} and on to a predicted $\ee$ symmetry of a hypothetical form $\lvtfe$
is described in (\cite{Unifi} chapter~9) and summarised in (\cite{Novel} sections 6 
and 7). Internal components of $\lvfs$, including elements transforming as Dirac 
spinors under the external $\sltc \subset \ese$ symmetry,   exhibit  transformation 
properties under the internal $\suth_c \times \uo_Q \subset \ese$ gauge group,
 incorporating  fractional charges and also a left-right asymmetry, consistent with 
the distinctive properties of the Standard Model for 
   a family of lepton and quark states, as summarised in (\cite{Novel} equation~66). 
These internal components of $\bv_{56}$, in addition to the internal gauge fields 
themselves, are interpreted as `matter fields' on the external spacetime $M_4$ and 
describe a non-trivial foothold in the structures of the Standard Model of particle 
physics as identified in the pattern of $\ese$ symmetry breaking for this theory.

 In Kaluza-Klein theory, as reviewed here in section~\ref{one3}, while a unified 
framework is provided for gravity and gauge  fields, equations~\ref{einfiek} and 
\ref{yangmk}, there are no spinor fields -- hence the matter fields for the leptons 
and quarks  are absent. These fields may be added by hand as sections of fibre bundles 
over $M_4$, associated to the principle bundle $P$, transforming as spinors under the 
external $\soot$ symmetry, via the $\sltc$ covering group as described after 
equation~\ref{gtoGam}, and in appropriately constructed representation multiplets of 
the internal gauge symmetry group. Coupling between the gauge fields and spinors may 
then be introduced through interaction terms, also added by hand for example via 
`minimal coupling' involving covariant derivatives, in the Lagrangian constructed for 
the theory. Another approach to introducing fermion states is through a supersymmetric 
extension of Kaluza-Klein theory, as discussed in (\cite{Unifi} section~5.4, see for 
example \cite{ChMaMa} section~VI).

  For the present theory  there is no need to postulate additional fields 
 in spacetime or attempt a supersymmetric extension; rather  it is the additional 
symmetries and components of the form $\lvh$, over and above those required to 
describe the external geometry of the base space $M_4$, that give rise to matter 
fields in 4-dimensional spacetime.
 That is,  out of the breaking of the full symmetry of the form $\lvh$ in  
  the geometric identification of the  external symmetry and spacetime manifold $M_4$, 
the residual internal gauge fields and surplus temporal components  collectively 
compose the apparent `matter' content of the world
 with the characteristic physical properties observed. 
  Through these structures the underlying one-dimensional flow of time is 
\textit{itself}  represented  as a flow of physical entities in an extended spacetime.  
 In the case of the $\slthc$ model in figure~\ref{mtogmaph}(b) 
  the vector ${\bv}_4 \inn \TM_4$ component of the temporal flow is projected onto the 
tangent space of the base manifold $M_4$ while  the residual temporal component $\psi$ 
transforms as a spinor under the external symmetry.
 In this model the internal $\uo$ symmetry acts on the spinor field, leading to 
interactions between the corresponding gauge field and spinor states through  dynamic 
expressions deriving from the constraints of the theory. In the extension to the 
$\esi$ symmetry as described for equation~\ref{esiext}, and onto the $\ese$ (and  
potential $\ee$) augmentation, the resulting elementary states and interactions 
closely resemble those of the Standard Model of particle physics.
 The quantum mechanical properties of the observed empirical particle states then 
remain to be accounted for.

\subsection{Quantum Theory}
\label{one53}

  While the construction presented in this paper has been developed from the 
properties of the \textit{local} geometry on the bundle space $P$ (in particular 
involving parallel transport as described for equations~\ref{covhor}--\ref{gamome} for 
example) the \textit{global} validity of equation~\ref{gchift} in principle should 
take into account the
  \textit{quantisation} of the theory (\cite{Unifi} chapters 10 and 11).
  In this case the local relation of equation~\ref{gchift} can be interpreted as 
describing cubic and quartic self-couplings in the non-Abelian gauge field components 
$A^{\alpha}_{\ph{\alpha}\mu}(x)$,  
  as a potential input for the set of local interaction terms in the structure of an 
effective quantum field theory. The global relation between the external and internal 
geometry may need to be reassessed in this context,
  potentially with a `renormalised' value for $\chi$ required in  
equation~\ref{gchift}, relating to the
  macroscopic empirical value for $\chi$ discussed after that equation.

  As also discussed towards the end of subsection~\ref{one42} the local relation of 
equation~\ref{gchift},
  together with the implied
   gauge field self-interactions $A \leftrightarrow A$ for a non-Abelian internal 
symmetry, might itself in principle be derived through a purely geometric argument 
without reference to the Einstein-Hilbert action.
  In addition
  interactions between the gauge and spinor fields of the form  $A \leftrightarrow 
\psi$  arise through the constraints of the present theory, for example in the terms 
of the expression $D_{\mu} L(\bvh) = 0$ in the symmetry breaking over $M_4$, without 
the need to introduce terms of the Standard Model Lagrangian by hand. The origin of 
all of these interaction terms for the present theory is described in (\cite{Unifi}  
section~11.1; see in particular equations 11.29 and 11.36--11.38, and with further 
discussion alongside table 15.1 in section 15.2).

  The structure of textbook quantum field theory itself is a critical guide for the 
quantisation of the present theory with for example the `optical theorem' of QFT 
(\cite{Unifi} section~10.5) providing the link with the calculation of process 
probabilities for the new theory (\cite{Unifi} section~11.2).
 As well as the structure of cross-section calculations in high energy physics 
experiments the nature of 
   the physical concepts of quantum and particle phenomena in general  need to be  
addressed (\cite{Unifi} section~11.3). 
 In the bigger picture the overall structure of a  unified framework for both quantum 
theory and gravitation can be developed (\cite{Unifi} section~11.4).
 With regard to the quantum field theory limit, as one element of the question of 
identifying empirically observable physical particle states
  in further work it will  be 
 necessary to incorporate  the `spin-statistics theorem', or a related argument, to 
account for the fermionic properties of states associated with the spinor fields 
identified in the elementary structure of the theory, as for $\psi(x)$ in 
figure~\ref{mtogmaph}(b) for the $\slthc$ model,  in contrast to the bosonic degrees 
of freedom of the gauge fields.

  The `Coleman-Mandula theorem'~\cite{ColMan} for any relativistic theory of 
interacting particles demonstrates that the only possible Lie group symmetry of a
 non-trivial \mbox{$S$-matrix} is locally isomorphic to the direct product of the 
Poincar\'e group and an internal symmetry group,  with any non-trivial combination of 
the external and internal symmetries prohibited.
   Since for the present theory the external and internal symmetries are identified as 
subgroups of the initial unifying simple group $\hG$, as the full isochronal symmetry 
of the  full temporal form $\lvh$, the compatibility of this approach with the 
Coleman-Mandula theorem will be addressed explicitly here.
  This compatibility rests on the nature of the 
   symmetry breaking structure for the present theory as described in 
subsection~\ref{one23} and also discussed in the two previous subsections.

  The essential point is that
  the full symmetry $\hG$ of $\lvh$ is \textit{not} a symmetry of the
  resulting QFT, rather a quantum field theory itself can only be constructed on the 
base space $M_4$ \textit{after} the full symmetry has been broken down to a direct 
product of the form $(\mbox{Lorentz} 
 \times G) \subset \hG$ as described for equation~\ref{dirprod}. That is, the full 
symmetry $\hG$ \textit{cannot} act on particle states in the quantum field theory 
limit of the theory since these latter structures are \textit{necessarily} constructed 
upon the extended spacetime manifold $M_4$, the identification of which through the 
isochronal and translational symmetries of $\lvh$ in turn \textit{necessarily} breaks 
the full symmetry. 
  From the starting point of the one-dimension of time only, the group $\hG$ describes 
the full mathematical isochronal symmetry  applying to the full multi-dimensional form 
of temporal flow $\lvh$, as derived for the general $n$-dimensional case in 
equation~\ref{lvn}. This symmetry $\hG$ is broken \textit{absolutely} to the form of 
equation~\ref{dirprod} in the identification of the physical 4-dimensional spacetime 
manifold $M_4$ and the associated partitioning of $\bvh$ into irreducible 
representations of the external Lorentz symmetry.

 This is unlike the case of symmetry breaking in a `Grand Unified Theory', such as an 
SU(5) model as originally described in~\cite{GeoGla}, for the unification of the 
internal gauge forces (or in the electroweak sector of the Standard Model),  for which 
a Lorentz scalar Higgs field is introduced together with suitable Lagrangian potential 
terms to break the symmetry for the field configuration in the ground state. For such 
cases the spontaneous symmetry breaking occurs below a certain energy scale, with the 
gauge bosons of the broken symmetry generators gaining a mass and suppressing the 
corresponding interactions at low energy;  leading for example to a prediction of 
proton decay that rules out the simplest SU(5) model. However for the symmetry 
breaking as depicted in figure~\ref{mtogmaph} the full symmetry $\hG$ is completely 
lost, and not just below a certain energy scale, but at all energy scales for all and 
any physics that can be defined on the 4-dimensional spacetime manifold, and in 
particular only the broken symmetry
 remnant of equation~\ref{dirprod} can apply for relativistic interacting particles.
   Again while further work is needed, the quantisation of the present theory  might  
not only be needed to fully establish the relation between 
 internal components of the temporal flow $\hat{\bv}$ and
 empirically observed physical particle states  but,
  following the discussion of equations~\ref{dirprod}--\ref{sudinsth},
   might itself  also guide the
 identification of the full internal symmetry $G$ acting upon these states, in a 
manner consistent with the properties of an effective quantum field theory.

  While appearing in the direct product decomposition of equation~\ref{sudinsth} the   
D(1) symmetry, associated with a non-compact generator of $\slthc$, was provisionally 
considered non-physical in subsection~\ref{one23} since it acts non-trivially on the 
external spacetime and hence does not belong to the subgroup
 $\stab \subset \slthc$. However this symmetry relates different types of Lorentz 
representation subspaces of $\bv_9 \inn \hthc$ via a dilation action rather than by 
mixing the components, and
  the possibility of such transformations having an important physical role in the 
very early universe has been speculated in (\cite{Unifi} section 13.2), with the 
projected magnitude $h(t)$ from equation~\ref{lorform2} considered to be dependent 
upon the cosmic time $t$ at the earliest epoch. (Potential cosmological applications 
for the present theory are described more generally in \cite{Unifi} chapters~12 and 
13).

  The possibility of identifying a \textit{compact} internal symmetry group $G$ in a 
direct product decomposition $(\mbox{Lorentz} 
 \times G) \subset \hG$ but with $G \nsubst \stab \subset \hG$ is also considered in   
  (\cite{Unifi} section 8.3 and chapter~9). There it is suggested that an internal
   $G = \sutw_L \times \uo_Y$ symmetry (proposed to be fully identified within the 
extension to $\hG = \ese$ or $\hG = \ee$) may impinge upon the external spacetime 
components of $\bv_4 \in \TM_4$, accounting for the properties of electroweak symmetry 
breaking in the Standard Model. 
 As a preliminary requirement in (\cite{Unifi} section~8.2) internal symmetries were 
first considered to be of the form $G \subset \stab \subset \hG$. However a more 
direct definition is given here in section~\ref{one23} in terms  actions 
  identified within the original full symmetry
 $\hG$ that respect the partitioning of the full temporal flow $\bvh$ into  vector, 
spinor or scalar objects  by the action of the external Lorentz symmetry, implying the 
direct product structure of equation~\ref{dirprod}. 

   Cases such as the internal $G=\uo$ symmetry of equation~\ref{suinsth},
 identified through its independent action on the three parts of 
equations~\ref{symb1}--\ref{symb3}, which \textit{also} have the form 
 $G \subset \stab \subset \hG$ may be associated with \textit{massless} gauge bosons 
such as the photon.
 This structure is augmented to the massless gluons and photons for the 
  $\suth_c \times \uo_Q \subset \stab \subset \esi$ symmetry under the extension of 
equation~\ref{esiext} and on to the $\ese$ case reviewed in the previous subsection.
 On the other hand, in the full theory, internal generators that impinge upon the 
components of $\bv_4 \in \TM_4$ will be associated with \textit{massive} gauge bosons, 
such as the $W^{\pm}$ and $Z^0$ states of an $\sutw_L \times \uo_Y$ electroweak theory 
(\cite{Unifi} subsection~8.3.3).
Again the possible structure of all such interactions may be both consistent with and 
guided by the Coleman-Mandula theorem and properties of the quantum field theory limit 
for the theory generally.

  Hence for the present theory there are two distinct elements to the breaking of the 
full symmetry $\hG$ with two distinct types of consequences for the corresponding 
phenomenology. The first element is the \textit{absolute} breaking of the full 
symmetry $\hG$ of the form $\lvh$ into the external and internal components of 
equation~\ref{dirprod}. The second element is the apparent breaking of an internal 
symmetry through an impingement on the external spacetime component
  $\bv_4 \inn \TM_4$ of $\bvh$, which is associated with the phenomena of electroweak 
symmetry breaking in the Standard Model. 
   In the discussion of (\cite{Unifi} equation~9.52) it is suggested that the vector  
$\bv_4 \equiv \bh_2 \inn \htwc$  could itself be composed of  spinors under an 
external $\sltc \subset \ee$ symmetry, with an internal $\sutw_L \times \uo_Y \subset 
\ee$ acting directly upon these as well as other spinor components of $\bv_{248}$. 
Hence the explicit structure of the full action of $\ee$ on the predicted form 
$\lvtfe$ may be needed to fully explore electroweak theory and the associated Higgs 
sector within the context of the present theory.

 All Grand Unified Theories, such as the SU(5) model alluded to above,
  involve two widely differing energy  scales and at least two multiplets of Higgs 
scalars with vacuum expectation values of around $10^{15}\,$GeV at the GUT scale and 
$10^{2}\,$GeV at the electroweak scale. Unavoidable interactions between the Higgses 
mixes the mass scales and destroys this hierarchy, with an aesthetically unappealing 
fine tuning of the potential parameters required at each order of perturbation theory 
to maintain the  light scalar Higgs.
  For the present theory while the resemblance between the `mock electroweak theory' 
described in (\cite{Unifi} section 8.3) and standard electroweak theory is
 provisional, the empirically observed light scalar Higgs is associated with local 
fluctuations in the scalar magnitude $\vert \bv_4 \vert = h$ of 
equation~\ref{lorform2}. However, the breaking of the full $\hG$ symmetry in the 
identification of the base space $M_4$,  proposed to ultimately provide the source of 
the full spectrum of Standard Model particle states, does \textit{not} resemble the 
Higgs mechanism.  Hence with two distinct mechanisms of symmetry breaking, and without 
any Higgs interactions associated directly with the breaking of $\hG$, there is no 
interference with the electroweak Higgs and in principle the `hierarchy problem' is 
avoided.  This possibility provides a further area of study for the future development 
of the theory.

  In this paper we have  explored and developed the elementary geometric structure and 
physical interpretation  associated with symmetries of a higher-dimensional  form of 
time $\lvh$
 when broken in the identification of the external  spacetime  manifold $M_4$.
 We  note that beginning with a \textit{single dimension} is perhaps the simplest 
conceivable starting point for a physical theory, and marks a significant contrast 
with Kaluza-Klein theories which generally begin with a complementary motivation in 
positing extra spatial dimensions over and above the 4-dimensional spacetime manifold.  
Unlike the unobserved extra dimensions of a Kaluza-Klein theory here
  founding the theory upon the one-dimensional flow of time offers both a conservative 
and an unambiguous starting point,
 with an interval of time modelled by an interval of the real line $\rrr$  which 
itself necessarily has zero intrinsic curvature.
  While setting out on a firm footing based on time as the 
   fundamental entity the development of this theory exhibits a significant overlap 
with the geometrical framework of non-Abelian Kaluza-Klein theories (including 
\cite{Cho,Kerner,ChMaMa,Orzalesi} and  
 \cite{Kopcz,OrzPau,Kalin,Katan}). The observations of these theories, adapted and 
reinterpreted for the present theory, help to motivate the construction of a linear 
connection $\acute{\Gamma}(p)$ as described for equation~\ref{gamsetkk} defined on a 
principle fibre bundle $P \equiv M_4 \times G$, which itself is identified through the 
broken symmetries of the full form of temporal flow $\lvh$. This structure in turn 
leads to the 
 relationship between the external and internal curvature described in 
equation~\ref{gchift}, derived here for a theory developed from the single initial  
dimension of time alone.


\pagebreak

\par}

\end{document}